\documentclass{article}
\usepackage{amsfonts,latexsym,eucal,amsmath,amsthm,amssymb}
\usepackage{pgf,tikz}
\usetikzlibrary{arrows}
\usepackage{fullpage}
\usepackage{enumerate}
\usepackage{comment}

\theoremstyle{definition}

\usepackage{hyperref}
\usepackage{bm}
\usepackage{subfig}
\usepackage{float}
\usepackage{xcolor}
\usepackage{epstopdf}

\newcommand{\bea}{\begin{eqnarray}}
\newcommand{\eea}{\end{eqnarray}}
\newcommand{\bsa}{\begin{subeqnarray}}
\newcommand{\esa}{\end{subeqnarray}}

\title{Drug release dynamics from a three-layer composite contact lens in the vial, eye wear with blinking, and blister pack settings}

\begin{document}

\author{D.M. Anderson\footnote{Correspondence: danders1@gmu.edu} \ and R.A. Luke \\
Department of Mathematical Sciences\\
George Mason University\\
Fairfax, VA 22030}

\maketitle

\begin{abstract}
In this work we design a multi-layer model of composite contact lens drug release. Such lenses have been designed by encapsulating drug-polymer films in contact lens hydrogels. Composite lenses can promote sustained discharge of drug and achieve near zero-order release kinetics, thus surpassing other ocular delivery methods that are limited by short residence times and an undesirable initial burst release. Our model is informed by \textit{in vivo} data, and includes three coupled partial differential equation layers to simulate the composite lens. We mathematically investigate the effect of composite contact lens design characteristics on the time to 50\% therapeutic drug release ($t_{50}$) in the vial,  eye, and blister pack settings.  In the eye setting, we incorporate our prior model that considers the effect of many blinks on the pre- and post-lens tear film drug concentrations. We simulate drug cumulative release profiles and study the variability of $t_{50}$ across: (1) the ratio of the drug-polymer film to hydrogel  diffusion coefficients, (2) the centerline of the polymer film within the hydrogel, and (3) the polymer film thickness. In the blister pack setting, we study storage questions that may inform future commercial design. This work may help medical professionals better understand the mechanics of contact lens drug delivery and predict targeted tissue transport of ophthalmic drugs.
\end{abstract}


\section{Introduction}

Eye drops are typically used to deliver ophthalmic drugs that treat acute
and chronic eye disorders like glaucoma. However, significant limitations of this method include the short residence time of the drug in the tear film due to drainage and resupply of fresh tears from a blink, and the need for frequent reapplication. Up to 95\% of the drug delivered to the eye via a drop is
lost through the canaliculi, lacrymal sac, nasolacrymal duct and bloodstream \cite{li2006modeling,Bengani_etal_2013,Carvalho_etal_2015}, which may introduce unintended drug side effects \cite{FM1987}. Drug bioavailability is further reduced by the difficulty of correct self-dosing by the patient. 

Drug-eluting contact lenses (CLs) are a potential alternative that can significantly increase drug residence time.  Such lenses have been studied mathematically and experimentally for decades,
but therapeutic options remain limited \cite{franco2021contact, zhao2023therapeutic}. Examples of design strategies include soaking conventional CLs in a drug solution before application (e.g., \cite{phan2016release}), 3-dimensional (3D) printing (e.g., \cite{mohamdeen2022development,garg20253d}), fabrication with an embedded drug-loaded region or element, such as microcavities  (e.g., \cite{manjeri2024hydrogel}) or nanoparticles (e.g., \cite{gulsen2005dispersion}), and encapsulating drug-polymer films in CL hydrogels (e.g., \cite{Ciolino_etal_2009, kudryavtseva2023drug,wei2020design}).   The soaking method greatly improves drug residence time compared to eye drops, but  much of the drug is discharged shortly after placement as a ``burst release'' (uncontrolled release at a nearly infinite initial slope), yielding  minimal extended drug release \cite{karlgard2003vitro, hui2014vitro}. Recently, 3D printing has been explored in proof-of-concept experiments as a drug-eluting lens design strategy to treat conditions such as glaucoma (see, e.g., \cite{mohamdeen2022development,garg20253d}).   As an emerging technology, limitations remain: the lenses 3D printed by Mohamdeen \textit{et al}. \cite{mohamdeen2022development} still exhibited burst release, and for the experiment by Garg \textit{et al}. \cite{garg20253d}, printing resolution restrictions yielded lenses that  are five times as thick as conventional CLs. 

The drug-eluting lens design strategy of encapsulating or ``sandwiching''
drug-polymer films in lens hydrogels addresses the burst release issue.
In this configuration, the drug-laden region takes an annular shape when
viewed from the top down. We refer to this design as a three-layer composite CL. Such a lens can achieve sustained drug release on a time scale of months and minimizes the initial burst release \cite{ciolino2014vivo,pimenta2016diffusion}. As an example, a series of studies by Ciolino and coauthors \cite{Ciolino_etal_2009,ross2019topical,bengani2020steroid,kuang2025overcoming} designed 
therapeutic CLs
with a drug-laden polymer annulus in a hydrogel, examining  cumulative drug release data collected in rabbits. The authors measured drug concentration across 7 days in various ocular compartments to investigate delivery to the retina as a target tissue, comparing the composite lens with eye
drops and drug-soaked CL wear \cite{ross2019topical}; the composite lens demonstrated sustained and nearly zero-order release kinetics (constant, steady release rate yielding a linear profile), taking five times as long to release all drug compared to the drug-soaked conventional lens. 

Several theoretical studies have modeled lens geometry or drug delivery separately in the context of solute or drug transport and drug cumulative release. Kim, Lin, and Radke \cite{kim2022central}  used a finite element solver to simulate  tension and oxygen permeation over the complex geometry introduced by semiconductor-based embedments in soft and scleral CLs; they did not characterize the positioning of the embedments by equations. In terms of drug transport, Pereira-da-Mota \textit{et al}. \cite{pereira2022testing}  proposed an ordinary differential equation (ODE) model for drug release from CLs, monitored the amount of drug in the tear film and the amount absorbed by the ocular tissue as functions of time, and presented many experimental examples of release profiles for
various systems, including both soaking and encapsulating drug loading strategies.
 Toffoletto \textit{et al}. \cite{toffoletto2023physiology}  designed a coupled partial differential equation (PDE) and ODE compartment model of
ophthalmic drug delivery from a uniformly soaked lens to the back of the eye,  comparing model solutions to \textit{in vivo} data from rabbits in \cite{ross2019topical}.
Recently, Carichino \textit{et al}. \cite{carichino2025meta} fit a linear diffusion model to experimental release profiles to estimate the drug diffusion coefficient and time to 50\% therapeutic drug release ($t_{50}$)  for over 50 experiments with lenses loaded using the soaking method.  Ferreira \textit{et al}. \cite{ferreira2010sustained,ferreira2011mathematical} designed a model for a lens comprising a drug-loaded polymer with entrapped drug in silicone particles, solving a linear diffusion equation using Laplace transforms and series expansions. They concluded that the introduction of particles delayed drug release, and studied model parameter influence on mean time to equilibrium, or ``therapeutic effect,'' finding a strong negative relationship between this output metric and diffusion coefficient.

Very few prior mathematical models have  simultaneously characterized  composite lens geometry and drug delivery. In combination with experiments, Pimenta \textit{et al}.  \cite{pimenta2016diffusion} designed a three-layered PDE  model of one-dimensional (in space) diffusion in the inner and outer lens layers in the vial setting. They examined composite lens features  and   varied, one at a time,  the inner layer thickness, diffusivity of the outer/coating layer, and mass transport resistance coefficient. The authors identified values of the aforementioned tunable lens properties that yielded nearly zero order release kinetics for a specific experimental situation. In a series of works, Gudnason and coauthors \cite{gudnason2017numerical,gudnason2018numerical,gudnason2021multi} designed multi-layer linear diffusion models for Franz diffusion cells in which a uniform CL comprised the inner layer. The models were solved using the finite element method, and some incorporated terms to characterize the drug binding and unbinding process or adsorption. The authors varied the diffusion coefficient of the loading compartment and the partition and mass transfer/permeability coefficients to determine their effects on release kinetics. 
The similarities and differences between our contact lens
diffusion model and those of these groups are identified in Section \ref{sec:vial} and examined in detail in Section~\ref{sec-model_discussion}.

To the best of our knowledge, no prior study has examined how simultaneously varying composite lens features  effects drug release kinetics and residence time, nor examined such lens properties in the blister pack or eye wear with blinking settings.
To address this gap, we design a three-layer model of CL drug release to mathematically investigate the effect of lens design characteristics on $t_{50}$ in the vial, eye, and blister pack settings. The eye wear case adapts our prior model \cite{anderson2024mathematical} and considers the effect of many blinks on drug release from the composite lens into pre- and post-lens tear film layers. In all three cases, we study the effect of individually or simultaneously varying (1) the ratio of the drug-polymer film to  hydrogel diffusion coefficients, (2) the centerline of the polymer film within the hydrogel, and (3) the polymer film thickness.

Our paper is organized as follows. In Section \ref{sec:vial_standard} we briefly revisit a model of 
release kinetics of a conventional contact lens. Section \ref{sec:vial} introduces the composite lens configuration in the vial setting, presents the model derivation and nondimensionalization, investigates the effect on $t_{50}$  of  diffusion coefficient ratio, polymer centerline positioning, and polymer film thickness. 
Section \ref{sec:eye} adapts the eye model with blinking of \cite{anderson2024mathematical} for the composite lens setting to study the effect on $t_{50}$ of the same lens parameters, and also considers corneal absorption. In Section \ref{sec:blister}, we study a simplified version of the eye model in the blister pack setting to explore storage-related questions. Section \ref{sec:discussion} includes further discussion, comparison to related work including that of Pimenta \textit{et al}. \cite{pimenta2016diffusion} and Gudnason \textit{et al}. \cite{gudnason2017numerical}, and avenues for future research, and Section \ref{sec:conclusion}  contains our conclusions.

\section{Drug release from a conventional lens: vial}
\label{sec:vial_standard}

Here we briefly revisit the simplest version of one-dimensional diffusion-driven drug release from a CL with uniform properties into perfect sink surroundings (see \cite{carichino2025meta,anderson2024mathematical} and others).
In dimensional form, the equations governing the drug concentration, $c(z,t)$, in the CL are
\bea
\label{eq:basic_dimensional}
\frac{\partial c}{\partial t} & = & D \frac{\partial^2 c}{\partial z^2}, \hspace{0.25in} \mbox{for $0 < z < H_{\rm cl}$ and $t >0$}, \\
c(z=0,t) & = & 0, \nonumber \\
c(z=H_{\rm cl},t) & = & 0, \nonumber \\
c(z,t=0) & = & c_{\rm load}, \nonumber
\eea
where $D$ is the diffusion coefficient of the drug in the CL, $z$ is the depth through the lens, $H_{\rm cl}$ is the thickness of the CL and $c_{\rm load}$ is the initial uniform concentration of the loaded drug.
Typical CLs have thickness $H_{\rm cl} = {\cal O}(100\mu$m$)$ and radius $R_{\rm cl} = {\cal O}(1$cm$)$.  Therefore, the relatively small ratio $H_{\rm cl}/R_{\rm cl} = {\cal O}(10^{-2})$ suggests
a dominance of one-dimensional diffusion in the $z$ direction.

Cumulative drug release dynamics from lenses are of particular interest -- initially in a `vial' setting but also during CL wear and during lens storage. The cumulative drug released as a function of time, $M(t)$, can be expressed as
\bea
M(t) & = & M_{\rm load} - A_{\rm cl} \int_0^{H_{\rm cl}} c(z,t) dz,
\label{eq:M_soak_dim}
\eea
where $M_{\rm load} = c_{\rm load} A_{\rm cl} H_{\rm cl}$ is the total mass of drug initially loaded and $A_{\rm cl}$ is the surface area of the lens.  Note that
in this setting the drug is released from both sides of the CL symmetrically.
The condition in which exactly half of the total drug has been released occurs at time $t=t_{50}$, called the time to 50\% therapeutic release, where $M(t_{50}) = \frac{1}{2} M_{\rm load}$.
This model has been studied and used extensively in the literature on CL drug delivery.  A recent survey by Carichino {\it \textit{et al}.} \cite{carichino2025meta} explored predictions of this type of model for a wide range of ophthalmic drugs and CL combinations.

If we introduce dimensionless variables
\bea
\bar{z} = \frac{z}{H_{\rm cl}},\quad
\bar{t} = \frac{Dt}{H_{\rm cl}^2},\quad
\bar{c} = \frac{c}{c_{\rm load}},
\eea
then the system of equations becomes
\bea
\label{eq:basic_dimensionless}
\frac{\partial \bar{c}}{\partial \bar{t}} & = & \frac{\partial^2 \bar{c}}{\partial \bar{z}^2}, \hspace{0.25in} \mbox{for $0 < \bar{z} < 1$ and $\bar{t} > 0$}, \\
\bar{c}(\bar{z}=0,\bar{t}) & = & 0, \nonumber \\
\bar{c}(\bar{z}=1,\bar{t}) & = & 0, \nonumber \\
\bar{c}(\bar{z},\bar{t}=0) & = & 1,\nonumber
\eea
where
\bea
\frac{M(t)}{c_{\rm load} A_{\rm cl} H_{\rm cl}} & = & 1 - \int_0^{1} \bar{c}(\bar{z},\bar{t}) d\bar{z}.
\eea
Note that in this framework, the value $t_{50}$ is given by
\bea
t_{50} & = & \left( \frac{H_{\rm cl}^2}{D} \right) \bar{t}_{50},
\eea
where the dimensionless quantity $\bar{t}_{50}$ is determined by the condition
\bea
\label{eq:bart50_condition}
\int_0^{1} \bar{c}(\bar{z},\bar{t}_{50}) d\bar{z} & = & \frac{1}{2}.
\eea
The value of $\bar{t}_{50}$ predicted from \eqref{eq:basic_dimensionless} and \eqref{eq:bart50_condition} is a `universal' quantity that does not depend on any system parameters; that is, the dimensionless combination $D t_{50}/H_{\rm cl}^2$ does not depend on any other parameters.
Numerical solution of the problem defined in \eqref{eq:basic_dimensionless} suggests $\bar{t}_{50} \approx 0.0492$, so
\bea
\label{eq:t50basic}
t_{50} \approx 0.0492 (H_{\rm cl}^2/D).
\eea
That is, the one-dimensional diffusion model in \eqref{eq:basic_dimensional} predicts that 
$t_{50}$ depends only on the ratio $H_{\rm cl}^2/D$ (it is linearly proportional to this ratio).
We use this result as a basis for comparison of the performance of the composite CL described below (there we refer to this result as $\bar{t}_{50} = \bar{t}_{50}^{\rm classic} \approx 0.0492$).

\section{Drug release from a composite contact lens: vial}
\label{sec:vial}

\begin{figure}[h!]
    \centering
    \includegraphics[width=0.75\linewidth]{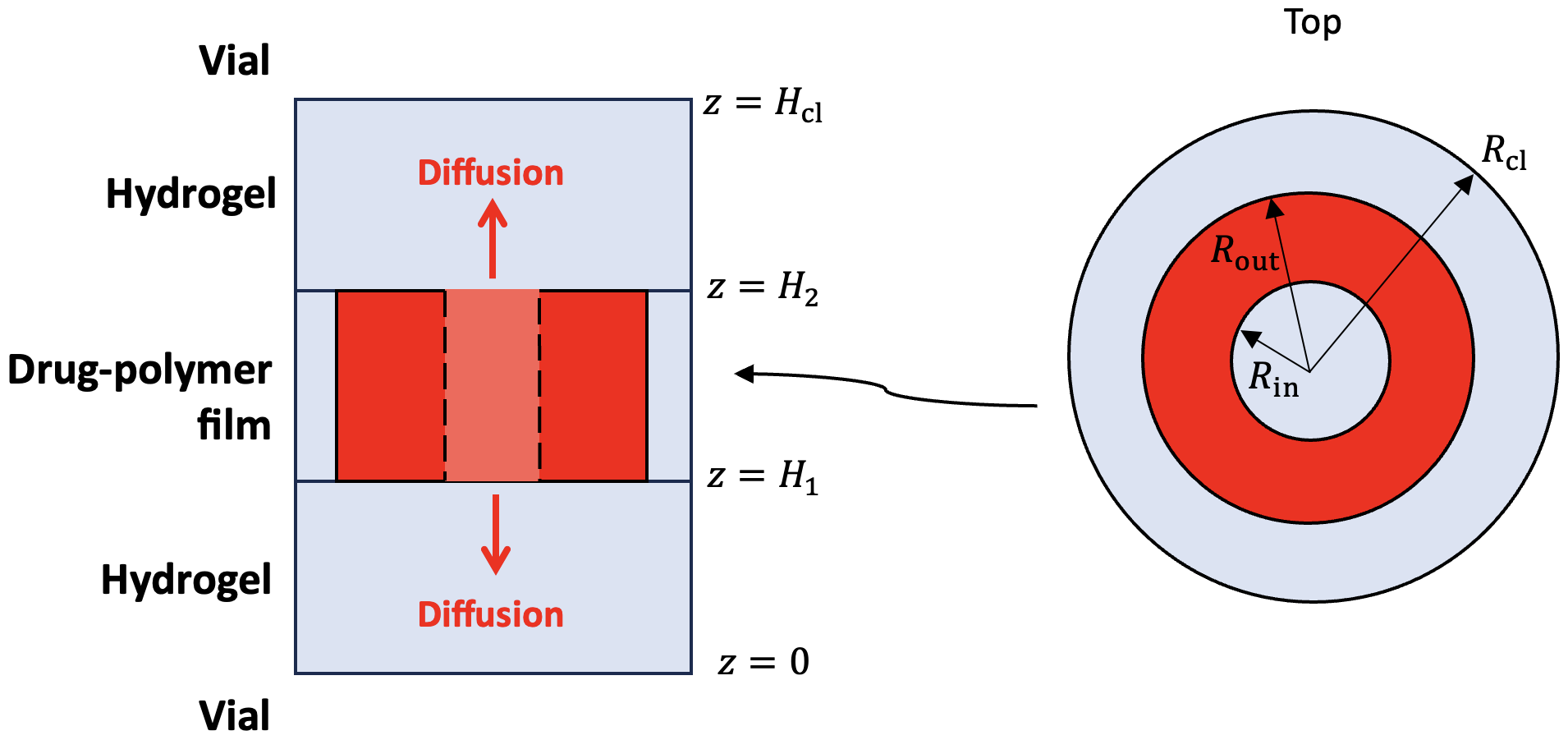}
    \caption{Schematic that describes the geometry of the composite lens model.}
    \label{fig:composite_schematic}
\end{figure}

As noted in the introduction, 
Pimenta \textit{et al}.~\cite{pimenta2016diffusion}  
examined a one-dimensional PDE-based model of diffusion in a three-layer contact lens, given by their equations (3.1)--(3.6). The model we present in this section is similar to theirs, as both versions assume the drug is initially located exclusively in the internal/middle layer, but there are a few relevant differences.   First,  the middle of the three layers, which we refer to as the polymer layer, is not assumed to be centered in the lens (anterior/posterior) in our model, and so we do not impose symmetry of the drug concentration across the lens.  An important consequence is that in our model the anterior and posterior drug release rates need not be the same.
Second, we assume continuity of drug concentration at the internal-layer boundaries 
in the contact lens.  This matches a limiting case of the more general internal boundary condition employed by Pimenta \textit{et al}. \cite{pimenta2016diffusion} that introduced a mass transfer, or permeability, parameter $\alpha$ to account for ``the resistance to mass transport''.
A third difference, with clinical but not technical significance in the context of a one-dimensional diffusion model across the contact lens, is that the cross-sectional area occupied by our polymer insert, $A_{\rm poly}$, need not be the full cross-sectional area of the contact lens, $A_{\rm cl}$ (see Figure~\ref{fig:composite_schematic} for a schematic of the geometry of our model).

The multi-layer diffusion model of Gudnason \textit{et al.} \cite{gudnason2017numerical} is also closely related to ours and that of Pimenta \textit{et al.} \cite{pimenta2016diffusion}. 
Their internal layer boundary conditions include mass transfer rate coefficients \cite{gudnason2017numerical}; we employ a simplified version as explained below.  Their study, and a related one \cite{gudnason2018numerical} that incorporates concentration dynamics of bound and unbound states, examines the drug transfer dynamics in Franz diffusion cells with donor and receptor regions separated by hydrogel/lens-type material and drug release from a pre-loaded lens into a receptor compartment.  In contrast to these donor/receptor-type configurations  in which the drug flux is primarily in a single direction \cite{gudnason2017numerical,gudnason2018numerical}, our configurations explore potentially asymmetric drug release from both sides of a pre-loaded internal layer.  

We outline the details of our model below.  
We then present results for drug release from this composite lens in the vial setting.  Specific connections to the work of Pimenta \textit{et al}. \cite{pimenta2016diffusion} and Gudnason \textit{et al}. \cite{gudnason2017numerical} are provided in a later discussion section \ref{sec-model_discussion}.  

\subsection{Model formulation}
\label{sec:vial_model}

We consider an idealized (hydrogel) CL in the form of a (flat) circular disk with outer radius $r=R_{\rm cl}$ with the bottom (posterior/post-lens) side of the contact
lens  at $z=0$ and the top (anterior/pre-lens) side of the CL at $z=H_{\rm cl}$.  The lens has an
annular insert (polymer) of inner radius $R_{\rm in}$ and outer radius $R_{\rm out}$  positioned between $z=H_1$ and $z=H_2$, where
$0 < H_1 < H_2 < H_{\rm cl}$ (see Figure~\ref{fig:composite_schematic} for a schematic of the geometry).  
The lens is manufactured with the drug mass initially loaded in the annular insert
with specified concentration $C_{\rm load}$.   The hydrogel portion of the lens surrounding the insert is assumed to contain zero drug mass initially.
This geometry and drug-loading configuration mimics the contact lenses with drug-containing polymer inserts examined by Ross {\it \textit{et al}.}~\cite{ross2019topical} and Bengani {\it \textit{et al}.}~\cite{bengani2020steroid}. 
We assume that the drug in the polymer insert is loaded in an axisymmetric and radially-uniform manner so that the drug concentration 
does not vary radially or azimuthally.  
Therefore, we pose a reduced one-dimensional model of this system relevant 
for $R_{\rm in} \le r \le R_{\rm out}$ (and $0 \le \theta \le 2\pi$), with
coupled drug concentration variables 
$C_2^{\rm post}(z,t)$ in the `post-lens' side of the hydrogel $0 < z < H_1$, 
$C_1(z,t)$ in the polymer layer $H_1 < z < H_2$, and
$C_2^{\rm pre}(z,t)$ in the `pre-lens' side of the hydrogel $H_2 < z < H_{\rm cl}$. The drug-polymer film has diffusion coefficient $D_1$; that of the hydrogel is $D_2$. 
In the `vial' setting as outlined below, the distinction between `pre-lens' and `post-lens' sides
of the hydrogel is typically not relevant as both sides are adjacent to a large and well-mixed fluid volume.  
However, the post-lens/pre-lens distinction is highly relevant for contact lenses during wear: in those conditions the pre-lens and post-lens fluid volumes are relatively small and physically separated from each other; for consistency we maintain this distinction in our model.

At the `post-lens'--contact lens interface, $z=0$, we have
\bea
C_2^{\rm post}(0,t) & = & k C_{\rm post}(t),
\eea
where $k$ is a partition coefficient.  
In the vial setting, $C_{\rm post}(t)$ is equivalent to $C_{\rm vial}(t)$ (the concentration of the drug in the bulk vial volume),  more generally understood as the drug concentration in the surrounding fluid at the CL.

In the posterior hydrogel layer of the lens, $0 < z < H_1$,
we have
\bea
\frac{\partial C_2^{\rm post}}{\partial t} & = & D_2 \frac{\partial^2 C_2^{\rm post}}{\partial z^2}.
\eea
\newpage
At the polymer layer -- posterior hydrogel layer interface, $z=H_1$, we assume continuous concentrations and fluxes 
\bea
\label{eq:C_continuity_at_H1}
C_2^{\rm post}(H_1,t) & = & C_1(H_1,t),\\
D_2 \left. \frac{\partial C_2^{\rm post}}{\partial z} \right|_{z=H_1} & = & D_1 \left. \frac{\partial C_1}{\partial z} \right|_{z=H_1}.
\eea

In the polymer layer, $H_1 < z < H_2$ we have
\bea
\frac{\partial C_1}{\partial t} & = & D_1 \frac{\partial^2 C_1}{\partial z^2}.
\eea

At the polymer layer -- anterior hydrogel layer interface, $z=H_2$, we assume continuous concentrations and fluxes
\bea
\label{eq:C_continuity_at_H2}
C_2^{\rm pre}(H_2,t) & = & C_1(H_2,t),\\
D_2 \left. \frac{\partial C_2^{\rm pre}}{\partial z} \right|_{z=H_2} & = & D_1 \left. \frac{\partial C_1}{\partial z} \right|_{z=H_2}.
\eea
The continuous concentration boundary conditions~(\ref{eq:C_continuity_at_H1}) and~(\ref{eq:C_continuity_at_H2}) can be viewed as a limiting case of a more general boundary condition used in related multi-layer diffusion models (e.g.~\cite{pimenta2016diffusion,gudnason2017numerical,gudnason2018numerical}).  Our discussion subsection~\ref{sec-model_discussion} provides further connections to such alternative boundary conditions.

In the anterior hydrogel layer of the lens 
$H_2 < z < H_{\rm cl}$,
we have
\bea
\frac{\partial C_2^{\rm pre}}{\partial t} & = & D_2 \frac{\partial^2 C_2^{\rm pre}}{\partial z^2}.
\eea

At the `pre-lens'--contact lens interface, $z=H_{\rm cl}$,
\bea
C_2^{\rm pre}(H_{\rm cl},t) & = & k C_{\rm pre}(t).
\eea

For the `vial' case currently under consideration, 
we assume that the CL is contained in a vial of volume $V_{\rm vial}$ that is filled with fluid/solution that remains well-mixed.
In this case, the above `pre-lens' and `post-lens' concentrations, $C_{\rm pre}(t)$ and $C_{\rm post}(t)$, are both equal to $C_{\rm vial}(t)$, where
\bea
\label{eq:vial_flux}
V_{\rm vial} \frac{d C_{\rm vial}}{dt} & = & \left( -D_2 \left. \frac{\partial C_2^{\rm pre}}{\partial z}\right|_{z=H_{\rm cl}} 
 + D_2 \left. \frac{\partial C_2^{\rm post}}{\partial z}\right|_{z=0}\right) A_{\rm poly}.
\eea
Here $A_{\rm poly} = \pi (R_{\rm out}^2 - R_{\rm in}^2)$ is the cross sectional area of the polymer inset.   The right-hand-side of \eqref{eq:vial_flux} accounts for the net flux of drug mass out of anterior and posterior sides of the CL into the vial volume.

We assume that all the drug is initially located in the polymer insert at a specified concentration of $C_{\rm load}$.
Then, initial conditions are $C_1(z,0) = C_{\rm load}$ and $C_2^{\rm pre}(z,0) = C_2^{\rm post}(z,0) = C_{\rm vial}(0)=0$.

A central quantity of interest in our model is the cumulative drug released from the lens as a function of time. 
This quantity is defined as
\bea
M(t) & = & M_{\rm load} - A_{\rm poly} \int_0^{H_{\rm cl}} C(z,t) dz,
\label{eq:cumulative_release}
\eea
where $M_{\rm load} = C_{\rm load} A_{\rm poly} (H_2 - H_1)$ represents the initial mass of drug loaded into the polymer insert 
and the notation $C(z,t)$ represents $C_2^{\rm pre}$, $C_1$, or $C_2^{\rm post}$ in the different layers of the contact lens.
The fact that $A_{\rm poly}$  multiplies  the integral term instead of $A_{\rm cl}$  reflects the one-dimensional drug transport in the CL assumption, and the fact that there is zero drug located outside of the annular region $R_{\rm in} \le r \le R_{\rm out}$ in the contact lens.

\begin{table}[h!]
\begin{center}
\begin{tabular}{lllll}
\tabcolsep=0.1in
Parameter & Units & Description  &  Value/Range &  Source  \\ \hline
  $ D_{\rm 1}$ & m$^2$/s & Drug-polymer diffusion coefficient & \begin{tabular}[c]{@{}l@{}} $1.4 \times 10^{-15} - $ \\  $7\times 10^{-12}$ \end{tabular}  & --  \\
 $ D_{\rm 2}$ & m$^2$/s & Hydrogel diffusion coefficient &$7 \times 10^{-13}$ & \cite{anderson2024mathematical} \\
 $H_{\rm cl}$ & $\mu$m & Lens thickness & 300     &\cite{ross2019topical} \\
$t_{50}$ & hours & 50\% therapeutic release time & $21.2 - 273.7$ & -- \\
$R_{\rm in}$ & mm & Drug-polymer inner radius & 3.7  &\cite{ross2019topical,bengani2020steroid} \\
$R_{\rm out}$ & mm & Drug-polymer outer radius & 5.85 &  \cite{ross2019topical,bengani2020steroid} \\
 $H_{\rm  poly} = H_2 - H_1$ & $\mu$m & Drug-polymer thickness & $15 - 285$ & -- \\ 
 $A_{\rm poly} = \pi(R_{\rm out}^2 - R_{\rm in}^2)$ & mm$^2$ & Drug-polymer surface area & 
 {{64.5}} & \cite{ross2019topical,bengani2020steroid} \\
 {{$V_{\rm poly} = A_{\rm poly} (H_2 - H_1) $}} & {{$\mu$L}} & {{Drug-polymer volume}} & {{5.3732}} & \cite{ross2019topical,bengani2020steroid} \\
{{$M_{\rm load}$}} & mg & Loading drug mass & 1.484   &    \cite{ross2019topical} 
\\   
 {{$C_{\rm load} = M_{\rm load}/V_{\rm poly}$}} & mg $\mu$L$^{-1}$ & Loading concentration & 
 {{$0.2762$}}  &  \cite{ross2019topical,bengani2020steroid}  
 \\
$k$ & -- & Partition coefficient  &  {{$0$}} & \cite{phan2021development}   \\
$V_{\rm vial}$ & mL & Vial volume & 5 &          \cite{ross2019topical} \\
 $\tau = H_{\rm cl}^2/D_2$ & hours & Time scale &  35.7   & \cite{ross2019topical,anderson2024mathematical}  \\
 {{$t_{\rm final}$}} & {} & {{Final Time}} &  {{\mbox{up to }$336$}}   &  -- \\
 \hline
\end{tabular}
\end{center}
\caption{{Parameters, descriptions and numerical values or ranges related to the composite lens drug delivery model for the vial setting. For the values of $V_{\rm poly}$ and $C_{\rm load}$ we use $H_2 - H_1 = 83.3$ $\mu$m estimated from \cite{ross2019topical}. Values without references are selected for use in our simulations, or computed as a result.}} 
\label{table-compA}
\end{table}

\subsection{Dimensionless system}
\label{sec-dimensionless_vial}

We introduce dimensionless variables
\bea
\bar{z} = \frac{z}{H_{\rm cl}},\quad
\bar{H}_1 = \frac{H_1}{H_{\rm cl}},\quad
\bar{H}_2 = \frac{H_2}{H_{\rm cl}},\quad
\bar{t} = \frac{t}{\tau},\quad
\bar{t}_{\rm final} = \frac{ t_{\rm final}}{\tau},\quad 
\bar{C} = \frac{C}{C_{\rm load}},
\eea
where the scaled concentration variables apply with all of the same subscripts and superscripts as used previously.  In the vial setting we choose the time scale $\tau = H_{\rm cl}^2/D_2$ which is the time scale associated with diffusion in the hydrogel material.  This is also the natural time scale that arises in the conventional hydrogel model (see Section \ref{sec:vial_standard}). 
In Section~\ref{sec:eye} for our model that involves blinking, a different choice will be made
for $\tau$ to represent the
typical time between blinks (e.g.~$10$ seconds).  The resulting dimensionless system for the vial model is outlined below.

At the `post-lens'--contact lens interface, $\bar{z}=0$ we have
\bea
\bar{C}_2^{\rm post}(0,\bar{t}) & = & k \bar{C}_{\rm post}(\bar{t}).
\label{eq:post_CL_BC}
\eea
In the posterior hydrogel layer of the lens, $0 < \bar{z} < \bar{H}_1$, 
\bea
\frac{\partial \bar{C}_2^{\rm post}}{\partial \bar{t}} & = & \frac{\partial^2 \bar{C}_2^{\rm post}}{\partial \bar{z}^2}.
\label{eq:C2post_diff}
\eea
At the polymer layer -- posterior hydrogel layer interface, $\bar{z}=\bar{H}_1$, we have
\bea
\bar{C}_2^{\rm post}(\bar{H}_1,\bar{t}) & = & \bar{C}_1(\bar{H}_1,\bar{t}),\\
\left. \frac{\partial \bar{C}_2^{\rm post}}{\partial \bar{z}} \right|_{\bar{z}=\bar{H}_1} & = & {\cal D} \left. \frac{\partial \bar{C}_1}{\partial \bar{z}} \right|_{\bar{z}=\bar{H}_1},
\label{eq:C2post_BC}
\eea
where ${\cal D} = D_1/D_2$ represents the diffusion coefficient ratio between the polymer insert and the surrounding hydrogel.
In the polymer layer, $\bar{H}_1 < \bar{z} < \bar{H}_2$, 
\bea
\frac{\partial \bar{C}_1}{\partial \bar{t}} & = & {\cal D} \frac{\partial^2 \bar{C}_1}{\partial \bar{z}^2}.
\label{eq:C1_diff}
\eea
At the polymer layer -- anterior hydrogel layer interface, $\bar{z}=\bar{H}_2$, we have
\bea
\bar{C}_2^{\rm pre}(\bar{H}_2,\bar{t}) & = & \bar{C}_1(\bar{H}_2,\bar{t}),\\
 \left. \frac{\partial \bar{C}_2^{\rm pre}}{\partial \bar{z}} \right|_{\bar{z}=\bar{H}_2} & = & {\cal D} \left. \frac{\partial \bar{C}_1}{\partial \bar{z}} \right|_{\bar{z}=\bar{H}_2}.
 \label{eq:C2pre_BC}
\eea
In the anterior hydrogel layer of the lens, $\bar{H}_2 < \bar{z} < 1$, 
\bea
\frac{\partial \bar{C}_2^{\rm pre}}{\partial \bar{t}} & = & \frac{\partial^2 \bar{C}_2^{\rm pre}}{\partial \bar{z}^2}.
\label{eq:C2pre_diff}
\eea
At the `pre-lens'--contact lens interface, $\bar{z}=1$,
\bea
\bar{C}_2^{\rm pre}(1,\bar{t}) & = & k \bar{C}_{\rm pre}(\bar{t}).
\label{eq:pre_CL_BC}
\eea
The vial concentration satisfies
\bea
\bar{V}_{\rm vial} \frac{d \bar{C}_{\rm vial}}{d\bar{t}} & = &  - \left. \frac{\partial \bar{C}_2^{\rm pre}}{\partial \bar{z}}\right|_{\bar{z}=1} 
 + \left. \frac{\partial \bar{C}_2^{\rm post}}{\partial \bar{z} }\right|_{\bar{z}=0},
\eea
where $\bar{V}_{\rm vial} = V_{\rm vial}/(A_{\rm poly} H_{\rm cl})$.  The well-mixed assumption implies $\bar{C}_{\rm pre}(\bar{t}) = \bar{C}_{\rm post}(\bar{t}) = \bar{C}_{\rm vial}(\bar{t})$.
The initial conditions simplify to $\bar{C}_1(z,0) = 1$ and $\bar{C}_2^{\rm pre}(z,0) =\bar{C}_2^{\rm post}(z,0) = \bar{C}_{\rm vial}(0)=0$.

A seemingly obvious choice for scaling the cumulative drug release is $M_{\rm load}$.  However, since $M_{\rm load} = C_{\rm load} A_{\rm poly} (H_2 - H_1)$ we see that  
for a fixed loading concentration $C_{\rm load}$ and fixed polymer annulus area $A_{\rm poly}$, the amount of loaded drug depends on the polymer thickness $H_2 - H_1$.
With this in mind, we instead enlist the length scale of the total CL thickness, $H_{\rm cl}$, and introduce a dimensionless cumulative drug release quantity $\bar{\cal M} = M/(C_{\rm load} A_{\rm poly} H_{\rm cl})$.  It then follows that
\bea
\label{eq:calMbar}
\bar{\cal M}(\bar{t}) & = & \bar{H}_2 - \bar{H}_1 -  \left( 
\int_0^{\bar{H}_1} \bar{C}_2^{\rm post}(\bar{z},\bar{t}) d\bar{z}
+ \int_{\bar{H}_1}^{\bar{H}_2} \bar{C}_1(\bar{z},\bar{t}) d\bar{z}
+ \int_{\bar{H}_2}^1 \bar{C}_2^{\rm pre}(\bar{z},\bar{t}) d\bar{z}
\right), \nonumber \\
 & = & \int_{\bar{H}_1}^{\bar{H}_2} \left( 1 - \bar{C}_1(\bar{z},\bar{t}) \right) d\bar{z} 
 - \int_0^{\bar{H}_1} \bar{C}_2^{\rm post}(\bar{z},\bar{t}) d\bar{z}
 -  \int_{\bar{H}_2}^1 \bar{C}_2^{\rm pre}(\bar{z},\bar{t}) d\bar{z}.
\eea
 This formulation reveals more explicitly the dependence on the geometry of the polymer layer through the dimensionless parameters $\bar{H}_1$ and $\bar{H}_2$.
 
 It will also be convenient to introduce dimensionless thickness and midpoint quantities
\bea
\label{eq:Delta_barH_and_barH_mid}
\Delta \bar{H} \equiv \bar{H}_2 - \bar{H}_1,\quad
\bar{H}_{\rm mid} & = & \frac{\bar{H}_1 + \bar{H}_2}{2} = \bar{H}_1 + \frac{1}{2} \Delta \bar{H},
\eea
where we note that the geometric constraint $0 < \bar{H}_1 < \bar{H}_2 < 1$ can be expressed as
\bea
0 < \frac{1}{2} \Delta \bar{H} < \bar{H}_{\rm mid} < 1 - \frac{1}{2} \Delta \bar{H} < 1.
\eea

The assumption that no drug remains in the CLin the limit $\bar{t} \rightarrow \infty$ (i.e. $\bar{C}_2^{\rm pre} \rightarrow 0$, $\bar{C}_1 \rightarrow 0$,
and $\bar{C}_2^{\rm post} \rightarrow 0$) implies that $\bar{\cal M} \rightarrow \Delta \bar{H}$ as $\bar{t} \rightarrow \infty$.
As noted earlier, a quantity of interest is the time at which 50\% of the drug
has been released from the contact lens, $t_{50}$.  In dimensional terms this condition, stated as $M(t = t_{50}) = \frac{1}{2} M_{\rm load}$, defines the time $t_{50}$.
In terms of dimensionless quantities, we derive that
\bea
M(t=t_{50}) & = & \frac{1}{2} M_{\rm load}, \nonumber \\
(C_{\rm load} A_{\rm poly} H_{\rm cl}) \bar{{\cal M}}(\bar{t}_{50};\bar{H}_{\rm mid},\Delta \bar{H},{\cal D}) & = & \frac{1}{2} C_{\rm load} A_{\rm poly} (H_2 - H_1),\nonumber \\
\bar{{\cal M}}(\bar{t}_{50};\bar{H}_{\rm mid},\Delta \bar{H},{\cal D}) & = & \frac{1}{2} \Delta \bar{H}.
\eea
This reveals that under the perfect sink assumption where the partition coefficient, $k$, does not play a role, $\bar{t}_{50}$ depends only on $\bar{H}_{\rm mid}$, $\Delta \bar{H}$ and ${\cal D}$.  
It follows that the dimensional value $t_{50}$ takes the form
\bea
\label{eq:t50_formula}
t_{50} & = & \bar{t}_{50}(\bar{H}_{\rm mid},\Delta \bar{H},{\cal D}) \left( \frac{H_{\rm cl}^2}{D_2} \right).
\eea
It will be of interest to determine the value of $\bar{t}_{50}$ for different values of $\bar{H}_{\rm mid}$, $\Delta \bar{H}$, and ${\cal D}$ especially in comparison
to the result in \eqref{eq:t50basic} identified for drug release from a conventional lens (no polymer insert).
While quantities $M_{\rm load}$, $C_{\rm load}$ and $A_{\rm poly}$ do not appear in the expression determining the value $\bar{t}_{50}$, 
it is important to recognize their relationship to $\Delta \bar{H}$, which is
\bea
\frac{M_{\rm load}}{C_{\rm load} A_{\rm poly} H_{\rm cl}} & = & \Delta \bar{H}.
\eea
That is, if one wanted to address a situation in which $M_{\rm load}$, $C_{\rm load}$, $A_{\rm poly}$, and $H_{\rm cl}$ were fixed, 
then the value of $\Delta \bar{H}$ is not adjustable in this context.

 \subsection{Numerical investigation}

 We discretize the diffusion
equation in space in the three layers of the CL and solve in time using the method of lines via
Matlab’s \verb|ode15s| solver. Second-order accurate finite difference approximations using equally-spaced points are implemented for the spatial derivatives.  We begin with comparisons to experimental drug release data from the composite lens system of Ross~{\it \textit{et al}.}~\cite{ross2019topical}.

\subsubsection{Comparison with experiments from Ross \textit{et al.} (2019)}

\begin{figure}[h]
\centering
\subfloat[][Drug concentration in three-layered lens]{\includegraphics[width=0.45\linewidth]{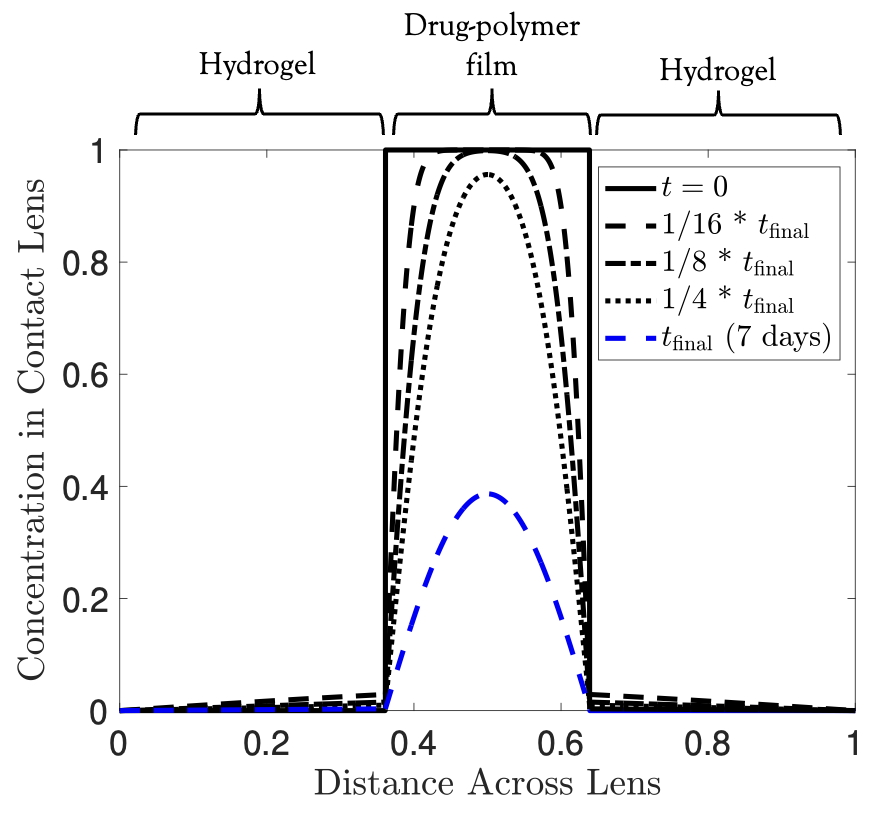}
\label{fig:base_case_CL}} \subfloat[][Cumulative drug release]{\includegraphics[width=0.495\linewidth]{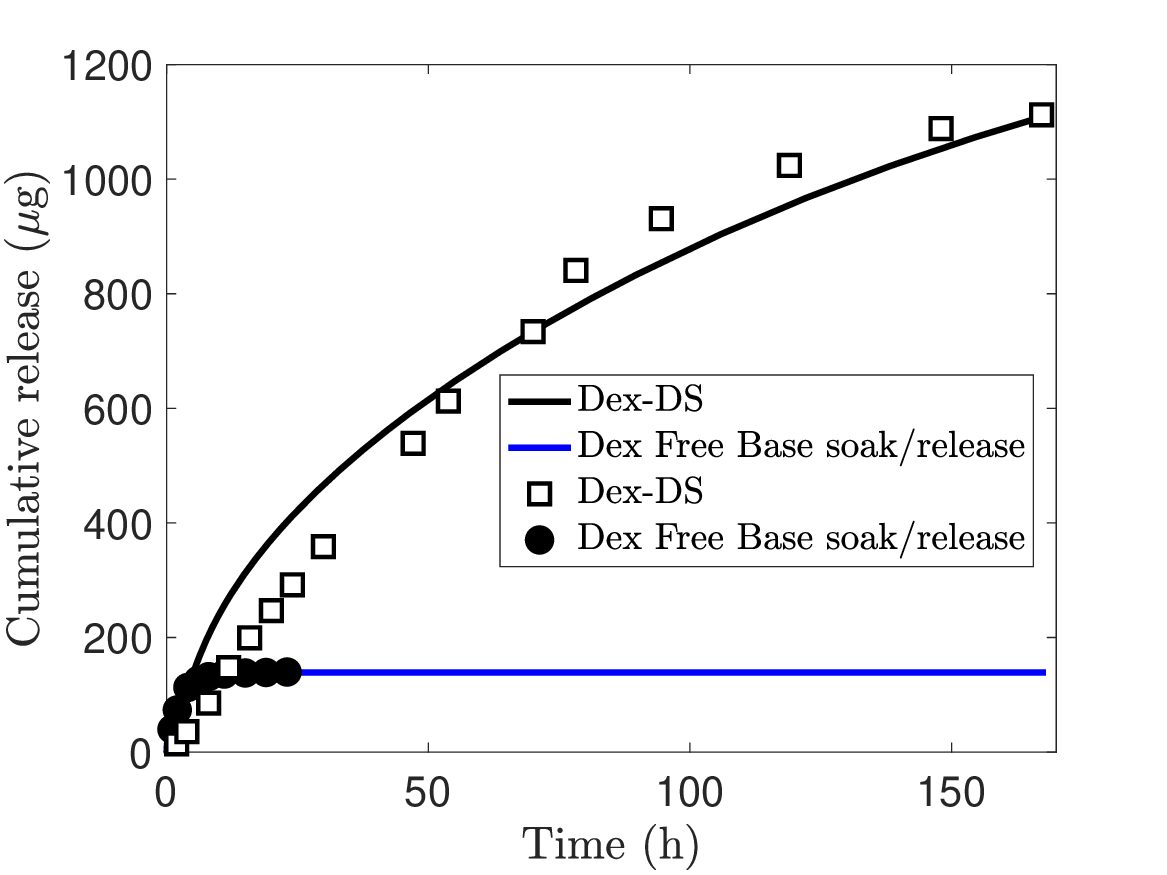}
\label{fig:base_case_release}}
\caption{Vial setting: simulated drug concentration and release profiles calibrated to fit data from Ross \textit{et al}.  \cite{ross2019topical} for a uniformly loaded conventional hydrogel CL (Dex Free Base soak/release) and from the encapsulated drug configuration (Dex-DS). In the Dex Free Base soak/release case, the initial mass of drug loaded is $M_{\rm load} = 139$ $\mu$g and the  lens surface area is $A_{\rm cl} = \pi R_{\rm cl}^2 \approx 186.3$ mm$^2$ using $R_{\rm cl} = 7.7$ mm \cite{ross2019topical}.  In the Dex-DS case, 
$M_{\rm load} = 1484$~$\mu$g and the relevant surface area for drug release is $A_{\rm poly} \approx 64.5$ mm$^2$ (see Table~\ref{table-compA}).  }
\label{fig:base_case}
   \end{figure}

It is first instructive to examine solutions corresponding to the system in Ross {\it \textit{et al}.} \cite{ross2019topical}.  Numerical solutions of the model are computed with representative parameters corresponding to experiments ($ k = 0, t_{\rm final} = 7$ days, $\Delta \bar{H} \approx 0.278, \bar{H}_{\rm mid}=0.5, M_{\rm load} = 1484 \ \mu$g) and are shown in Figure~\ref{fig:base_case}. The nondimensional concentration in the lens is shown in Figure~\ref{fig:base_case_CL} at several time points up to the final dimensional time of 7 days. 
The profiles are symmetric about the centerline of the lens, as we have centered the drug-polymer film in this example. Further, a substantial amount of drug remains in the drug-polymer film at the final time, but the concentration in the hydrogel layer is very small. This observation is consistent with the relative sizes of the diffusion coefficients of the two lens components, discussed next.

In Figure~\ref{fig:base_case_release} we simulate release over 7 days for our three-layered model and for a one-layer model of ``soak and release'' given by \eqref{eq:basic_dimensional} for
a conventional hydrogel, for comparison. Cumulative release for the conventional hydrogel is computed via \eqref{eq:M_soak_dim}. Our primary comparison is with the experimental rabbit data from Ross \textit{et al}.  \cite{ross2019topical}, shown as white squares or black circles in Figure~\ref{fig:base_case_release}.  They reported cumulative drug release from the two aforementioned lens types.\footnote{The `Dex-DS' corresponds to what we refer to as the three-layer composite lens and the `Dex Free Base soak/release' corresponds to a conventional lens.} Data are obtained from Ross \textit{et al.}'s Figure 2 \cite{ross2019topical} via \verb|Matlab|'s image analysis software \verb|grabit.m| (MathWorks, Natick, MA, USA).  Model parameters, namely diffusion coefficients, are hand-tuned to fit this data in Figure~\ref{fig:base_case_release}; we find $D_{\rm 2} = 7 \times 10^{-13}$ m$^2$/s (hydrogel) and $D_{\rm 1} = 1.4 \times 10^{-15}$ m$^2$/s (drug-polymer film). Note that this yields the ratio $\mathcal{D} = D_1/D_2 = 0.002$.

\begin{figure}[h!]
\centering
\subfloat[][Narrower film (x 1/2)]{\includegraphics[width=0.33\linewidth]{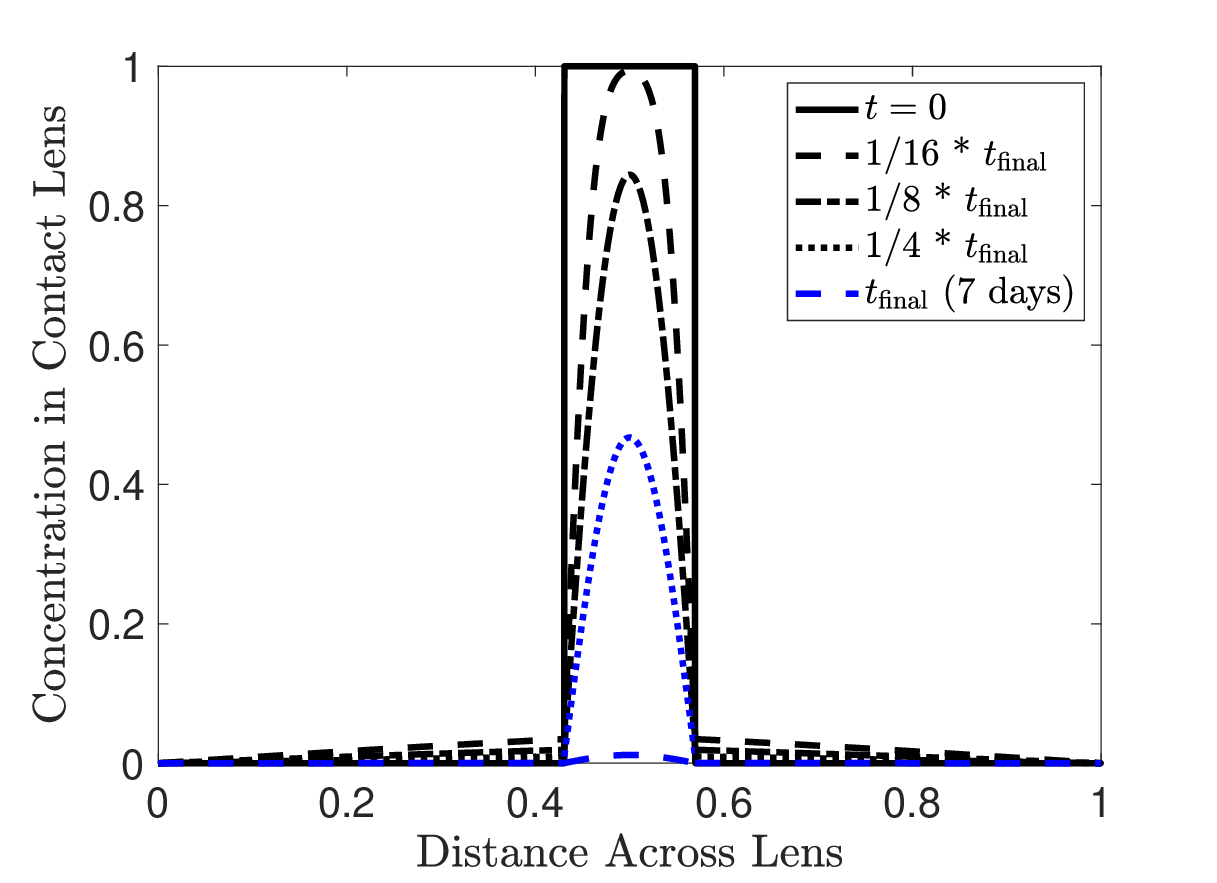}\label{fig:narrower_film}}
\subfloat[][Wider film (x 2)]{\includegraphics[width=0.32\linewidth]{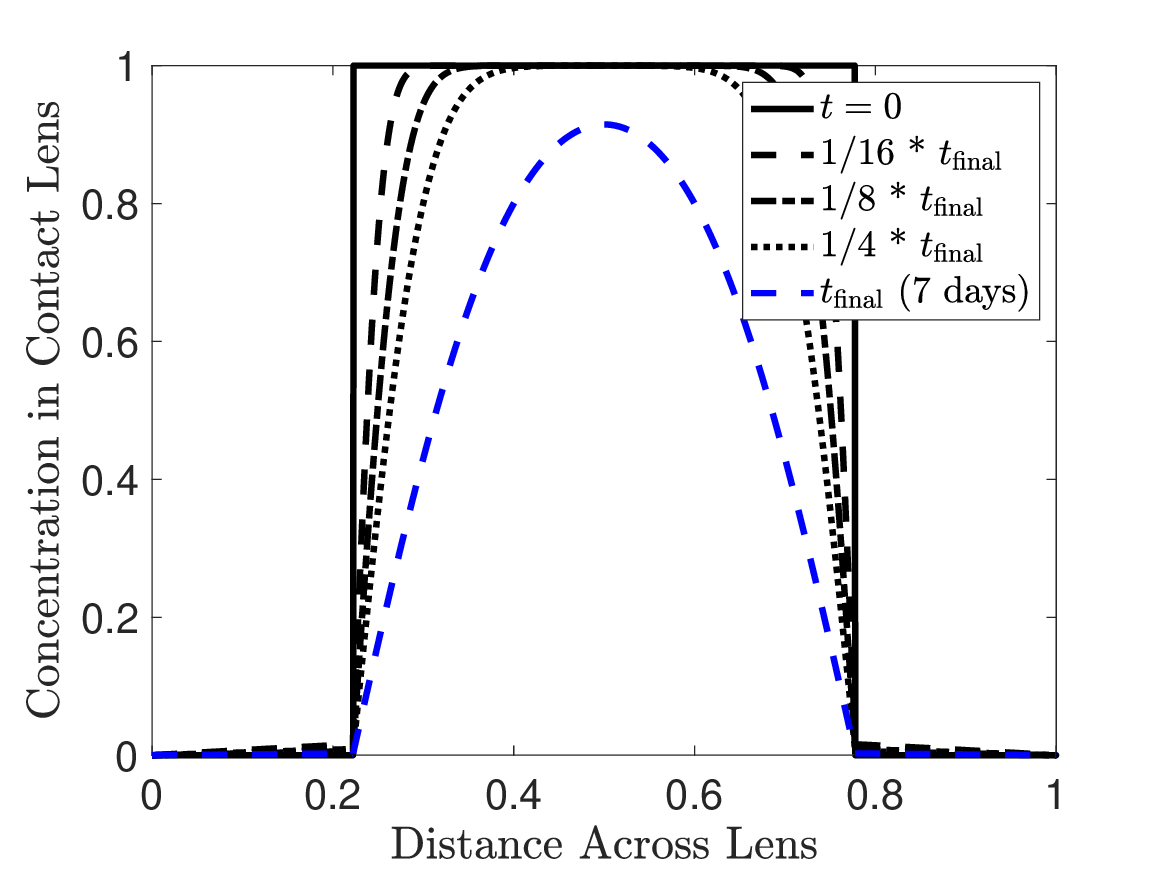}\label{fig:wider_film}}
\subfloat[][Vary film thickness]{\includegraphics[width=0.32\linewidth]{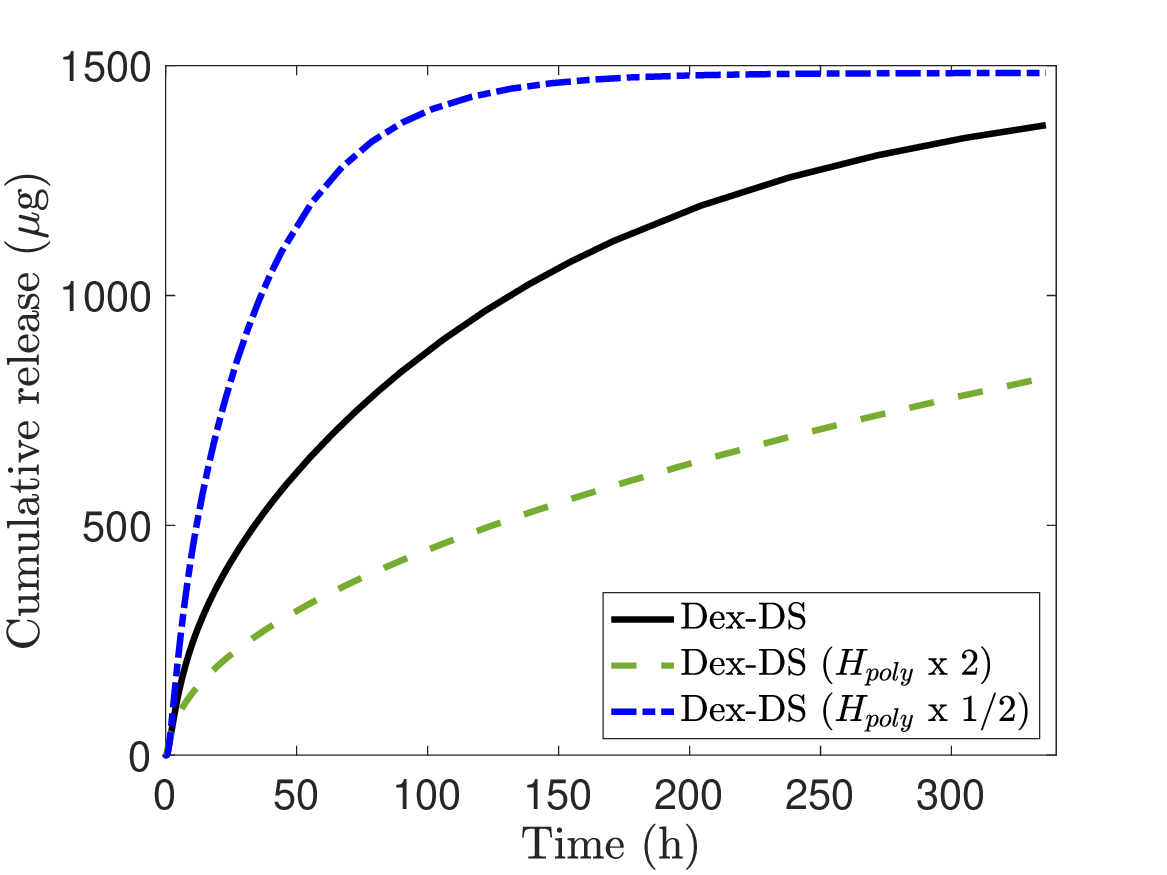} \label{fig:release_vary_thickness}}
\caption{Vial setting: simulated drug concentration in the lens for drug-polymer films that are (a) half the width of that in Figure~\ref{fig:base_case} and (b) twice the width of that in Figure~\ref{fig:base_case}. (c) Effect of varying drug-polymer thickness on cumulative drug release.
}
\label{fig:vary_lens}
 \end{figure}

We continue to use parameter values pertaining to Ross \textit{et al.} \cite{ross2019topical}, but investigate the effect  of changing model parameters on drug concentration and cumulative release.
Figure \ref{fig:vary_lens} shows the effect of increasing and decreasing the drug-polymer film thickness, $H_{\rm poly} = H_2 - H_1$, by a factor of two on the drug concentration in the lens. A narrower drug-polymer film ($\Delta \bar{H} = 0.139$, Figure~\ref{fig:narrower_film})  releases nearly all drug after 7 days, while a wider drug-polymer film ($\Delta \bar{H} = 0.555$, Figure~\ref{fig:wider_film}) retains much of the drug in the lens after 7 days.  
Figure~\ref{fig:release_vary_thickness} compares the corresponding cumulative drug release for the base case to the narrower and wider drug-polymer film situations. To motivate a thorough investigation of the dependence of $t_{50}$ on lens parameters in the next subsection, we compare the values for the situations in Figure~\ref{fig:release_vary_thickness}. We find that for the base case Dex-DS (solid black curve), $t_{50} = 71.5$ hours. If the drug-polymer film is widened by a factor of 2 (dashed green curve), $t_{50} = 273.7$ hours; if it is narrowed by a factor of 2 (dash-dot blue curve), $t_{50} = 21.2$ hours. Thus, there is an interesting,  positive, nonlinear relationship between drug-polymer thickness and $t_{50}$ for the ${\cal D}=0.002$ case. 

Recall that in dimensional terms we have $M_{\rm load} = C_{\rm load} A_{\rm poly} (H_2 - H_1)$. Thus, comparing two lenses with different polymer thicknesses, one can hold the total loaded mass $M_{\rm load}$ fixed (in which case either $C_{\rm load}$ and/or $A_{\rm poly}$ would differ between
these two lenses), or, hold the concentration, $C_{\rm load}$, fixed (in which case either $M_{\rm load}$ and/or $A_{\rm poly}$ would differ between
these two lenses).

 \begin{figure}[h!]
\centering
\subfloat[][Vary loaded mass of drug]{\includegraphics[width=0.48\linewidth]{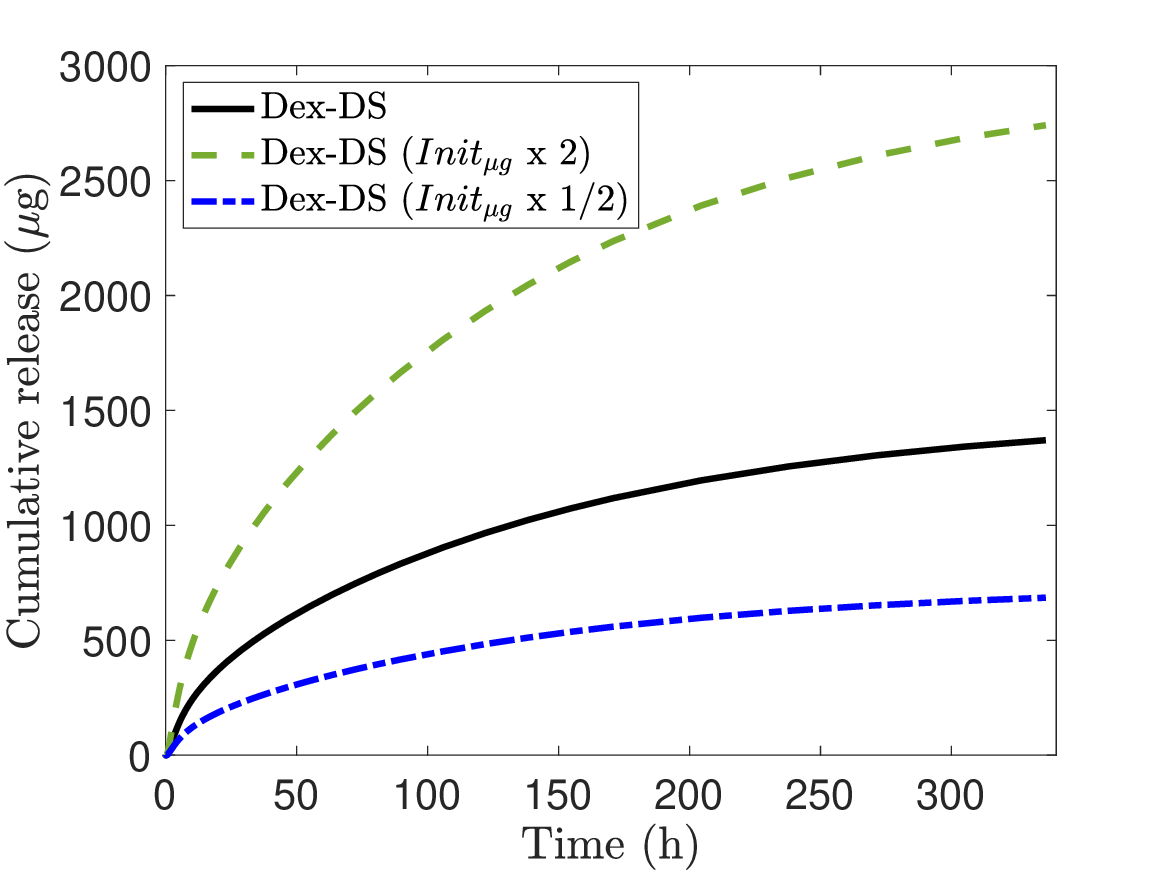}\label{fig:release_vary_load}} \hspace{-5mm}
\subfloat[][Vary thickness and loaded drug]{\includegraphics[width=0.48\linewidth]{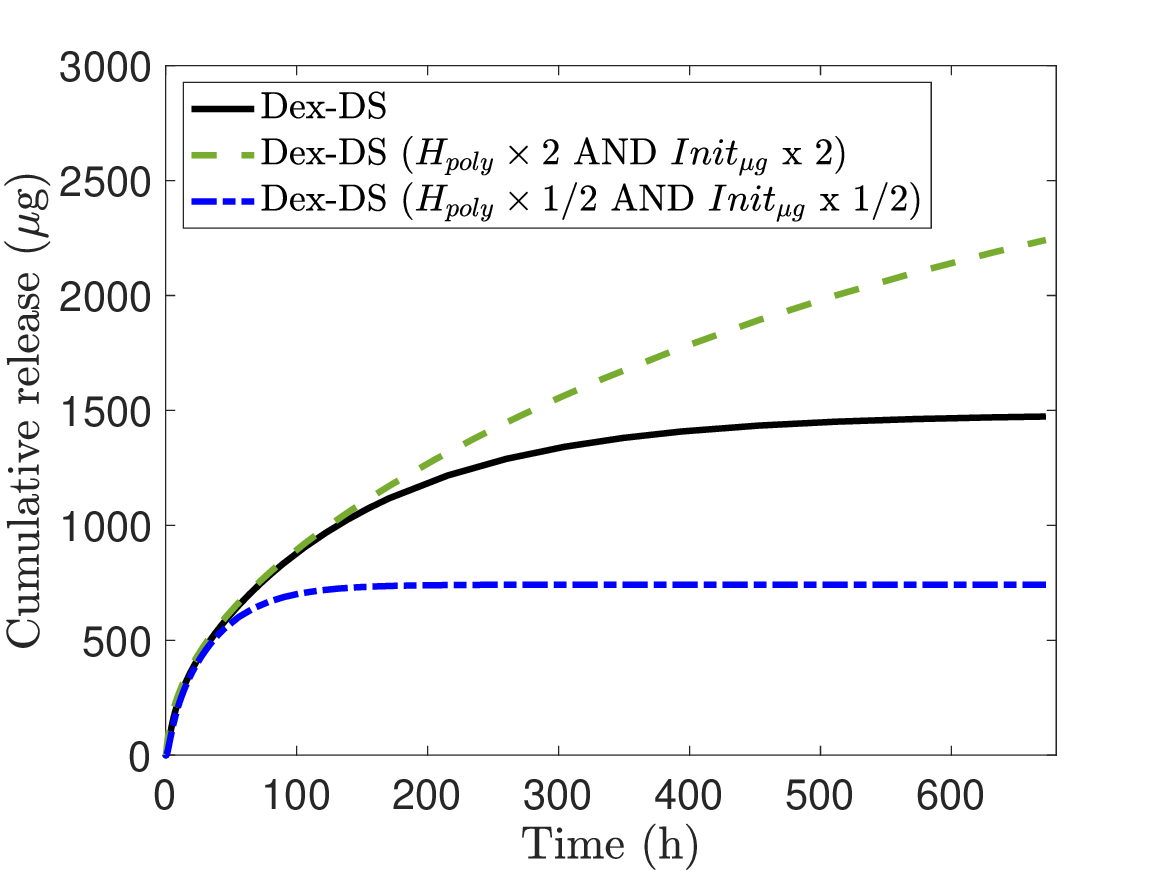}\label{fig:release_vary_both}}
\caption{Vial setting: cumulative drug release profiles. Effects of varying  (a)  the initial drug mass loaded and (b)  both drug-polymer thickness and initial drug mass loaded. }
\label{fig:vary_drug}
 \end{figure}

 In Figure~\ref{fig:vary_drug}, which shows cumulative release over two and four week periods (336 and 672 hours), we study the effect on cumulative release of varying (a) the initial drug loading amount and (b) both the loading amount and drug-polymer film thickness. 
 Figure~\ref{fig:release_vary_load} shows that a lens loaded with half 
 or twice the original mass releases nearly all drug in two weeks. 
 As expected, changing the initial drug loading amount does not affect $t_{50}$, since the cumulative release scales with the load value.
 Figure~\ref{fig:release_vary_both} shows that a narrower drug-polymer film and lens loaded with half the drug releases all of the drug in a week, whereas a wider film and lens loaded with twice the drug is still releasing drug after a month.

\subsubsection{Effect of model parameters on time to 50\% therapeutic drug release}

\begin{figure}[h!]
\begin{center}
\vskip 0.05in
\includegraphics[width=.49\linewidth]{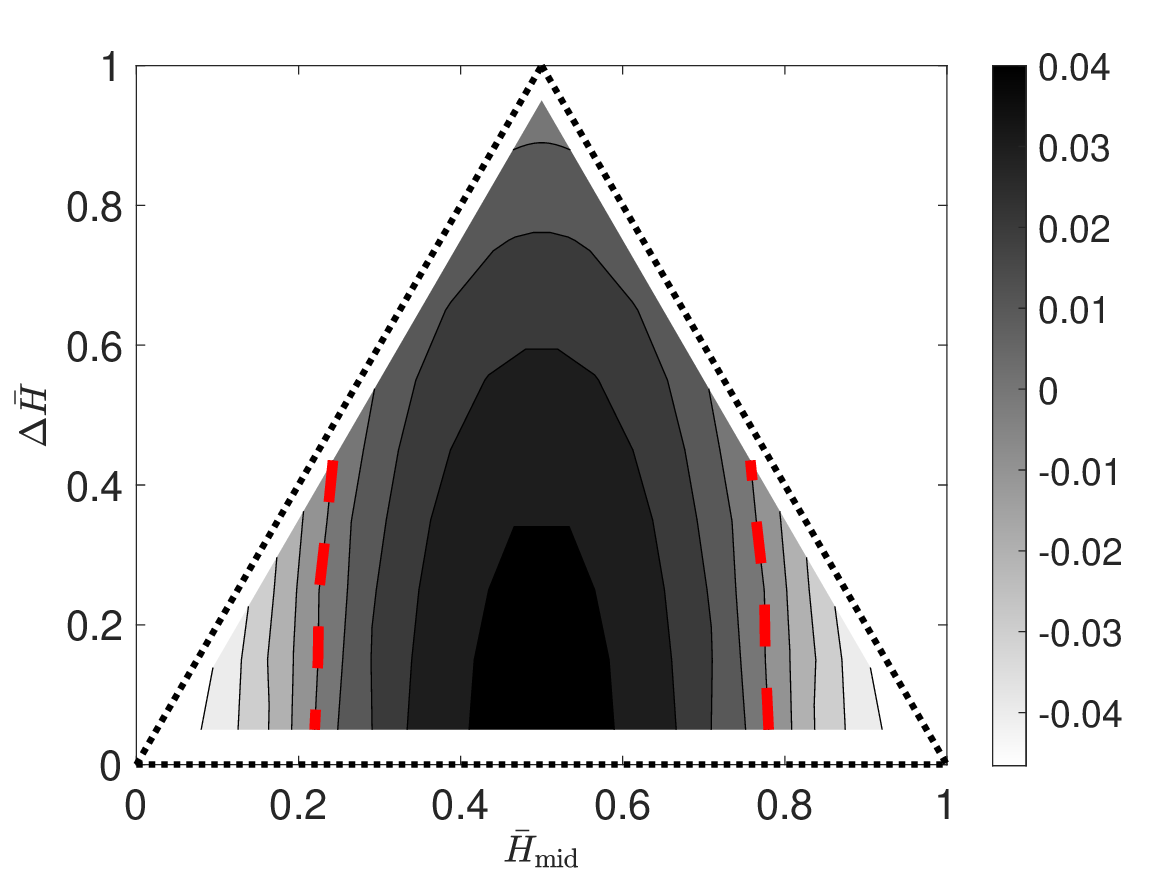}
\caption{Vial setting: contour plot of $\bar{t}_{50} - \bar{t}_{50}^{\rm classic}$ for the case ${\cal D}=1$ corresponding to the polymer and hydrogel 
regions having the same diffusion coefficient (recall $\bar{t}_{50}^{\rm classic} \approx 0.0492$).   
The thick red-dashed lines indicate the zero contour where $\bar{t}_{50} = \bar{t}_{50}^{\rm classic}$.
The dotted lines indicate the extreme values of the parameters $\bar{H}_{\rm mid}$
and $\Delta \bar{H}$ corresponding to zero thickness polymer layer (bottom of triangle), zero thickness postlens hydrogel layer (left side of triangle),
and zero thickness prelens hydrogel layer (right side of triangle).  The shaded portion shows the region of parameter space for which we have computed
solutions. }
\label{fig_303_Fig}
\end{center}
\end{figure}

In the previous subsection, we explored the dependence of $t_{50}$ on $\Delta \bar{H}$ for cases with ${\cal D}=0.002$ and $\bar{H}_{\rm mid}=0.5$.
In \eqref{eq:t50_formula} we observed that the dimensionless
$\bar{t}_{50}$ value depends on 
$\bar{H}_{\rm mid}$, $\Delta \bar{H}$, and ${\cal D}$; here we explore the dependence as all three parameters are varied.
As a first general observation, the mathematical region in parameter space $(\bar{H}_{\rm mid},\Delta \bar{H})$ relevant to the present lens geometry is the interior space of a triangle
whose base is $\Delta \bar{H}=0$ for $0 \le \bar{H}_{\rm mid}  \le 1$ and whose apex has $(\bar{H}_{\rm mid},\Delta \bar{H}) = (1/2,1)$.   An example is shown
in Figure~\ref{fig_303_Fig} for the case in which the diffusion coefficient in the polymer layer matches that in the hydrogel layer, ${\cal D}=1$.
For our three-layered model each layer must have nonzero thickness, which forces the parameter values to be chosen inside this triangle.  
Presumably, design parameters for actual lenses may be further restricted by minimum thickness requirements of any individual layer. For the present purposes, we take the example in Figure~\ref{fig_303_Fig} to contain
the feasible design space.
For interpretation, however, parameter values near the bottom of the triangle ($\Delta \bar{H}=0$) represent a `thin' polymer limit, 
parameter values near the left-side of the triangle ($\Delta \bar{H} = 2 \bar{H}_{\rm mid}$) represent a `thin' post-lens hydrogel layer, and
parameter values near the right-side of the triangle ($\Delta \bar{H} = 1 - 2 \bar{H}_{\rm mid}$) represent a `thin' pre-lens hydrogel layer.

In the vial case with perfect sink conditions, the drug release dynamics are symmetric with respect to the line $\bar{H}_{\rm mid} = 0.5$,
since the pre-lens and post-lens hydrogel layers  have the same material properties. 
Since ${\cal D}=1$ in Figure~\ref{fig_303_Fig}, this case allows the direct assessment of the geometric positioning of a drug insert on the drug release dynamics.  
The two red-dashed lines in this plot show the parameter locations for which the predicted $\bar{t}_{50}$ value exactly matches that of a uniformly loaded, conventional contact lens.    
The darker shaded regions indicate parameter space where the drug release is relatively slower and the lighter shaded regions indicate parameter space where the drug 
release is relatively faster.  
We observe that when ${\cal D}=1$ a drug-polymer layer centered in the lens give the slowest release rates for any given polymer thickness $\Delta \bar{H}$.  While this may not seem surprising, we demonstrate below that centering the polymer (i.e.~$\bar{H}_{\rm mid}=0.5$) does not always yield the largest value of $\bar{t}_{50}$ for other values of ${\cal D}$.  Also, the drug is released at a faster rate than that of  a conventional lens only when a thin polymer layer is located
near the post-lens side (lower left portion of triangle) or near the pre-lens side (lower right portion of the triangle).
A final observation for the ${\cal D}=1$ example is that for a centered polymer layer ($\bar{H}_{\rm mid}=0.5$) the predicted $\bar{t}_{50}$ value {\it decreases} with increasing polymer thickness $\Delta \bar{H}$.

A careful reader will notice that two \textit{different} trends in $\bar{t}_{50}$ 
result from varying the polymer layer thickness, $\Delta \bar{H}$, for a symmetrically-placed polymer layer ($\bar{H}_{\rm mid}=0.5$).
On one hand, 
in Figure~\ref{fig:release_vary_thickness} where ${\cal D}=0.002$, 
increasing $\Delta \bar{H}$ when $\bar{H}_{\rm mid}=0.5$ leads to an {\it increase} in the predicted value of $\bar{t}_{50}$.  On the other hand, in 
Figure~\ref{fig_303_Fig} where ${\cal D}=1$, we just observed that increasing $\Delta \bar{H}$ when $\bar{H}_{\rm mid}=0.5$ leads to a {\it decrease} in the predicted value of $\bar{t}_{50}$.  This suggests a non-trivial interplay between the three main parameters: $\bar{H}_{\rm mid}$, $\Delta \bar{H}$ and ${\cal D}$, on the predicted value of $\bar{t}_{50}$.  We explore this further below.

\begin{figure}[h!]
\begin{center}
\vskip 0.05in
\includegraphics[width=.47\linewidth]{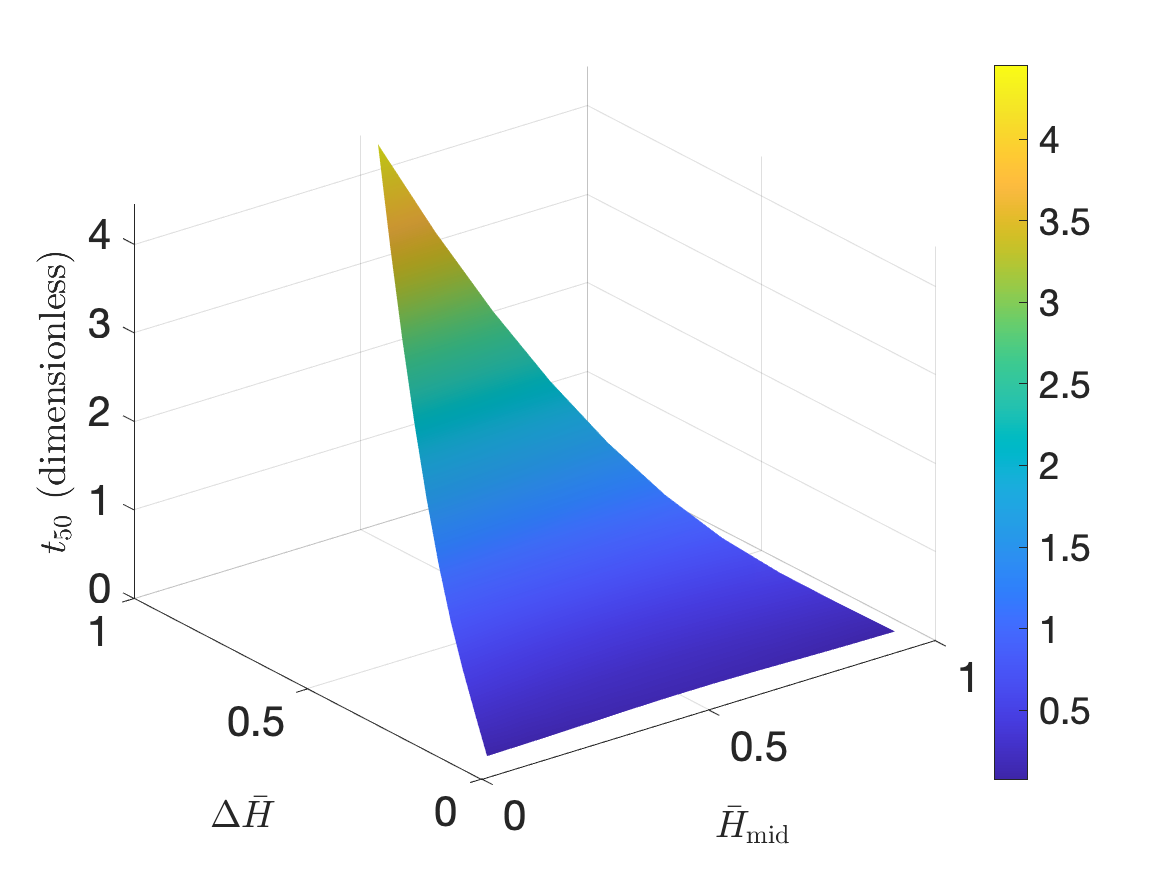}
\includegraphics[width=.47\linewidth]{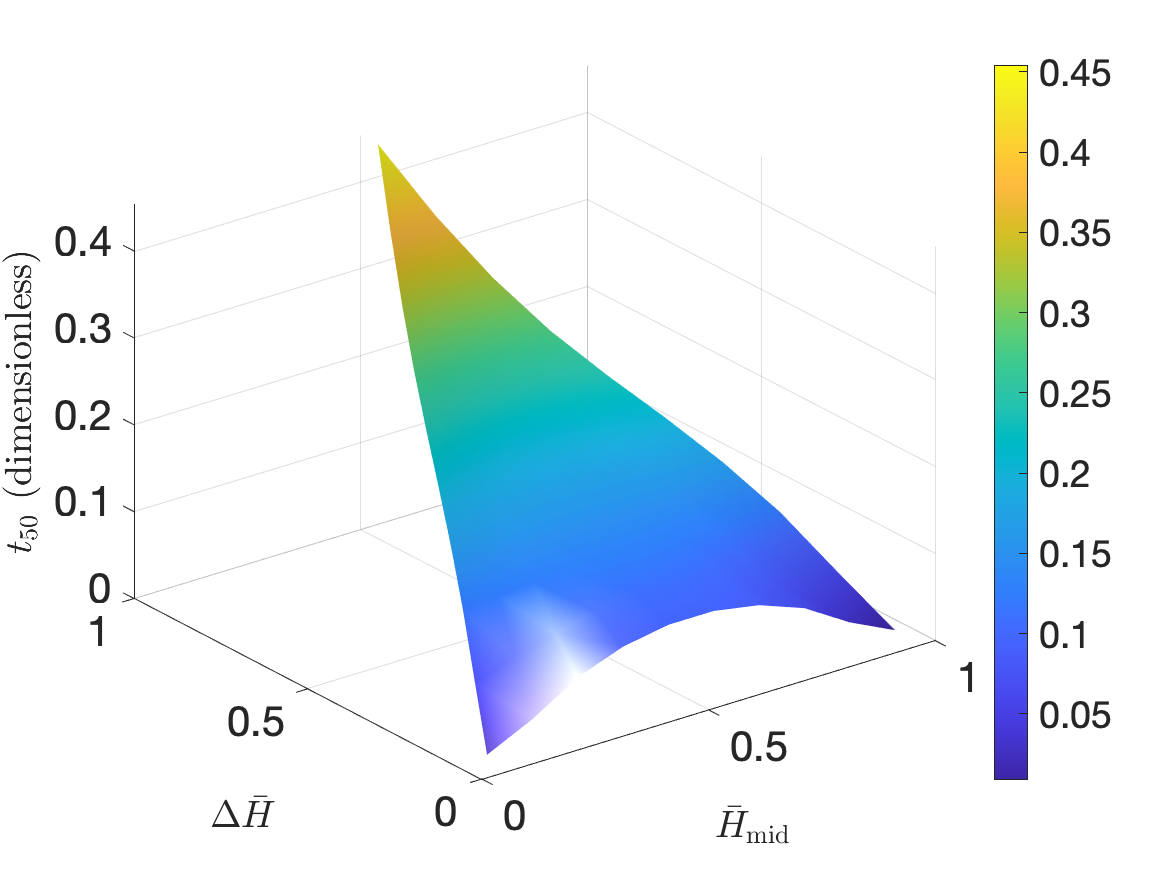} \\
\includegraphics[width=.47\linewidth]{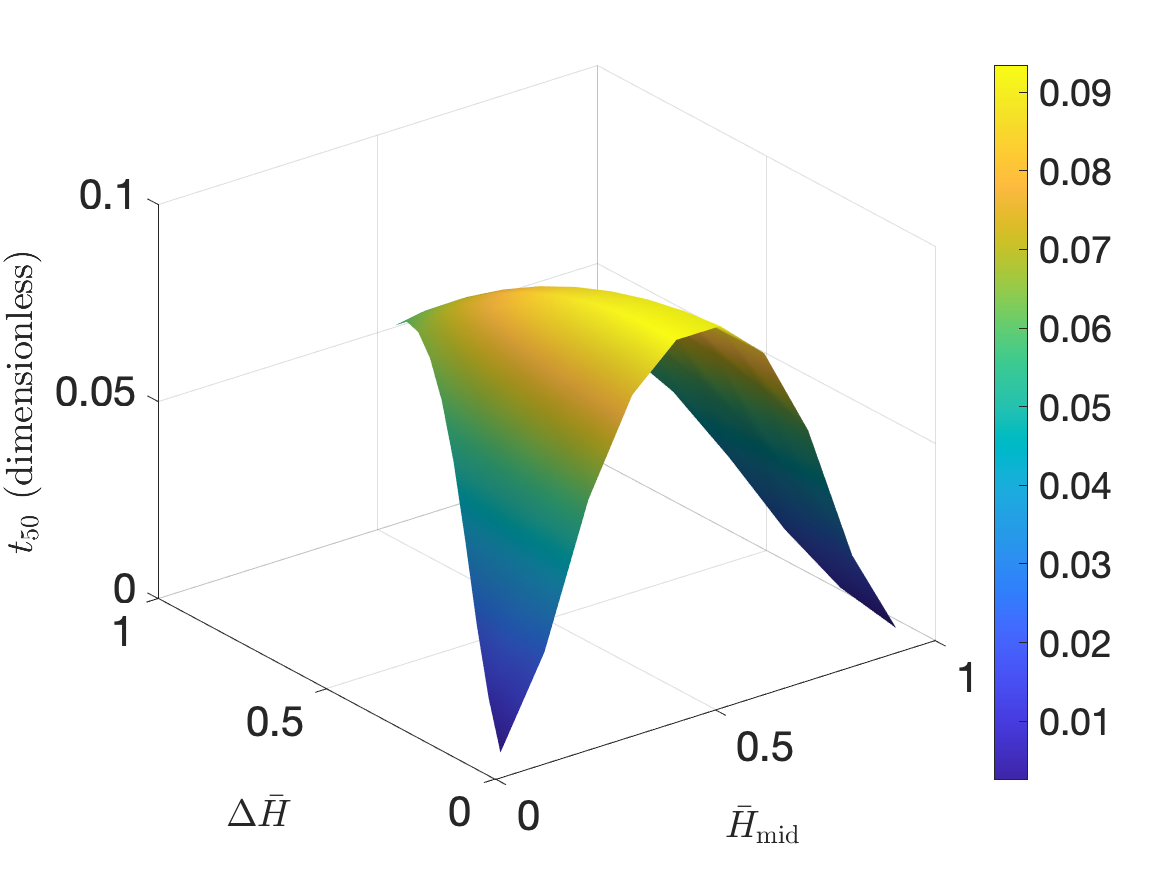}
\includegraphics[width=.47\linewidth]{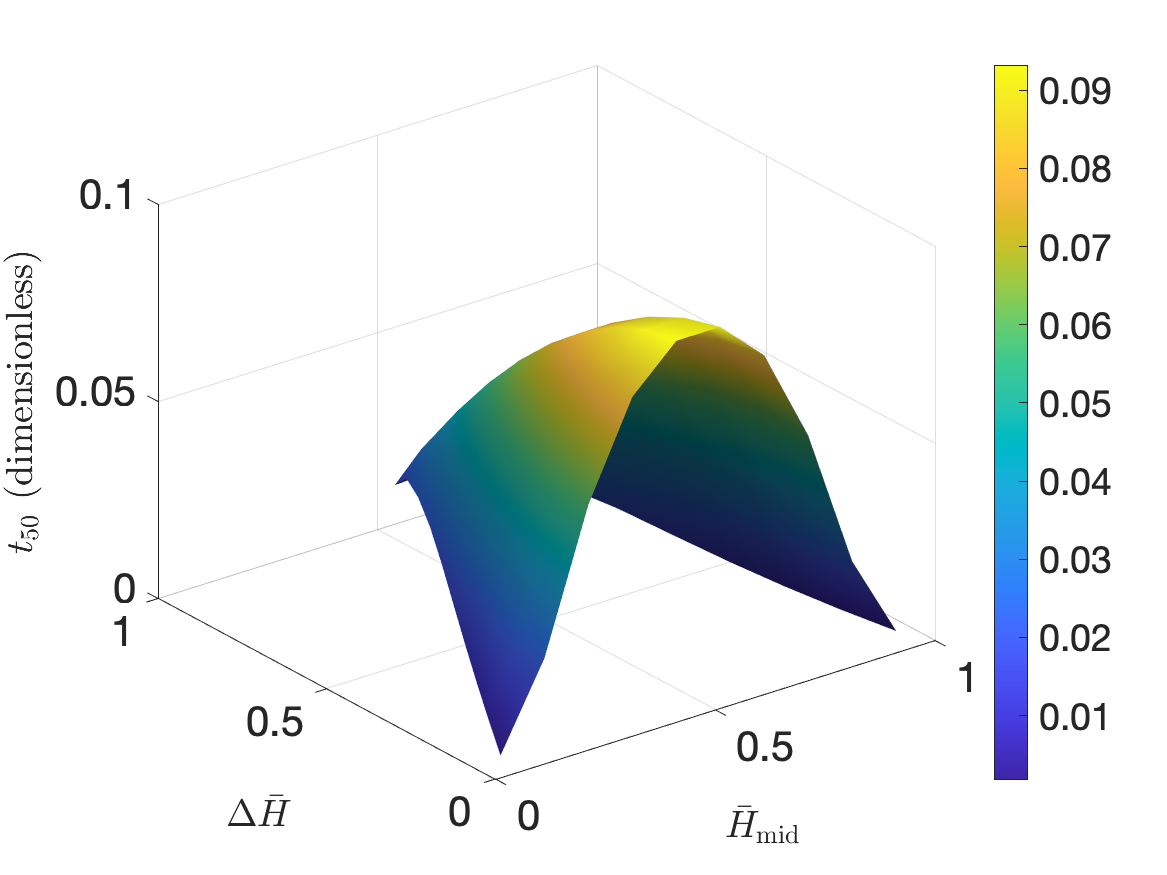}
\caption{Vial setting: 
dimensionless $\bar{t}_{50}$ plotted as functions of $\Delta \bar{H}$ and $\bar{H}_{\rm mid}$
for four different values of ${\cal D}$: $10^{-2}$ (upper left), $10^{-1}$ (upper right), $1$ (lower left), $10$ (lower right); recall  
$\bar{t}_{50}^{\rm classic} \approx 0.0492$. Note the differences in $\bar{t}_{50}$-axis and color bar scales across the four subplots. }
\label{fig_t50_plots103}
\end{center}
\end{figure}

Figure~\ref{fig_t50_plots103} shows computed values of $\bar{t}_{50}$ appearing in \eqref{eq:t50_formula} 
as the parameters $\bar{H}_{\rm mid}$ and $\Delta \bar{H}$ are varied for four fixed values of ${\cal D}$: $10^{-2}$ (upper left), $10^{-1}$ (upper right), $1$ (lower left), $10$ (lower right). 
For the smallest ${\cal D}$ value shown (${\cal D} = 10^{-2}$), all feasible choices of the geometric parameters $\bar{H}_{\rm mid}$ and $\Delta {\bar H}$ lead to larger $\bar{t}_{50}$ as compared to $\bar{t}_{50}^{\rm classic}$ for a standard hydrogel lens.  Parameters $\bar{H}_{\rm mid} =0.5$ and $\Delta \bar{H} \rightarrow 1$ yield the slowest release;  in this configuration, the vast majority of the composite lens has significantly smaller diffusion coefficient
compared to the narrow region of hydrogel, and this dictates the dynamics.  As ${\cal D}$ is increased by a factor of $10$ in the sequence of plots shown in
Figure~\ref{fig_t50_plots103}, we see this region of slower release rate become increasingly restricted near the lens midline ($\bar{H}_{\rm mid} = 0.5$).
Generally, we observe an overall increase in the predicted 
$\bar{t}_{50}$ as ${\cal D}$ decreases.
Keep in mind that the dimensional $t_{50}$
is a product of the dimensionless $\bar{t}_{50}(\bar{H}_{\rm mid}, \Delta \bar{H}, {\cal D})$
and the factor $H_{\rm cl}^2/D_2$ so trends
in the dimensional $t_{50}$ can be deduced
from the dimensionless one, $\bar{t}_{50}$, most easily when the CL thickness $H_{\rm cl}$ and the hydrogel diffusion coefficient $D_2$ are fixed.
When ${\cal D}$ is relatively small (e.g.~cases ${\cal D}=10^{-2}$ and ${\cal D}=10^{-1}$ in Figure~\ref{fig_t50_plots103}), the predicted $\bar{t}_{50}$ increases along the `midline' $\bar{H}_{\rm mid}=0.5$ when $\Delta \bar{H}$ increases.  In the ${\cal D}=1$ and ${\cal D}=10$ cases in Figure~\ref{fig_t50_plots103}, we see exactly the opposite trend when $\bar{H}_{\rm mid}=0.5$ and $\Delta \bar{H}$ increases.  \textit{An important conclusion based on the present model is that until the diffusion coefficient ratio is known for a composite lens drug delivery system, it is not possible to know whether increasing the thickness of a symmetrically-placed polymer insert will increase or decrease the predicted $50$\% therapeutic drug release time.}

\begin{figure}[h!]
\begin{center}
\vskip 0.05in
\includegraphics[width=.47\linewidth]{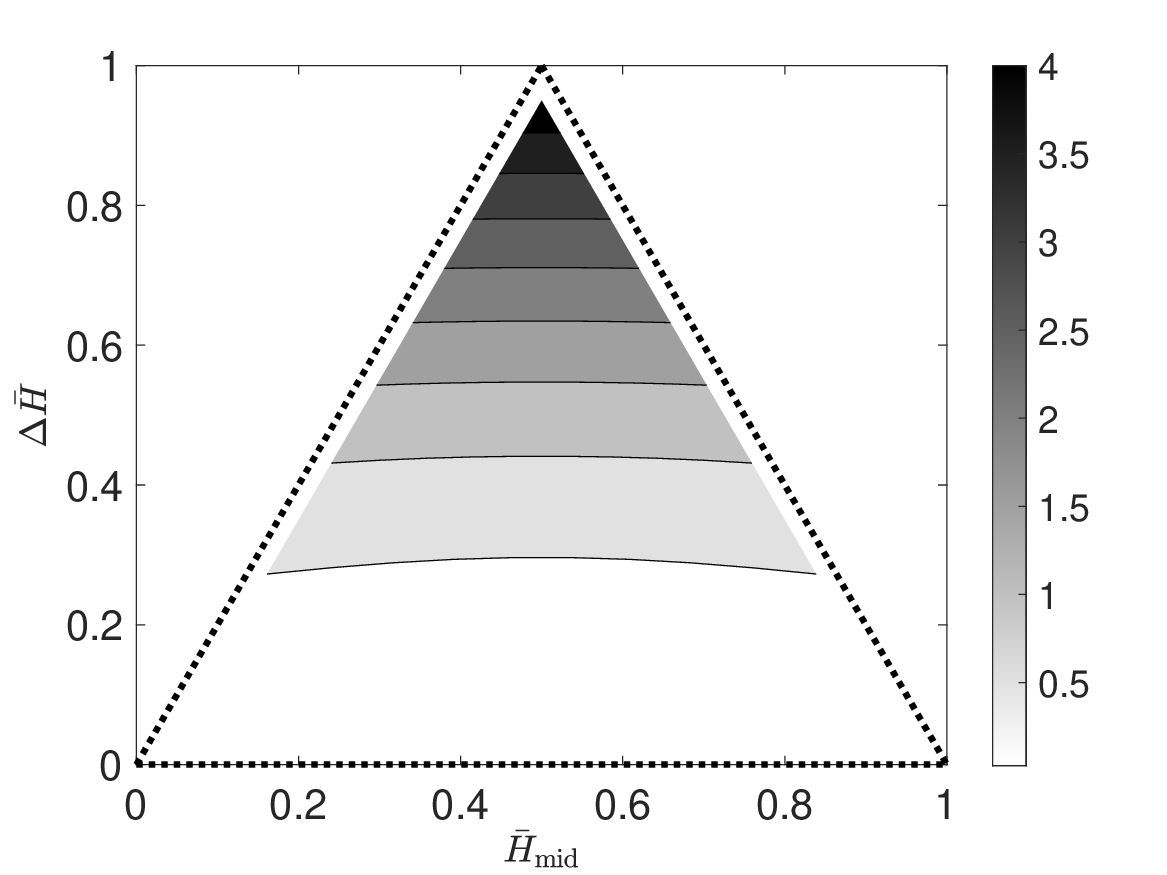}
\includegraphics[width=.47\linewidth]{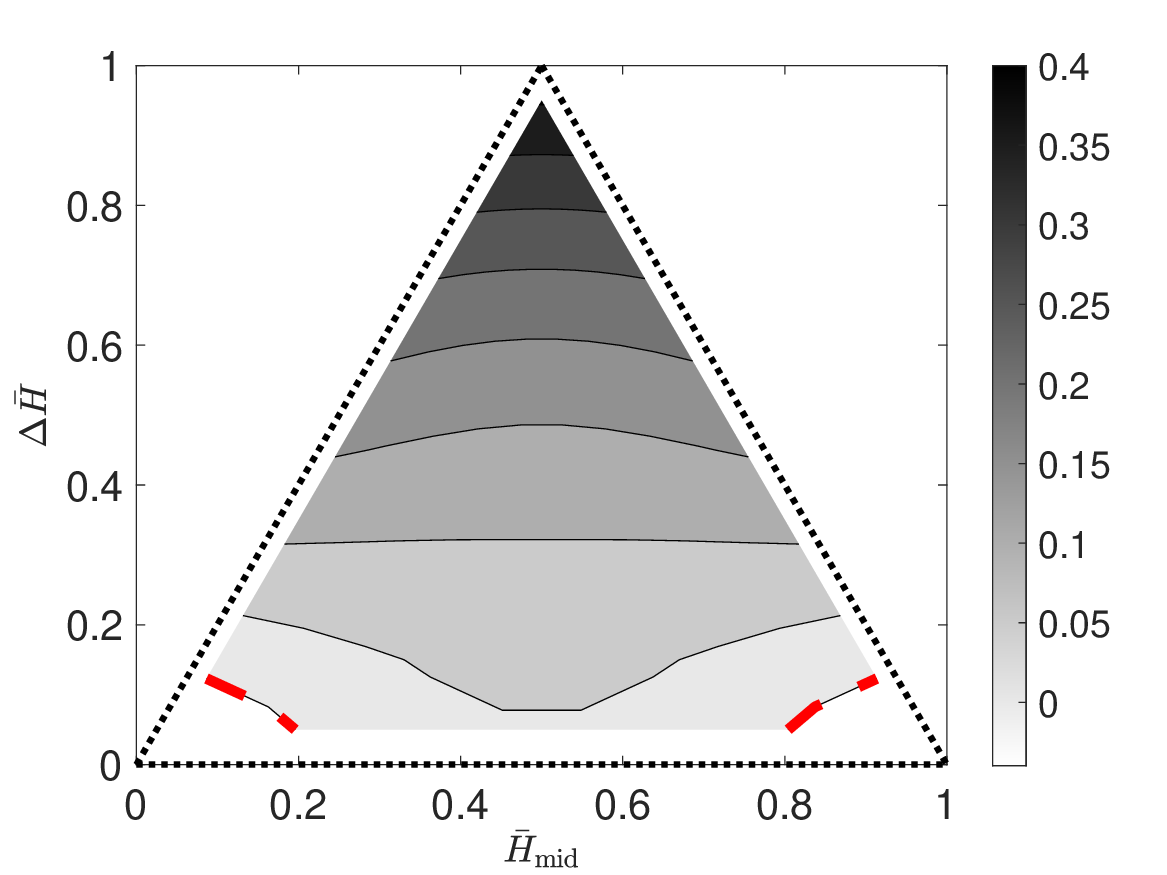} \\
\includegraphics[width=.47\linewidth]{figures/caseC22_303dash.eps}
\includegraphics[width=.47\linewidth]{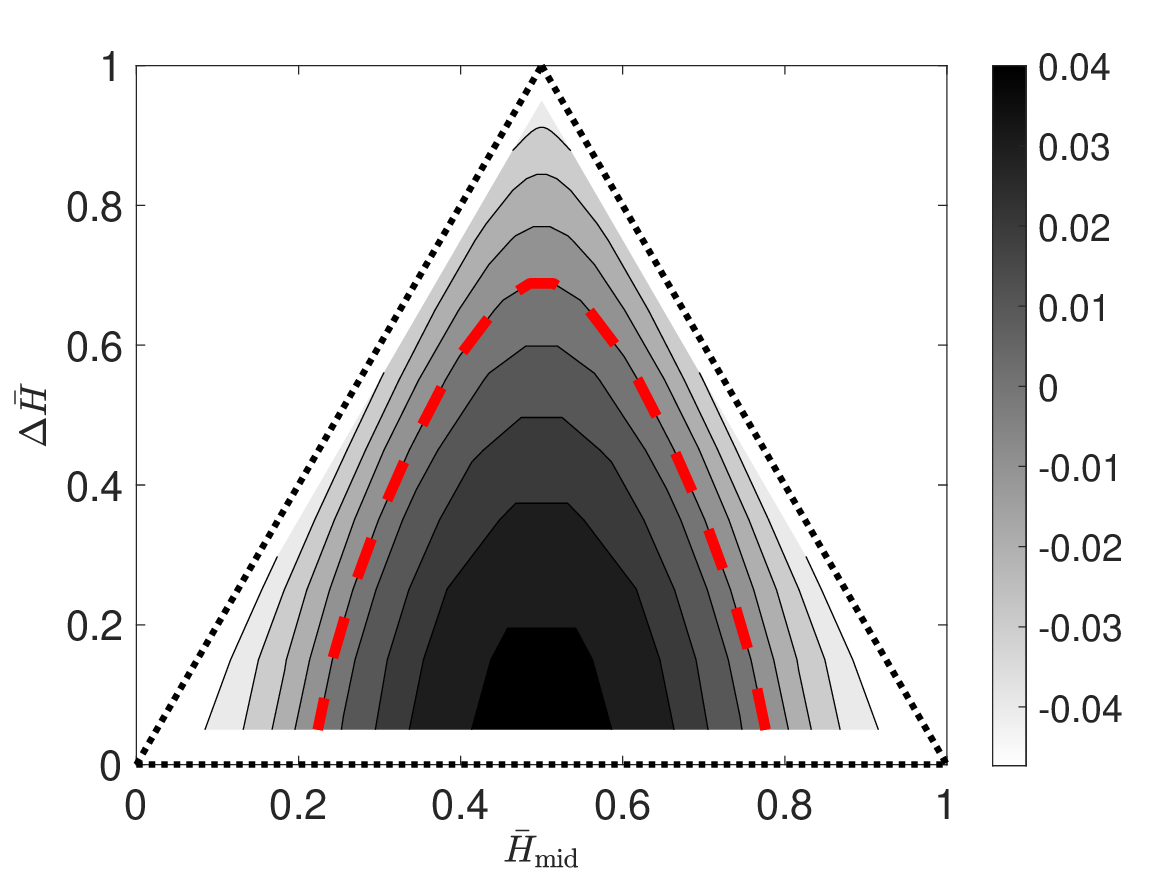}
\caption{Vial setting: contour plots of $\bar{t}_{50} - \bar{t}_{50}^{\rm classic}$ as functions of $\Delta \bar{H}$ and $\bar{H}_{\rm mid}$
for four different values of ${\cal D}$: $10^{-2}$ (upper left), $10^{-1}$ (upper right), $1$ (lower left), $10$ (lower right). 
Thick red-dashed lines 
indicate the zero contour where $\bar{t}_{50} = \bar{t}_{50}^{\rm classic}$. Note the differences in color bar scales across the four subplots.}
\label{fig_303all_Fig}
\end{center}
\end{figure}

Figure~\ref{fig_303all_Fig} displays our predictions in the form of contour plots of $\bar{t}_{50}  - \bar{t}_{50}^{\rm classic}$.  The red-dashed lines show the zero
contour marking the parameter values at which $\bar{t}_{50} = \bar{t}_{50}^{\rm classic}$ (i.e.~there is no difference in the predicted $\bar{t}_{50}$ between
the conventional and composite lens cases).  
In the case with ${\cal D}=10^{-1}$ shown in the upper right plot, 
observe that for a fixed polymer width near $\Delta \bar{H} \approx 0.1$, for example, moving $\bar{H}_{\rm mid}$ off the centerline ($\bar{H}_{\rm mid}=0.5$) results in a decrease in $\bar{t}_{50}$.  This is a similar trend that appears in the lower two plots of 
Figure~\ref{fig_303all_Fig} with ${\cal D}=1$ and ${\cal D}=10$; moving the polymer insert off center in these cases leads to faster drug release.  However, for ${\cal D}=10^{-1}$ and a larger value of $\Delta \bar{H} \approx 0.5$, we see that moving $\bar{H}_{\rm mid}$ off the centerline results in an {\it increase} in the predicted $\bar{t}_{50}$ value (e.g.~upper right plot of Figure~\ref{fig_303all_Fig}).
A slight increase in $\bar{t}_{50}$ for `off center' polymer inserts is also observable in the ${\cal D}=10^{-2}$ case in Figure~\ref{fig_303all_Fig} (upper left plot).  \textit{The slightly counter-intuitive conclusion is that increased proximity of the polymer insert to an external boundary of the CL in the vial release setting does not guarantee faster drug release; there is a non-trivial dependence on the ratio of diffusion coefficients between the polymer and the hydrogel layers.}  We observe this diffusion rate/geometry interplay also in the context of drug release from a composite lens during blinking in the next section.

In the vial geometry with perfect sink conditions, our results show a sensitivity of the predicted 50\% drug release time to the geometric parameters associated with polymer layer positioning $\bar{H}_{\rm mid}$ and thickness $\Delta \bar{H}$ and the diffusion coefficient ratio $\mathcal{D}$.  In general, there is symmetry with respect to midline positioning; the predicted $\bar{t}_{50}$ depends on the quantity $|\bar{H}_{\rm mid} - 0.5|$ but not on the sign of $\bar{H}_{\rm mid} - 0.5$.  Interestingly, however, our predictions show that $\bar{t}_{50}$ can either increase or decrease with $|\bar{H}_{\rm mid} - 0.5 |$ depending on 
$\Delta \bar{H}$ and 
${\cal D}$.
We have also seen that $\bar{t}_{50}$ can either increase or decrease as $\Delta \bar{H}$ increases with fixed $\bar{H}_{\rm mid}=0.5$, depending on the diffusion coefficient ratio ${\cal D}$. In the next section, we explore the drug release dynamics during wear of a composite lens where, in general, this mid-plane symmetry is broken owing to the distinct dynamics of the pre-lens and post-lens tear films.

\section{Composite contact lens drug release: simulated eye wear}
 \label{sec:eye}

 In this section we adapt our previous model (Anderson \& Luke \cite{anderson2024mathematical}) to simulate the dynamics of drug release from the
 composite CL during wear and blinking.   The model matches Anderson \& Luke \cite{anderson2024mathematical}, except  that here the diffusion within the lens is governed by the 3-layer composite lens model. 

 \subsection{Dimensionless model formulation}

 In addition to the drug concentration within the composite lens, the simulated eye wear model accounts for the drug concentration
 in the pre-lens tear film, the post-lens tear film, and in an upper eyelid compartment.  Additionally, both pre-
 and post-lens tear film thickness are dependent variables, nondimensionalized as $\bar{h}_{\rm pre}$ and $\bar{h}_{\rm post}$ with respect to CL thickness $H_{\rm cl}$.
 As in the Anderson \& Luke \cite{anderson2024mathematical} model,
 we have interblink intervals, during which the drug concentrations evolve in the contact lens, pre-lens, post-lens, and lid regions, and 
 blink reset conditions, which model drug transport that occurs as a result of the blink.  
 There are corresponding time-dependent drug concentration variables in the pre-lens tear film, $\bar{C}_{\rm pre}$, the post-lens tear film, $\bar{C}_{\rm post}$, 
 and upper lid drug concentration, $\bar{C}_{\rm lid}$. 
 These are in addition to the concentration variables in the three layers of the composite contact lens: $\bar{C}_2^{\rm pre}$, $\bar{C}_1$, and
 $\bar{C}_2^{\rm post}$ which are functions of space and time. The concentrations $\bar{C}_{\rm pre}(t)$ and $\bar{C}_{\rm post}(t)$ evolve in distinct and nontrivial ways when the contact
lens is worn on the eye; here the corresponding volumes of the pre-lens 
and post-lens are much smaller than typical vial volumes.
 The numerical results in the present study will exclude certain mechanisms such as drug transport into the upper lid and pre-lens tear film evaporation for simplicity, but for completeness
 we include these in the model formulation following Anderson \& Luke \cite{anderson2024mathematical}.
 The details are outlined in dimensionless form below.
 
 {\bf Interblink:} In the pre-lens tear film we have
 \bea
 \frac{d \bar{h}_{\rm pre}}{d \bar{t}} & = & - \bar{J}_{\rm E}, \\
 \frac{d \left( \bar{h}_{\rm pre} \bar{C}_{\rm pre} \right) }{d \bar{t}} & = & - \bar{D}_2 \left. \frac{\partial  \bar{C}_2^{\rm pre}}{\partial \bar{z}} \right|_{\bar{z}=1} -  
 \bar{k}_{\rm lid}  \bar{C}_{\rm pre},
 \eea
 where the dimensionless evaporation parameter, the dimensionless hydrogel diffusion coefficient, and dimensionless lid permeability are given by
 \bea
 \bar{J}_{\rm E} = \frac{J_{\rm E} \tau}{H_{\rm cl} A_{\rm cl} }, \quad
 \bar{D}_2 = \frac{D_2 \tau}{H_{\rm cl}^2},\quad
 \bar{k}_{\rm lid} = \frac{k_{\rm lid} \tau}{H_{\rm cl}} \frac{A_{\rm overlap}}{A_{\rm cl}}.
 \eea
 Here, $A_{\rm overlap}$ gives the overlap area of the eyelid and the contact lens, which is relevant for any drug uptake. The dimensionless time is $\bar{t} = t/\tau$.  In constrast to the vial model where we chose $\tau = H_{\rm cl}^2/D_2$, we instead choose $\tau = t_{\rm blink} = 10$ seconds,
 corresponding to a typical time between blinks.
 
 The pre-lens dynamics are coupled to those in the CL via the dimensionless model already presented in Section~\ref{sec-dimensionless_vial}, but with changed dimensionless diffusion equation coefficients due to our new choice of time scaling.  
 For clarity, we restate those equations here. 
 At the `pre-lens'--contact lens interface, $\bar{z}=1$,
\bea
\bar{C}_2^{\rm pre}(1,\bar{t}) & = & k \bar{C}_{\rm pre}(\bar{t}).
\eea
In the anterior hydrogel layer of the lens, $\bar{H}_2 < \bar{z} < 1$ (e.g.~pre-lens side), 
\bea
\frac{\partial \bar{C}_2^{\rm pre}}{\partial \bar{t}} & = & \bar{D}_2 \frac{\partial^2 \bar{C}_2^{\rm pre}}{\partial \bar{z}^2}.
\eea
At the polymer layer -- anterior hydrogel layer, $\bar{z}=\bar{H}_2$, we have
\bea
\bar{C}_2^{\rm pre}(\bar{H}_2,\bar{t}) & = & \bar{C}_1(\bar{H}_2,\bar{t}),\\
\bar{D}_2 \left. \frac{\partial \bar{C}_2^{\rm pre}}{\partial \bar{z}} \right|_{\bar{z}=\bar{H}_2} & = & \bar{D}_1 \left. \frac{\partial \bar{C}_1}{\partial \bar{z}} \right|_{\bar{z}=\bar{H}_2},
\eea
where $\bar{D}_1 = D_1 \tau / H_{\rm cl}^2$.
In the polymer layer, $\bar{H}_1 < \bar{z} < \bar{H}_2$, we have
\bea
\frac{\partial \bar{C}_1}{\partial \bar{t}} & = & \bar{D}_1 \frac{\partial^2 \bar{C}_1}{\partial \bar{z}^2}.
\eea
At the polymer layer -- posterior hydrogel layer, $\bar{z}=\bar{H}_1$, we have
\bea
\bar{C}_2^{\rm post}(\bar{H}_1,\bar{t}) & = & \bar{C}_1(\bar{H}_1,\bar{t}),\\
\bar{D}_2 \left. \frac{\partial \bar{C}_2^{\rm post}}{\partial \bar{z}} \right|_{\bar{z}=\bar{H}_1} & = & \bar{D}_1 \left. \frac{\partial \bar{C}_1}{\partial \bar{z}} \right|_{\bar{z}=\bar{H}_1}.
\eea
In the posterior hydrogel layer of the lens, $0 < \bar{z} < \bar{H}_1$ (e.g.~post-lens side), we have
\bea
\frac{\partial \bar{C}_2^{\rm post}}{\partial \bar{t}} & = & \bar{D}_2 \frac{\partial^2 \bar{C}_2^{\rm post}}{\partial \bar{z}^2}.
\eea
At the `post-lens'--contact lens interface, $\bar{z}=0$ we have
\bea
\bar{C}_2^{\rm post}(0,\bar{t}) & = & k \bar{C}_{\rm post}(\bar{t}).
\eea
In the post-lens tear film we have
 \bea
 \frac{d \bar{h}_{\rm post}}{d \bar{t}} & = & 0, \\
 \frac{d \left( \bar{h}_{\rm post} \bar{C}_{\rm post} \right) }{d \bar{t}} & = & \bar{D}_2 \left. \frac{\partial  \bar{C}_2^{\rm post}}{\partial \bar{z}} \right|_{\bar{z}=0} 
 -  \bar{k}_{\rm C}  \bar{C}_{\rm post} ,
 \eea
 where the dimensionless corneal permeability is given by
 \bea
\bar{k}_{\rm C} = \frac{k_{\rm C} \tau}{H_{\rm cl}}.
  \eea
  New dimensional parameters in this section are summarized in Table \ref{table-eye_params}.
  Note that there are no mechanisms considered here that result in post-lens thickness changes during the interblink (when the eyelids are not moving).  This model assumes that
  any motion of the lens occurs during the blink and resultant transport that occurs is accounted for in blink/reset conditions explained in more
  detail below.  In the more
  general model of Anderson \& Luke \cite{anderson2024mathematical}, the possibility of other fluxes active during the interblink were discussed.
  In the upper eyelid during the blink, we have
  \bea
  \frac{d \bar{C}_{\rm lid}}{d\bar{t}} & = & \bar{k}_{\rm lid} V_{\rm ratio}^{\rm lid} \bar{C}_{\rm pre},
  \eea
  where the lens to lid volume ratio is given by
  \bea
  V_{\rm ratio}^{\rm lid} = \frac{V_{\rm cl}}{V_{\rm lid}} = \frac{H_{\rm cl} A_{\rm cl}}{H_{\rm lid} A_{\rm lid}}.
  \eea
  
 These governing equations for drug transport in the CL and in the pre-lens tear film, post-lens tear film, and upper eyelid 
 are solved on a time interval $\bar{t} \in [0, \bar{t}_{\rm blink}]$, then
 $\bar{t} \in [\bar{t}_{\rm blink}, 2\bar{t}_{\rm blink}]$, $\ldots$, 
 $\bar{t} \in [N \bar{t}_{\rm blink}, (N+1) \bar{t}_{\rm blink}]$, $\ldots$, etc.
 with blink reset 
 conditions applied after each interval that account for blink-induced drug transport and to establish new initial conditions for the next interblink interval.

{\bf Blink/reset:} We use the same blink/reset conditions as in Anderson \& Luke \cite{anderson2024mathematical} (see their equations (30a)--(30f))
with a notational update to reflect $\bar{C}_2^{\rm pre}$, $\bar{C}_1$ and $\bar{C}_2^{\rm post}$ 
for a composite lens. In particular,
\bea
\label{eq:reset1}
\bar{C}_2^{\rm pre}(z,\bar{t}_{\rm blink}^+) & = & \bar{C}_2^{\rm pre}(z,\bar{t}_{\rm blink}^-), \\
\label{eq:reset2}
\bar{C}_1(z,\bar{t}_{\rm blink}^+) & = & \bar{C}_1(z,\bar{t}_{\rm blink}^-), \\
\label{eq:reset3}
\bar{C}_2^{\rm post}(z,\bar{t}_{\rm blink}^+) & = & \bar{C}_2^{\rm post}(z,\bar{t}_{\rm blink}^-), \\
\label{eq:reset4}
\bar{h}_{\rm pre}(\bar{t}_{\rm blink}^+) \bar{C}_{\rm pre}(\bar{t}_{\rm blink}^+) & = & (1-p) \bar{h}_{\rm pre}(\bar{t}_{\rm blink}^-) \bar{C}_{\rm pre}(\bar{t}_{\rm blink}^-), \\
\label{eq:reset5}
\bar{h}_{\rm pre}(\bar{t}_{\rm blink}^+) & = & \bar{h}_{\rm pre}^{\rm init},\\
\label{eq:reset6}
\bar{C}_{\rm post}(\bar{t}_{\rm blink}^+) & = & \left\{ \begin{array}{l}
\bar{C}_{\rm post}^{\rm slide} \\
\bar{C}_{\rm post}^{\rm squeeze}
\end{array} \right. \\
\label{eq:reset7}
\bar{h}_{\rm post}(\bar{t}_{\rm blink}^+) & = & \bar{h}_{\rm post}^{\rm init},\\
\label{eq:reset8}
\bar{C}_{\rm lid}(\bar{t}_{\rm blink}^+) & = & \bar{C}_{\rm lid}(\bar{t}_{\rm blink}^-),
\eea
where $\bar{t}_{\rm blink}^+$ represents the instant just after the blink and $\bar{t}_{\rm blink}^-$ represents the instant just before the blink.
The conditions \eqref{eq:reset1}--\eqref{eq:reset3} represent the assumption that no change occurs in the drug concentration profiles
in the CL as a consequence of a blink.  A similar assumption for the lid concentration is represented in \eqref{eq:reset8}.
The conditions \eqref{eq:reset5} and \eqref{eq:reset7} indicate that the pre-lens thickness and post-lens thickness reset to their (original)
initial values after each blink.
The condition \eqref{eq:reset4} introduces a parameter $p \in [0,1]$ that represents a proportional change in the drug mass in the pre-lens tear film resulting from a blink.
 When $p=0$, no pre-lens drug mass is lost as the result of a blink and the post-blink concentration matches the pre-blink concentration.
 When $p=1$, all of the pre-lens drug mass is lost; the  blink supplies a completely fresh, drug-free, pre-lens tear film.
 For cases in which the pre-lens tear film thickness does not vary during the interblink (e.g., when no 
 evaporation is present) the parameter $p$ gives the proportional change in pre-lens drug concentration.
 As in Anderson \& Luke \cite{anderson2024mathematical}, we treat the parameter, $p$, as adjustable.
The condition \eqref{eq:reset6} models the post-lens drug lost as a result of CL motion during the blink.
Anderson \& Luke \cite{anderson2024mathematical} gave two versions of this drug loss; one associated with a sliding motion of the lens that generates a 
Couette-type flow in the post-lens tear fluid, and one associated with a squeezing motion of the lens that generates a squeeze-film
type flow of the post-lens tear fluid.
 Expressions for $\bar{C}_{\rm post}^{\rm slide}$ and $\bar{C}_{\rm post}^{\rm squeeze}$ can be found in equations~(20) and~(13), respectively,
 of Anderson \& Luke \cite{anderson2024mathematical}.  In this study we report only on the slide-out scenario, in which case equation~(20) of
 Anderson \& Luke \cite{anderson2024mathematical} gives
 \bea
 \label{eq:Cpost_slideout}
 \bar{C}_{\rm post}^{\rm slide} & = & \bar{C}_{\rm post}(\bar{t}_{\rm blink}^-) \left( 1 - \frac{3 \Delta X_{\rm cl}}{2\pi R_{\rm cl}^{\rm eff} } \right).
 \eea
 Here, $\Delta X_{\rm cl}$ is the change in position of the center of the lens due to translational (sliding) motion of the lens during a blink and $R_{\rm cl}^{\rm eff}$
 is the effective radius of the lens as measured along an arc length from the lens center to the edge of the lens.
 We adopt a specific value for the ratio $\Delta X_{\rm cl}/R_{\rm cl}^{\rm eff}$ previously used in Anderson \& Luke \cite{anderson2024mathematical}.
 Note that the blink reset conditions \eqref{eq:reset4} and \eqref{eq:reset6} provide separate
 models for how the pre- and post-blink drug concentrations in the pre- and post-lens tear films evolve.  This can be viewed as the direct source in the model of the symmetry-breaking with respect to the centerline positioning of $\bar{H}_{\rm mid}$ between 
 pre-lens and post-lens drug concentrations.  The simulations presented below will show this asymmetry.

  As in the vial setting, we study the time for which 50\% of the drug is released from the CL in this setting, $t_{50}$.
  We monitor the dimensionless cumulative drug release which can be expressed as
  \bea
\label{eq:calMbar_rescaled_blink}
\bar{\cal M}(\bar{t};\bar{H}_{\rm mid},\Delta \bar{H},\bar{D}_1,\ldots) 
 & = & \int_{\bar{H}_1}^{\bar{H}_2} \left( 1 - \bar{C}_1(\bar{z},\bar{t}) \right) d\bar{z} 
 - \int_0^{\bar{H}_1} \bar{C}_2^{\rm post}(\bar{z},\bar{t}) d\bar{z}
 -  \int_{\bar{H}_2}^1 \bar{C}_2^{\rm pre}(\bar{z},\bar{t}) d\bar{z},
\eea
where the dimensional mass of drug in the CL is given by $M = (C_{\rm load} A_{\rm poly} H_{\rm cl} )\bar{\cal M}$.
Recall also that $\Delta \bar{H} = \bar{H}_2 - \bar{H}_1$ and $\bar{H}_{\rm mid} = (\bar{H}_1 + \bar{H}_2)/2$.  
Note that $\bar{\cal M} =0$ at time zero and the total dimensionless drug mass initially loaded in the lens is $\Delta \bar{H}$.
In the argument list of $\bar{\cal M}$, we denote by `$\ldots$' the collection of parameters we plan to fix in this context (e.g.~$\bar{D}_2$, $k$, $\bar{k}_{\rm C}$, $\bar{k}_{\rm lid}$, etc.)
to focus on the parameters associated most directly with the composite lens polymer insert (e.g.~$\bar{H}_{\rm mid}$, $\Delta \bar{H}$, and $\bar{D}_1$).
The dimensionless $\bar{t}_{50}$ is reached when
\bea
\bar{{\cal M}}(\bar{t}_{50};\bar{H}_{\rm mid},\Delta \bar{H},\bar{D}_1,\ldots) & = & \frac{1}{2} \Delta \bar{H}.
\eea
It follows that the dimensional value $t_{50}$ takes the form
\bea
\label{eq:t50_formula_blink}
t_{50} & = & \bar{t}_{50}(\bar{H}_{\rm mid},\Delta \bar{H},\bar{D}_1,\ldots) \tau,
\eea
where we recall the time scale is $\tau = t_{\rm blink}$ (e.g.~$t_{\rm blink}=10$ s).  
Therefore, in the present setting, each dimensionless time unit corresponds to one blink.  
In our simulations we fix the value of $\bar{D}_2$ and give model interpretations in terms of the polymer/hydrogel ratio ${\cal D} = \bar{D}_1/\bar{D}_2$.

\begin{table}[t!]
\begin{center}
\begin{tabular}{llll}                       Parameter (units) & Description  &  Value/Range &  Source  \\ \hline
$h_{\rm pre}^{\rm init}$ ($\mu$m) & Initial pre-lens height  & 5 & \cite{nichols2005} \\
$h_{\rm post}^{\rm init}$ ($\mu$m) & Initial post-lens height  & 5 & \cite{nichols2005}\\
$\tau = t_{\rm blink}$ (s) & (Blink) time scale &   10 & \cite{phan2021development} \\
$k$ & Partition coefficient &    2 & \cite{phan2021development}  \\
{{$\bar{k}_{\rm C}$}} & {{dimensionless corneal permeability}} & {{$0$, $0.4$}} & {{\cite{braun2018}}} \\
{{$p$}} & {{pre-lens drug loss parameter}} & {{$0$ -- $1$}} & \cite{anderson2024mathematical} \\
{{$\Delta X_{\rm cl}/R_{\rm cl}^{\rm eff}$}} & {{Dimensionless slide out displacement}}  & {{$0.00757$}}  & \cite{anderson2024mathematical}  \\
{{$\bar{k}_{\rm lid}$}} & {{dimensionless eyelid permeability}} & {{$0$}} & \cite{anderson2024mathematical} \\
$\bar{J}_{E}$ & Evaporation rate  & 0 & -- \\ \hline \hline
$V_{\rm B}$ (mL) & Blister pack volume & $0.2$ -- $5$ & \cite{hamilton2007patent} \\
$C_B^{\rm load}$ (mg $\mu$L$^{-1}$) & Blister pack concentration & $0.0028$ -- $0.0552$ & \cite{ross2019topical,bengani2020steroid} \\
$\bar{C}_B^{\rm load}$ & Relative blister pack concentration & $0.01$ -- $0.2$ & -- \\
$\bar{V}_B$ & Blister pack/polymer volume ratio & 10.3 -- 258 & -- \\ 
 \hline
\end{tabular}
\end{center}
\caption{Parameters, descriptions and values or ranges used in the eye wear with blinking and blister pack models (the latter introduced in Section~\ref{sec:blister}) in addition to those in Table~\ref{table-compA}. Values without references are selected for use in our simulations, or computed as a result.
}
\label{table-eye_params}
\end{table}

\subsection{Results}
 
 Anderson \& Luke \cite{anderson2024mathematical} explored many features of drug release dynamics during blinking from a conventional lens.  As our focus
 is the composite lens, rather than report an exhaustive parameter study, we fix most of the parameters in this model to a representative set. For the simulations reported in this section we neglect evaporation ($\bar{J}_{\rm E} =0$) and drug transport into the upper eyelid ($\bar{k}_{\rm lid} = 0$).  
 In most cases we neglect drug transport into the cornea ($\bar{k}_{\rm C}=0$) but briefly make a comparison to cases with nonzero $\bar{k}_{\rm C}$.
 We fix the value of the partition coefficient at $k=2$.
 If the partition coefficient is zero, the drug release predictions during CL wear would exactly match those of the `perfect sink' vial, as the drug diffusion out of the CL in that case is independent of processes associated with the blink and those associated with the pre-lens and post-lens tear film dynamics.  
 Additionally, we focus only on the slide-out mechanism during the blink but note that Anderson \& Luke \cite{anderson2024mathematical} found similar dynamics also for the squeeze-out mechanism. 
In this context we specifically use
$\Delta X_{\rm cl}/R_{\rm cl}^{\rm eff} = 0.00757$, which was one of the lens displacement
values explored in Anderson \& Luke \cite{anderson2024mathematical}.  With this blink-induced lens displacement, in view of equations~(\ref{eq:reset6}) and~(\ref{eq:Cpost_slideout}) one can associate 
a corresponding effective reduction factor of 
drug concentration in the post-lens due to the slide 
out mechanism, $p_{\rm slide}^{\rm eff} = 3 (0.00757)/(2 \pi) \approx 0.0036$, which means that the consequence of a single blink is approximately a $0.4$\% reduction in the drug concentration in the post-lens tear film.

In terms of parameters specifically connected to the 
composite lens geometry, we  fix the value of
the polymer layer thickness at $\Delta \bar{H}=0.2$. 
Our simulations below focus on the influence
of the polymer/hydrogel diffusion coefficient ratio, ${\cal D}=D_1/D_2 = \bar{D}_1/\bar{D}_2$ and the midline position, $\bar{H}_{\rm mid}$.  We
 present simulations for a range of values of $p$, which measures the percentage of drug lost from the pre-lens tear film during each blink. 
The hydrogel diffusion coefficient is $D_2 = 7 \times 10^{-13}$ m$^2$ s$^{-1}$.  With $H_{\rm cl} = 300$ $\mu$m, this gives a diffusive time scale of
$H_{\rm cl}^2/D_2 \approx 35.7$ hours, and $\bar{D}_2 = 7.78 \times 10^{-5}$.

\begin{figure}[t!]
\begin{center}
\vskip 0.05in
\includegraphics[width=.32\linewidth]{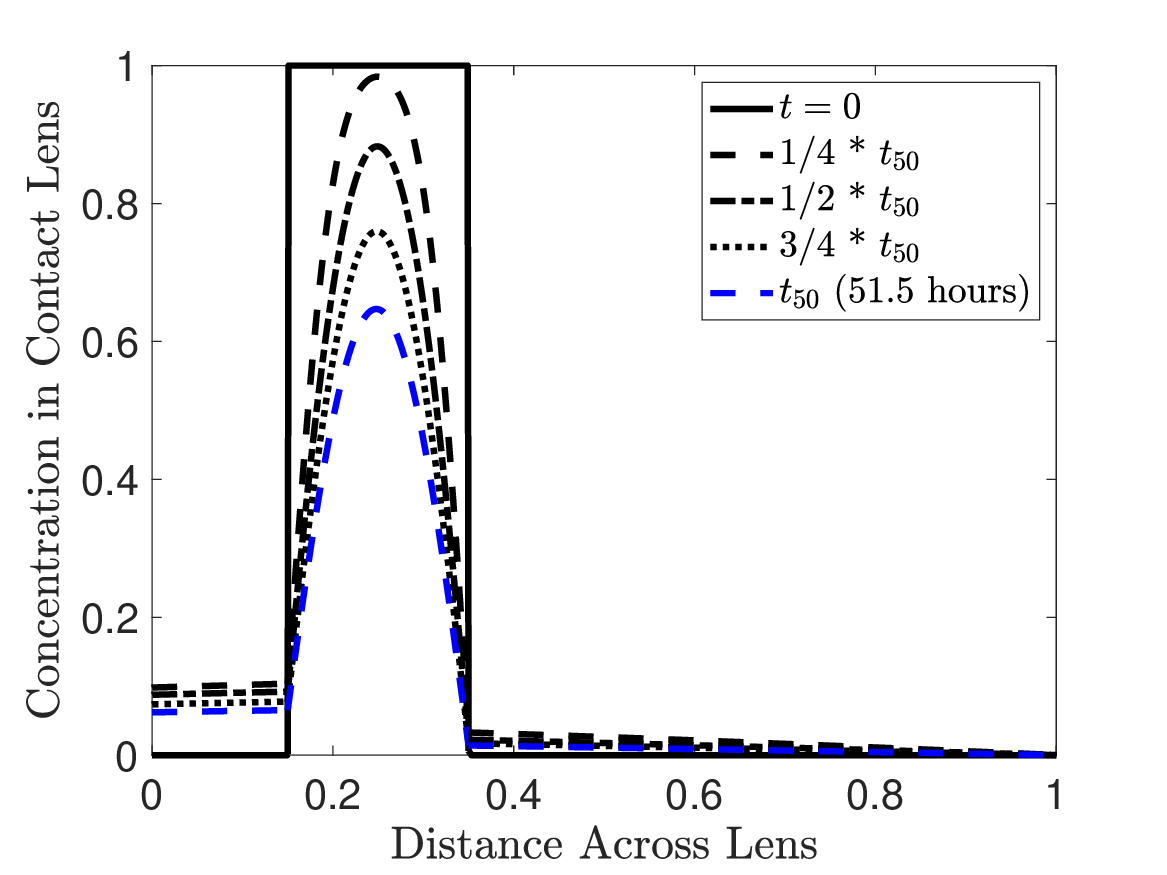}
\includegraphics[width=.32\linewidth]{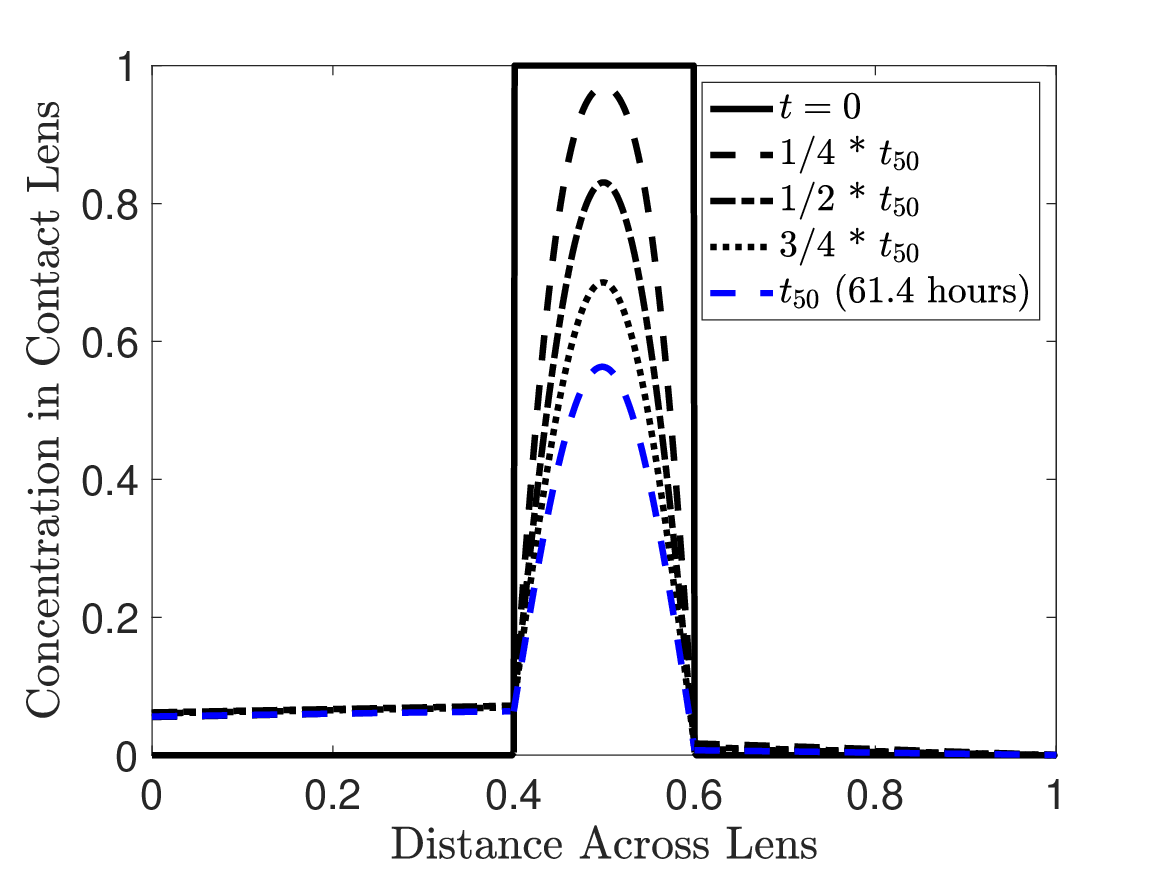}
\includegraphics[width=.32\linewidth]{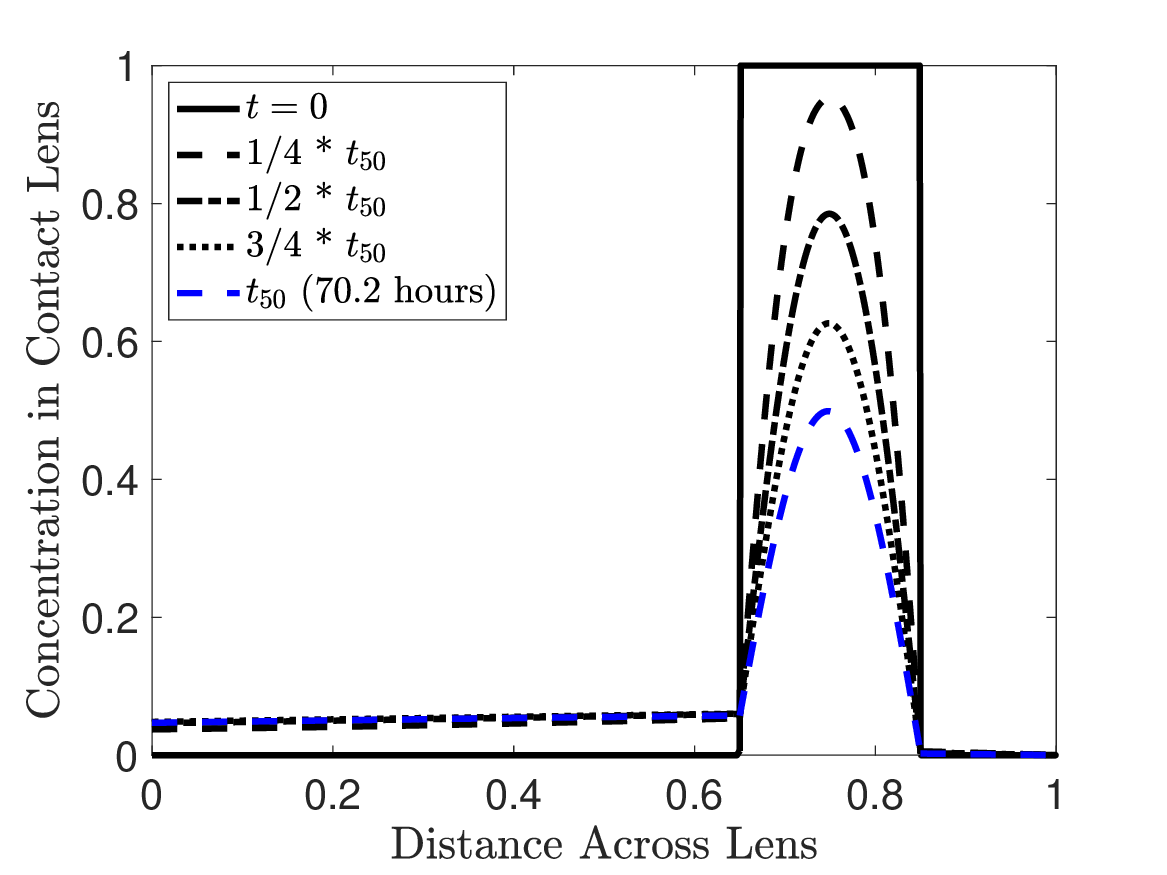} \\
\includegraphics[width=.32\linewidth]{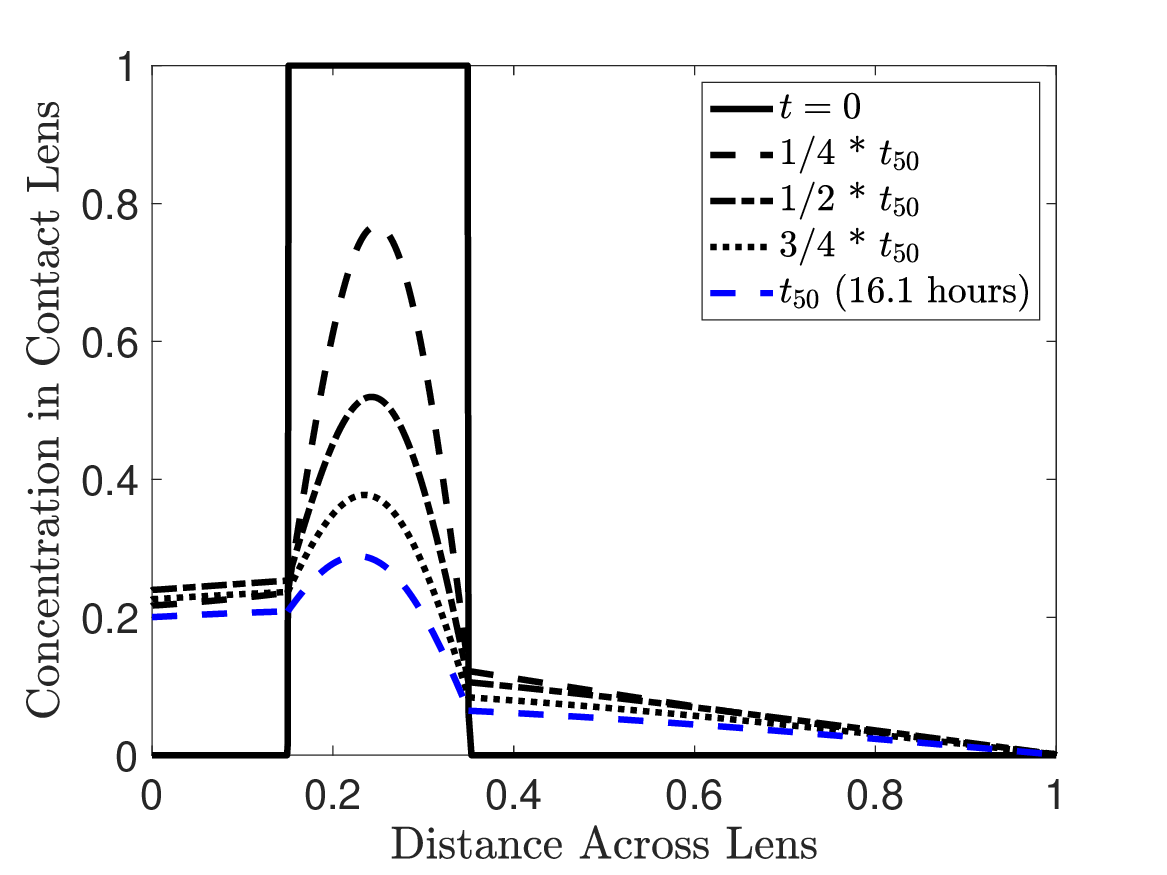}
\includegraphics[width=.32\linewidth]{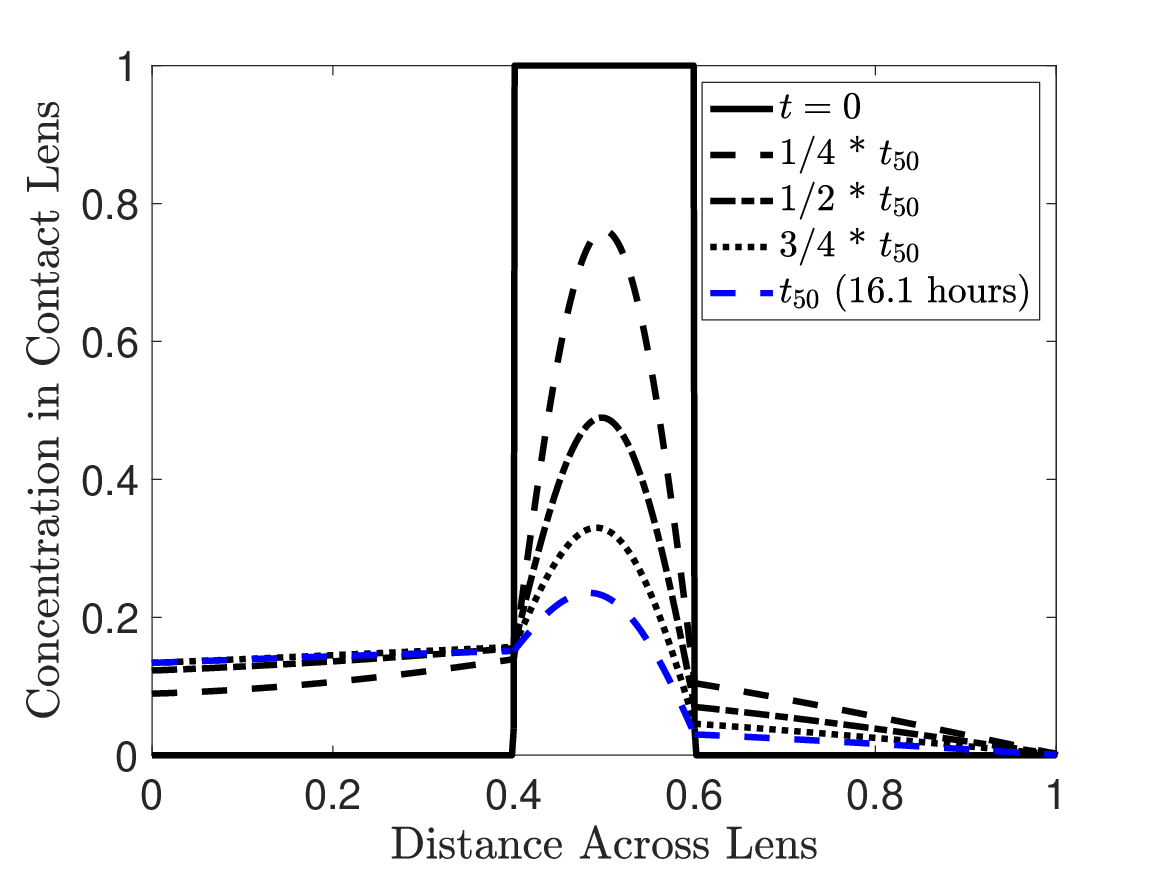}
\includegraphics[width=.32\linewidth]{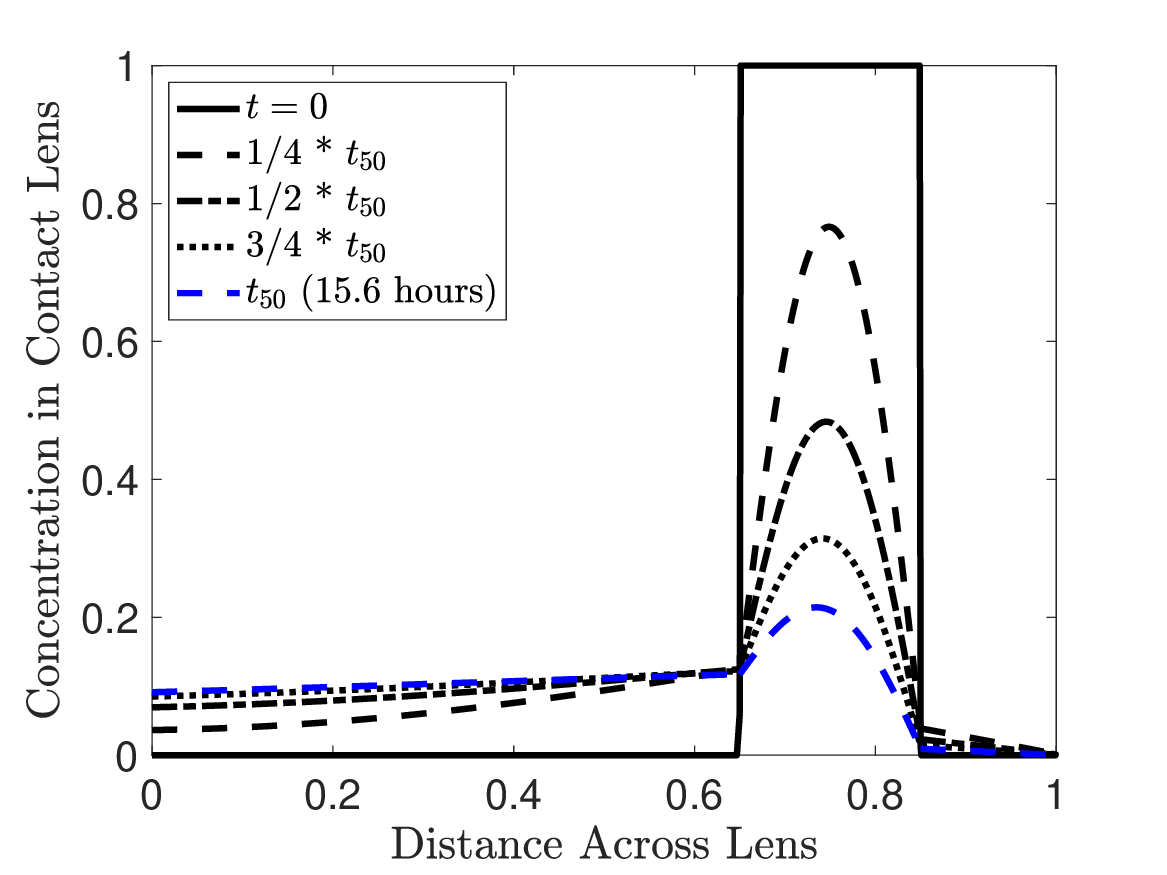} \\
\includegraphics[width=.32\linewidth]{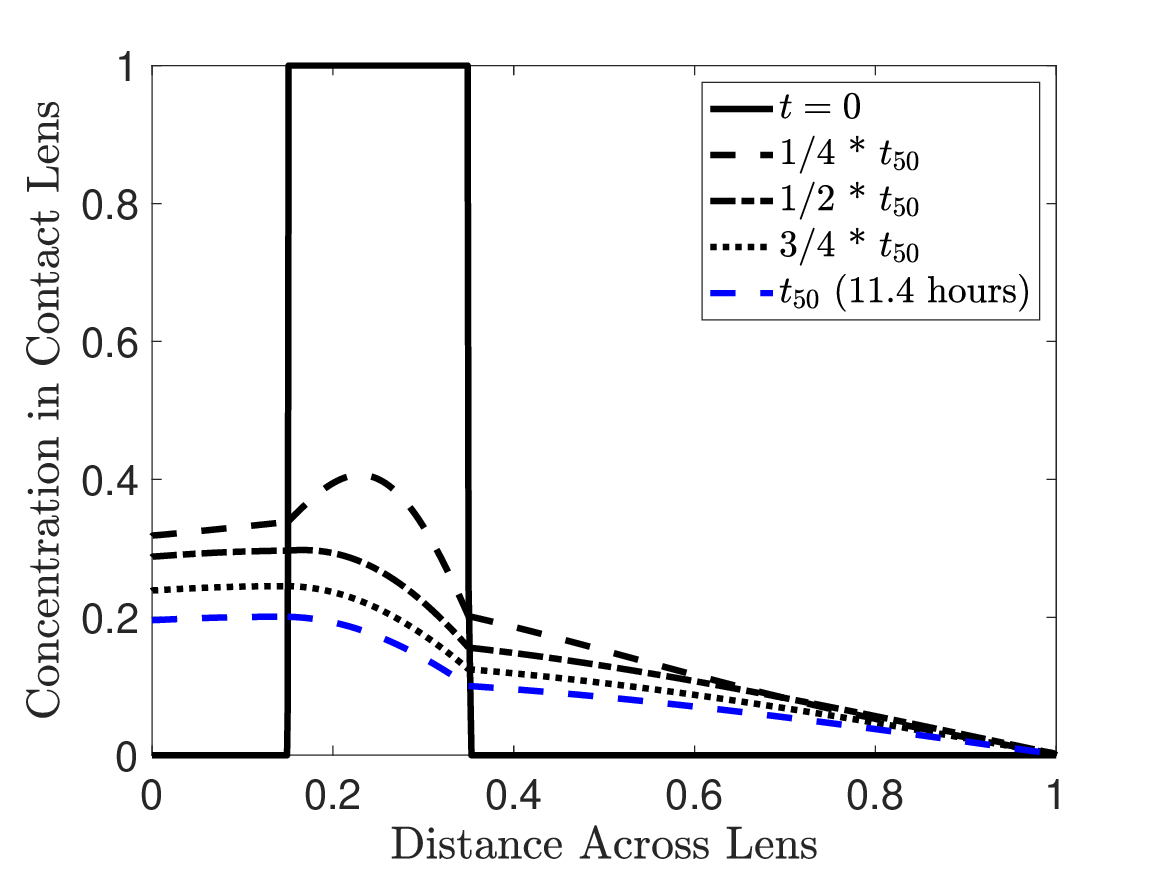}
\includegraphics[width=.32\linewidth]{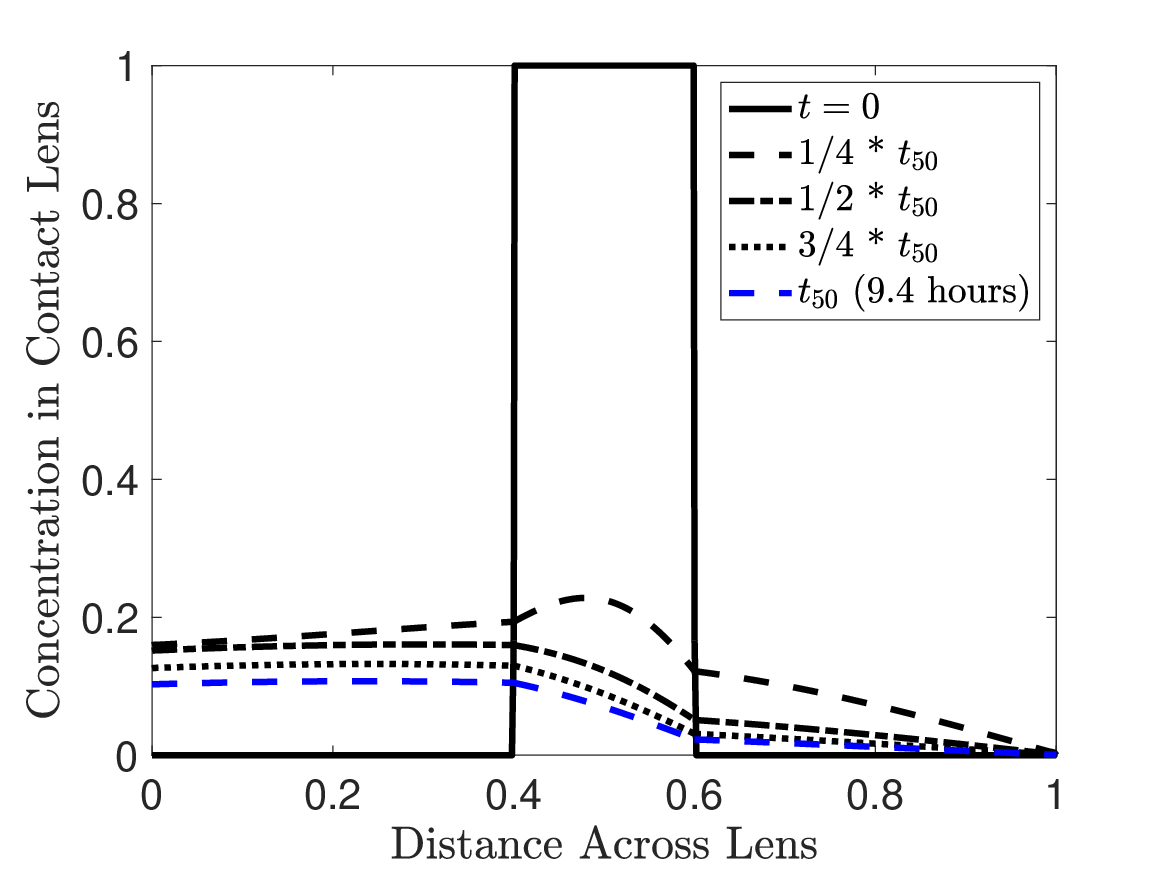}
\includegraphics[width=.32\linewidth]{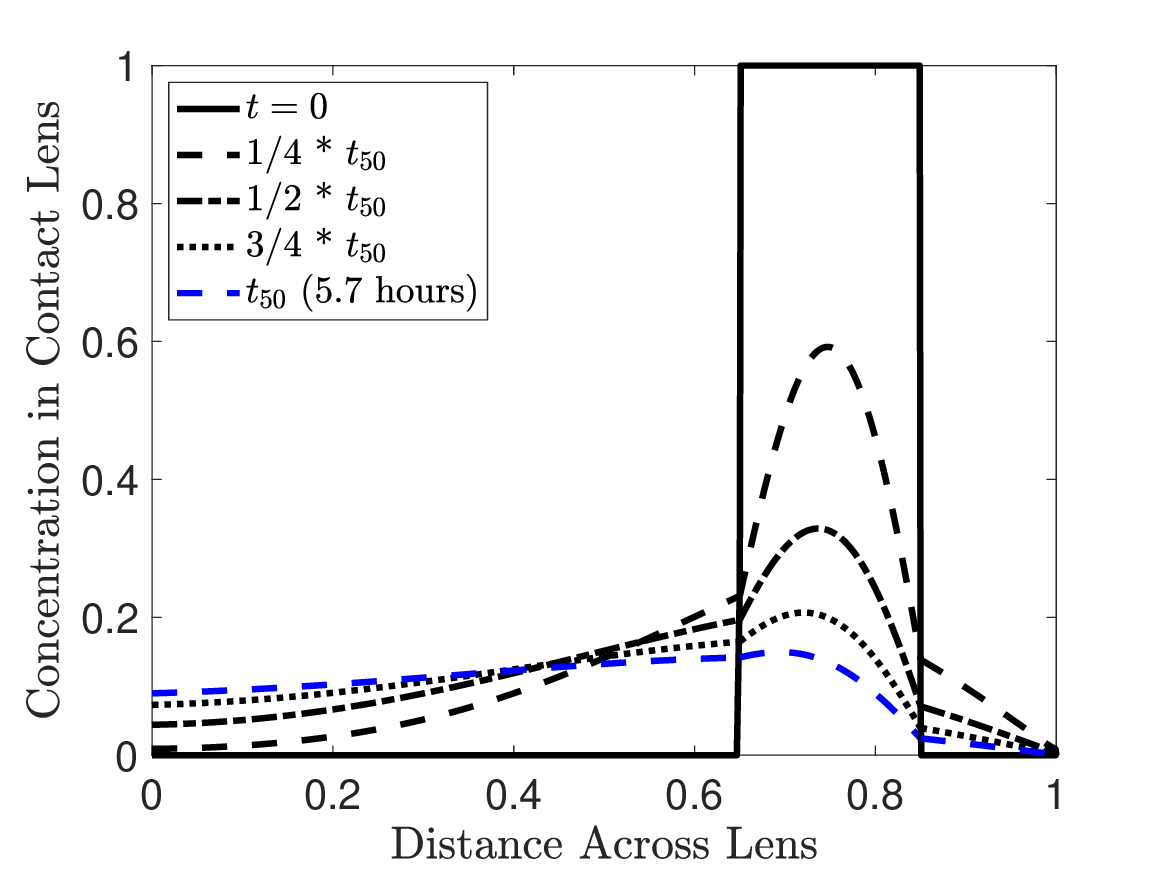}
\caption{
Eye wear with blinking setting: contact lens drug concentration profiles  from the post- to pre-lens side (posterior to anterior) for a range of equally-spaced time values, $t \in [0,t_{50}]$. 
All cases use $\Delta \bar{H} = 0.2$, $p=1$, and $\bar{k}_{\rm C}=0$.
The upper, middle, and lower rows use ${\cal D}=0.002$,
${\cal D}=0.02$, 
and  ${\cal D} = 0.1$.  
The left, middle, and right columns use $\bar{H}_{\rm mid}=0.25$, 
 $\bar{H}_{\rm mid} =0.5$, 
and  $\bar{H}_{\rm mid}=0.75$.  
The values for $t_{50}$ are reported in the legends and in Table~\ref{t50_table_kcZERO_kcNOTZERO}.
}
\label{fig_CL_Profiles_Hmid_variesP1}
\end{center}
\end{figure}

Figure~\ref{fig_CL_Profiles_Hmid_variesP1} shows CL drug concentration profiles for a representative set of parameter values. The pre-lens drug reduction parameter is $p=1$ (all pre-lens drug lost each blink) and the corneal permeability is $\bar{k}_{\rm C}=0$.
Since the rate of post-lens 
drug loss due to blinking is relatively small ($p_{\rm slide}^{\rm eff} \approx 0.0036$) and there is no drug transfer to the cornea, all scenarios shown in this figure correspond to relatively rapid drug loss on the pre-lens side and relatively slow drug loss on the post-lens side.
The top, middle, and bottom rows use
${\cal D}=0.002$, ${\cal D}=0.02$, and ${\cal D}=0.1$, respectively.
The left, middle, and right columns use
$\bar{H}_{\rm mid}=0.25$, $\bar{H}_{\rm mid}=0.5$, and
$\bar{H}_{\rm mid}=0.75$, respectively.
The corresponding values of $t_{50}$ are indicated 
in the legends and also reported in the upper portion of Table~\ref{t50_table_kcZERO_kcNOTZERO}.  

Spatial asymmetry in the drug concentration profiles develops because the mechanisms for drug transport out
of the pre-lens and post-lens differ substantially in this case, with relatively slow transport out of the post-lens compared to
the pre-lens.  This asymmetry in the concentration occurs even with a symmetrically-placed
polymer insert at $\bar{H}_{\rm mid} = 0.5$. 
The $t_{50}$ values are smaller with the polymer drug insert placed closer to the pre-lens side for ${\cal D}=0.1$; one might associate this reduction in $t_{50}$  with `proximity to an escape route' out the pre-lens side of the contact lens.
This trend is reversed when ${\cal D}=0.002$; here one might associate this behavior with an `entrapment effect,' whereby the drug in the polymer film that initially diffuses into the post-lens side of the CL (where $0 < \bar{z} < \bar{H}_1$) is temporarily trapped by the post-lens tear film on one side and the polymer insert on the other. 
At any given time, the drug mass in the post-lens hydrogel 
depends on both the thickness 
(i.e.~$H_1$) and concentration of that layer, which also varies from case to case as seen in
Figure~\ref{fig_CL_Profiles_Hmid_variesP1}.  Since the blink-induced post-lens drug loss 
is proportional to the post-lens 
concentration at the time of the blink,
higher concentrations yield more drug loss
(which is the case here for smaller $\bar{H}_{\rm mid}$).
The 
polymer layer best  acts as a trapping barrier for the drug in the post-lens hydrogel when ${\cal D}$ is small; the post-lens tear film is the other barrier in this scenario.  
This trapping is clear from the right-most column of Figure~\ref{fig_CL_Profiles_Hmid_variesP1};
in the upper right case (${\cal D}=0.002$) at $t=t_{50}$, the maximum concentration in the polymer layer is considerably larger than that in the post-lens hydrogel layer. A similar comparison made in the lower right case (${\cal D}=0.1$) shows the polymer and post-lens hydrogel concentrations at nearly the same level.

\begin{table}[h!]
\begin{center}
\begin{tabular}{ll|ccc}
$\bar{k}_{\rm C}=0$   &    &  \multicolumn{3}{c}{$\bar{H}_{\rm mid}$}  \\
    &   & 0.25 & 0.5 & 0.75 \\ \hline
& $0.002$ 
&  $51.5$ hrs
&  $61.4$ hrs
&   $70.2$ hrs \\                
${\cal D}$
& $0.02$
&  $16.1$ hrs
&  $16.1$ hrs
&   $15.6$ hrs\\                
& $0.1$  
&  $11.4$ hrs
&  $9.4$ hrs
&   $5.7$ hrs\\                
   \hline 
 \end{tabular}\\
 \vspace{0.1in}
 \begin{tabular}{ll|ccc}
$\bar{k}_{\rm C}=0.4$   &    &  \multicolumn{3}{c}{$\bar{H}_{\rm mid}$} \\
   &    & 0.25 & 0.5 & 0.75 \\ \hline
& $0.002$ 
&  $40.6$ hrs
&  $39.2$ hrs
&  $40.6$ hrs \\                
${\cal D}$
& $0.02$
&  $8.3$ hrs
&  $7.7$ hrs
&   $8.3$ hrs \\                
& $0.1$  
&  $3.8$ hrs
&  $4.3$ hrs
&  $3.8$ hrs \\                
   \hline 
 \end{tabular}\\
\end{center}
\caption{Eye wear with blinking setting: dimensional ${t}_{50}$ values (in hours) 
predicted by the model.  The upper table corresponds to  
the results shown in Figure~\ref{fig_CL_Profiles_Hmid_variesP1}, which 
have $p=1$, $\Delta \bar{H}=0.2$, and $\bar{k}_{\rm C}=0$.   
The lower table corresponds to  
the results shown in Figure~\ref{fig_CL_Profiles_Hmid_variesP1_KCnotZERO}, which 
have $p=1$, $\Delta \bar{H}=0.2$, and $\bar{k}_{\rm C}=0.4$.
The asymmetry in $t_{50}$ values with respect to midpoint positioning of the polymer insert depends non-trivially on the diffusion ratio ${\cal D}$
and on $\bar{k}_{\rm C}$.}
\label{t50_table_kcZERO_kcNOTZERO}
\end{table}

\newpage

\begin{figure}[h!]
\begin{center}
\includegraphics[width=0.78\linewidth]{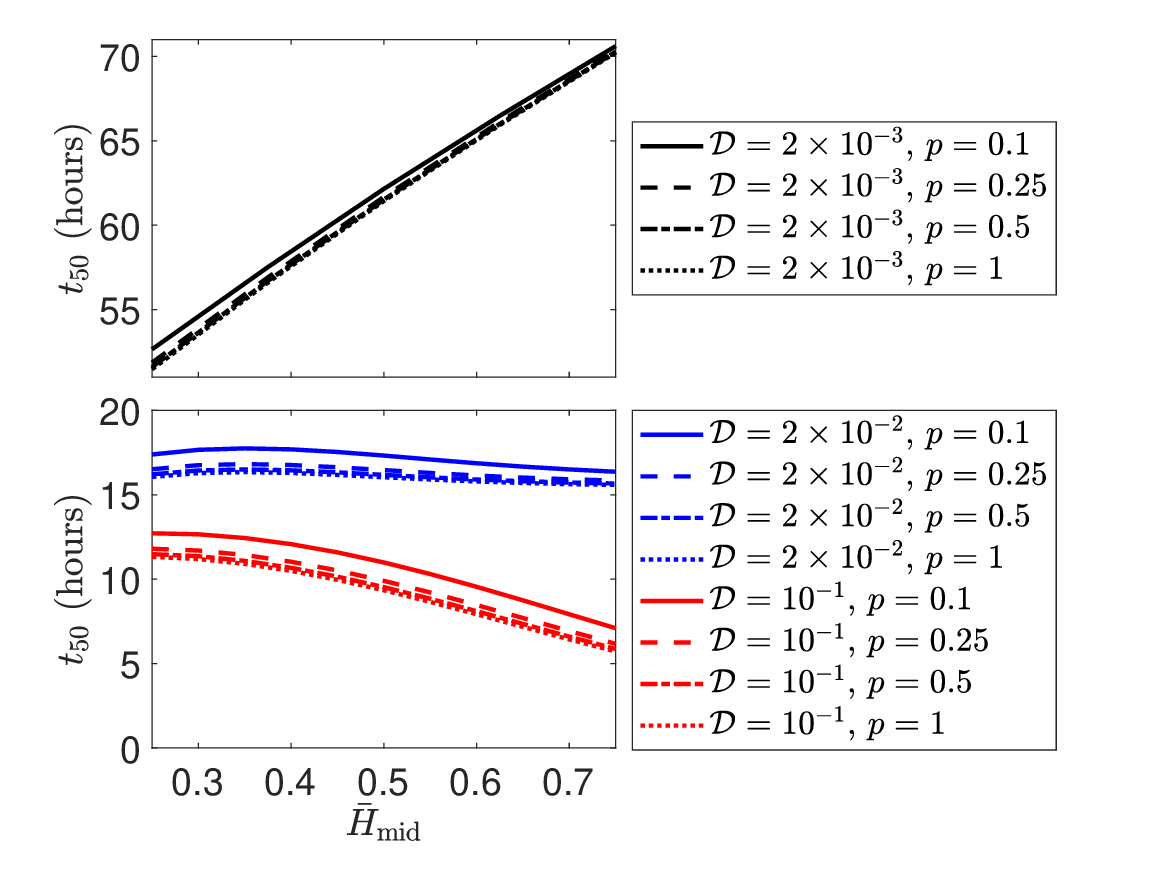}
\caption{Eye wear with blinking setting: dimensional $t_{50}$ values (in hours) for drug release from a composite lens during wear.   Predictions are shown for 
${\cal D} = 0.002$ (upper black curves), ${\cal D} = 0.02$ (middle blue curves) 
and ${\cal D}=0.1$ (lower red curves) and for four values of the pre-lens drug reduction parameter
$p$.  Note that there is a break in the vertical (time) axis between the upper and lower plots, but  the indicated time range covers 20 hours.
We use $\Delta \bar{H}=0.2$  and the range $\bar{H}_{\rm mid} \in [0.25,0.75]$. 
}
\label{fig_t50_blink}
\end{center}
\vspace{-2mm}
\end{figure}

Figure~\ref{fig_t50_blink} shows dimensional $t_{50}$ values computed for drug release
from a composite lens during eye-wear for parameter
ranges including those used in 
Figure~\ref{fig_CL_Profiles_Hmid_variesP1}.
The upper (black), middle (blue), and lower (red) sets of curves have ${\cal D}=0.002$,  
${\cal D}=0.02$,
and ${\cal D}=0.1$, respectively,
and dimensional $t_{50}$ values are reported over the range $\bar{H}_{\rm mid} \in [0.25,0.75]$.   We fix $\Delta \bar{H}=0.2$ and $\bar{k}_{\rm C}=0$.
In terms of computation time, a typical data point for the ${\cal D}=0.002$ cases (with $t_{50}$ in the 50  to 70 hour range) took approximately 8 to 12 hours on a standard laptop running {\tt Matlab}.\footnote{MacBook Pro with 2.3 GHz 8-Core Intel Core i9 Processor. Recall that one blink occurs every $10$ seconds (360 blinks every hour); e.g.~$5$ hours corresponds to $1,800$ blinks and $20$ hours corresponds to $7,200$ blinks. }

A first observation from Figure~\ref{fig_t50_blink} is that as the value of ${\cal D}$ is decreased,
corresponding to relatively slower diffusion in the polymer layer, $t_{50}$ values increase, consistent with trends observed during drug release in the vial setting. 
Specifically, we observe $t_{50}$ values in the $6$ to $12$ hour range for ${\cal D}=0.1$ and in the
$51$ to $71$ hour range for ${\cal D}=0.002$ (see also Table~\ref{t50_table_kcZERO_kcNOTZERO}).
As a reference, for the vial release example with ${\cal D}=0.1$, we estimate $\bar{t}_{50} = {\cal O}(0.1)$ for parameter values near $\Delta \bar{H}=0.2$ over the range $\bar{H}_{\rm mid} \in [0.25,0.75]$.  This yields a dimensional 
vial release $t_{50} \approx 0.1 (H_{\rm cl}^2/D_2)$ of 3 to 4 hours.

A second observation from Figure~\ref{fig_t50_blink} is that the dependence of $t_{50}$ on the polymer layer midline, $\bar{H}_{\rm mid}$, reveals three distinct trends for different ranges of $\cal{D}$: monotonically decreasing, non-monotonic, and monotonically increasing.
For relatively large values of ${\cal D}$ (e.g.~red curves with ${\cal D}=0.1$ in Figure~\ref{fig_t50_blink}), $t_{50}$  decreases as the 
midline is moved from the post-lens
side ($\bar{H}_{\rm mid} < 0.5$) towards the pre-lens side ($\bar{H}_{\rm mid} > 0.5$). 
As noted in the discussion of
Figure~\ref{fig_CL_Profiles_Hmid_variesP1},
for these values of ${\cal D}$, positioning the polymer insert farther from the pre-lens side of the CL leads to a larger $t_{50}$ (and slower drug release). 
Interestingly, however, this dependence of $t_{50}$ on $\bar{H}_{\rm mid}$ is exactly reversed in the case of relatively small values of ${\cal D}$ (e.g.~black curves with ${\cal D}=0.002$ in Figure~\ref{fig_t50_blink}).
In this situation,
$t_{50}$ {\it increases} as the 
midline is moved from the post-lens side ($\bar{H}_{mid} < 0.5$) towards the pre-lens side ($\bar{H}_{\rm mid} > 0.5$). This counter-intuitive result arises as a consequence of a trapping effect, already noted in the discussion of Figure~\ref{fig_CL_Profiles_Hmid_variesP1}, that occurs for the drug in the post-lens hydrogel. 
An intermediate case  
(e.g., blue curves with ${\cal D}=0.02$ in Figure~\ref{fig_t50_blink}) arises
for which $t_{50}$ varies non-monotonically with $\bar{H}_{\rm mid}$.
Here, the dependence of $t_{50}$ on $\bar{H}_{\rm mid}$ is relatively weak, with values falling in the $15$  to $16$ hour range.

A final observation about 
Figures~\ref{fig_CL_Profiles_Hmid_variesP1} and~\ref{fig_t50_blink} relates to an anterior--posterior invariance.  
As the corneal and lid permeabilities are set to zero and the pre- and post-lens thicknesses are 
equal, the values of $p$, $p_{\rm slide}^{\rm eff}$, and $\bar{H}_{\rm mid}$ are 
what mathematically distinguish the post- and pre-lens tear films.  In particular, the predictions of Figures~\ref{fig_CL_Profiles_Hmid_variesP1} and~\ref{fig_t50_blink} are invariant under the transformations: $p \rightarrow p_{\rm slide}^{\rm eff}$,  $p_{\rm slide}^{\rm eff} \rightarrow p$, and $\bar{H}_{\rm mid} \rightarrow 1 - \bar{H}_{\rm mid}$; i.e., an anterior--posterior `flip' of the configuration.
For example, the conclusions drawn from Figure~\ref{fig_t50_blink} regarding a trapping effect for ${\cal D}=0.002$ and a positive correlation of $t_{50}$ with $\bar{H}_{\rm mid}$ when the pre-lens 
clearance rate greatly exceeds that of the post-lens ($p \gg p_{\rm slide}^{\rm eff}$) are reversed 
when the post-
and pre-lens 
clearance rates  are exchanged: then, $t_{50}$ negatively correlates with $\bar{H}_{\rm mid}$.  
While in reality, nonzero corneal  
and/or lid permeabilities yield more complicated 
 pre- and post-lens tear film  drug clearance rates than represented in 
Figures~\ref{fig_CL_Profiles_Hmid_variesP1} and~\ref{fig_t50_blink}, 
the transformation $\bar{H}_{\rm mid} \rightarrow 1 - \bar{H}_{\rm mid}$ provides an approximate understanding of  the reversed roles of the pre- and post-lens tear films.

\begin{figure}[h!]
\begin{center}
\vskip 0.05in
\includegraphics[width=5.0cm]{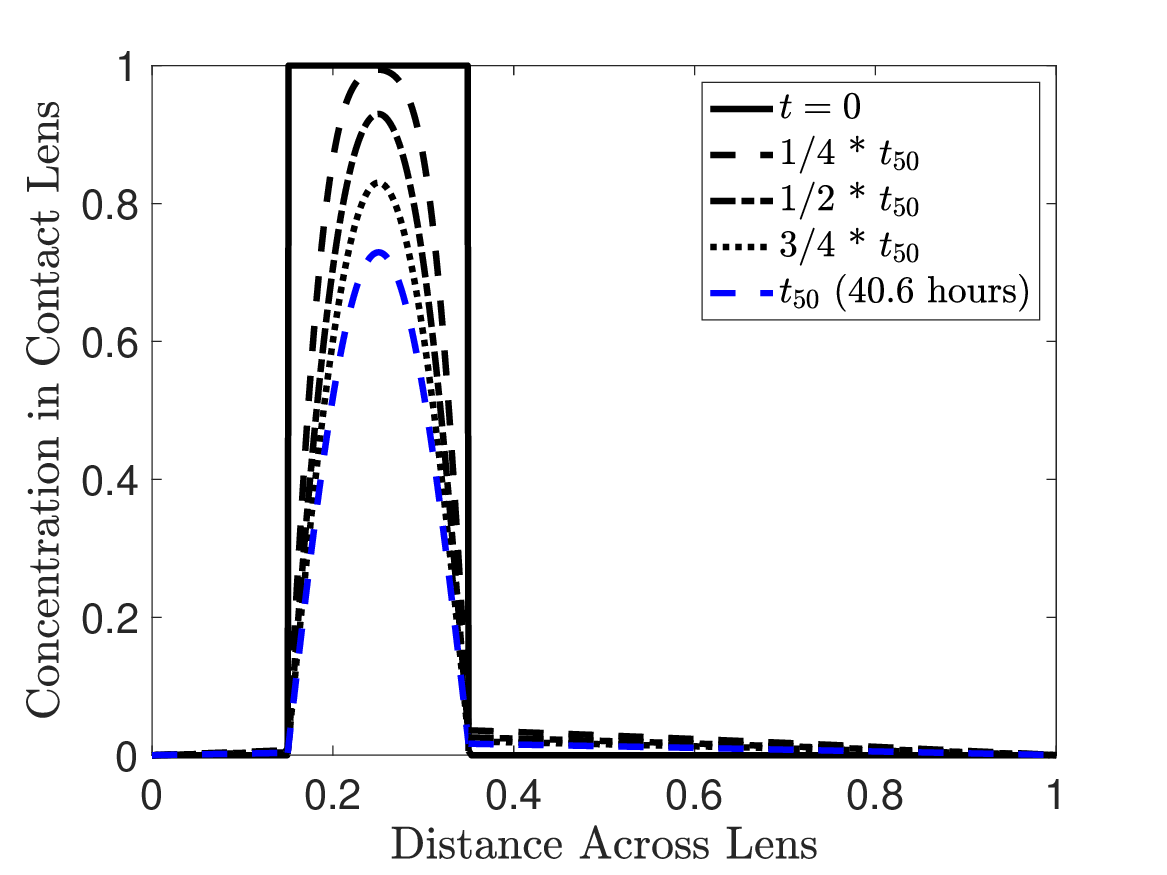}
\includegraphics[width=5.0cm]{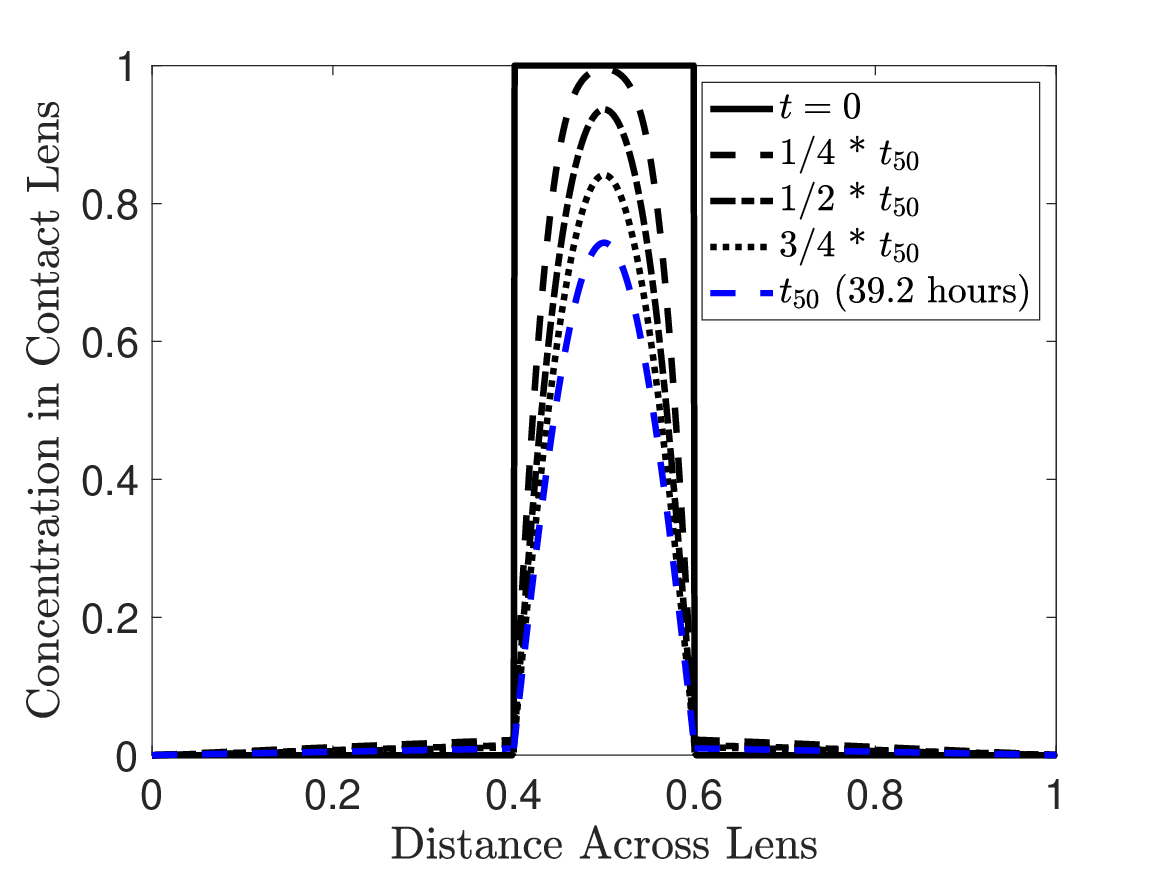}
\includegraphics[width=5.0cm]{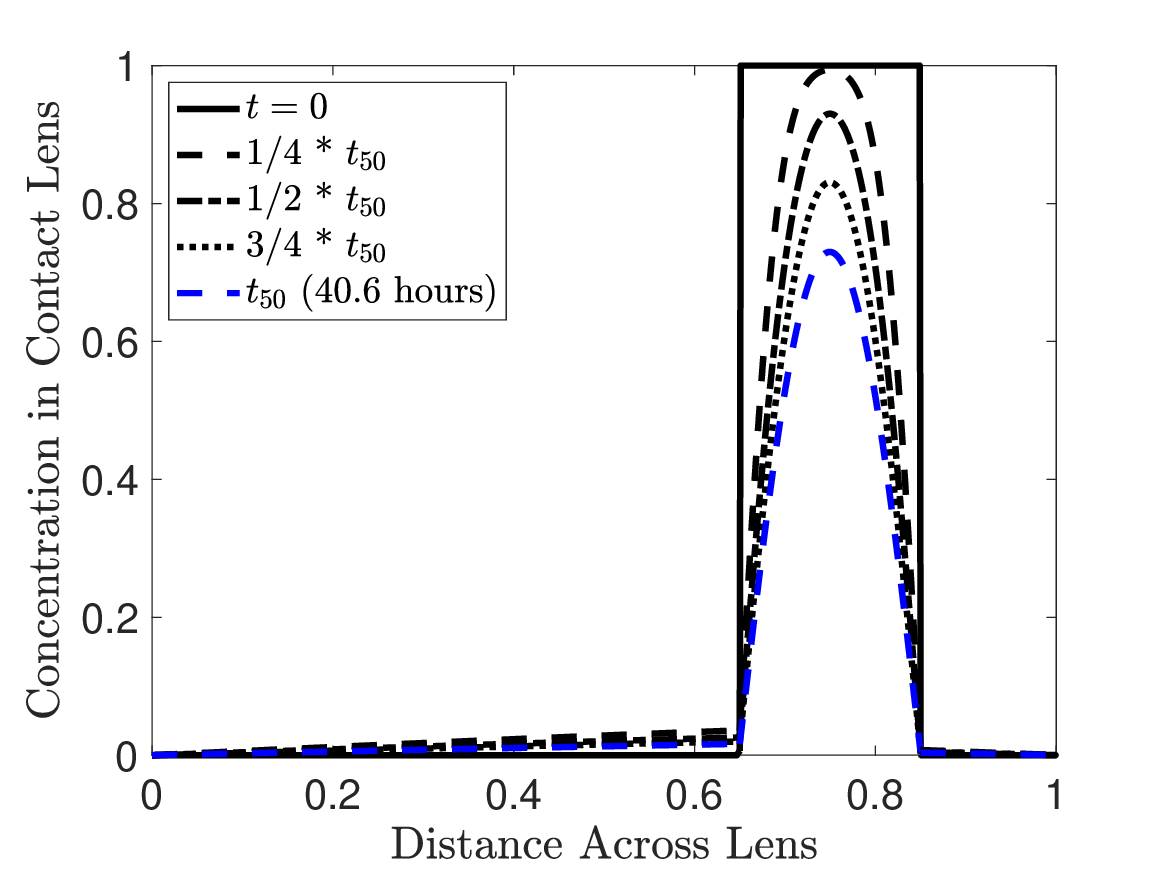} \\
\includegraphics[width=5.0cm]{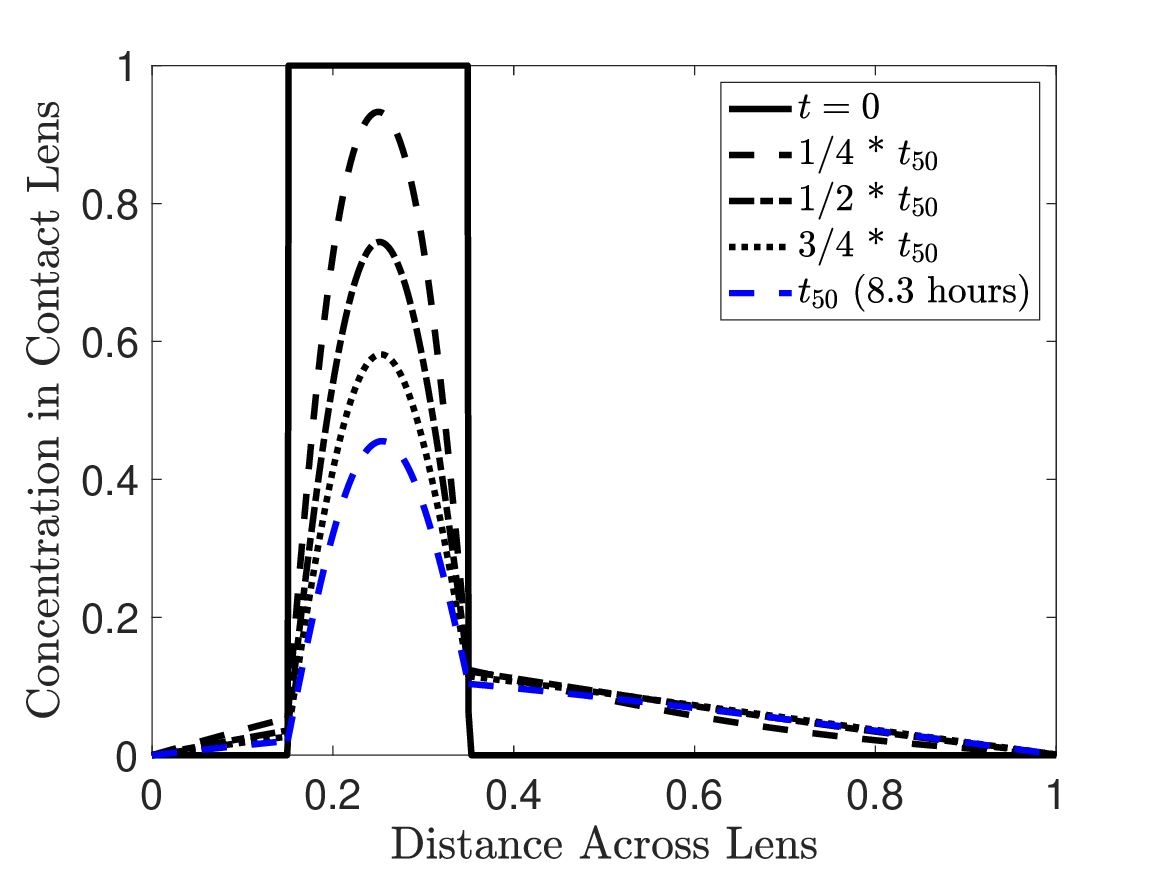}
\includegraphics[width=5.0cm]{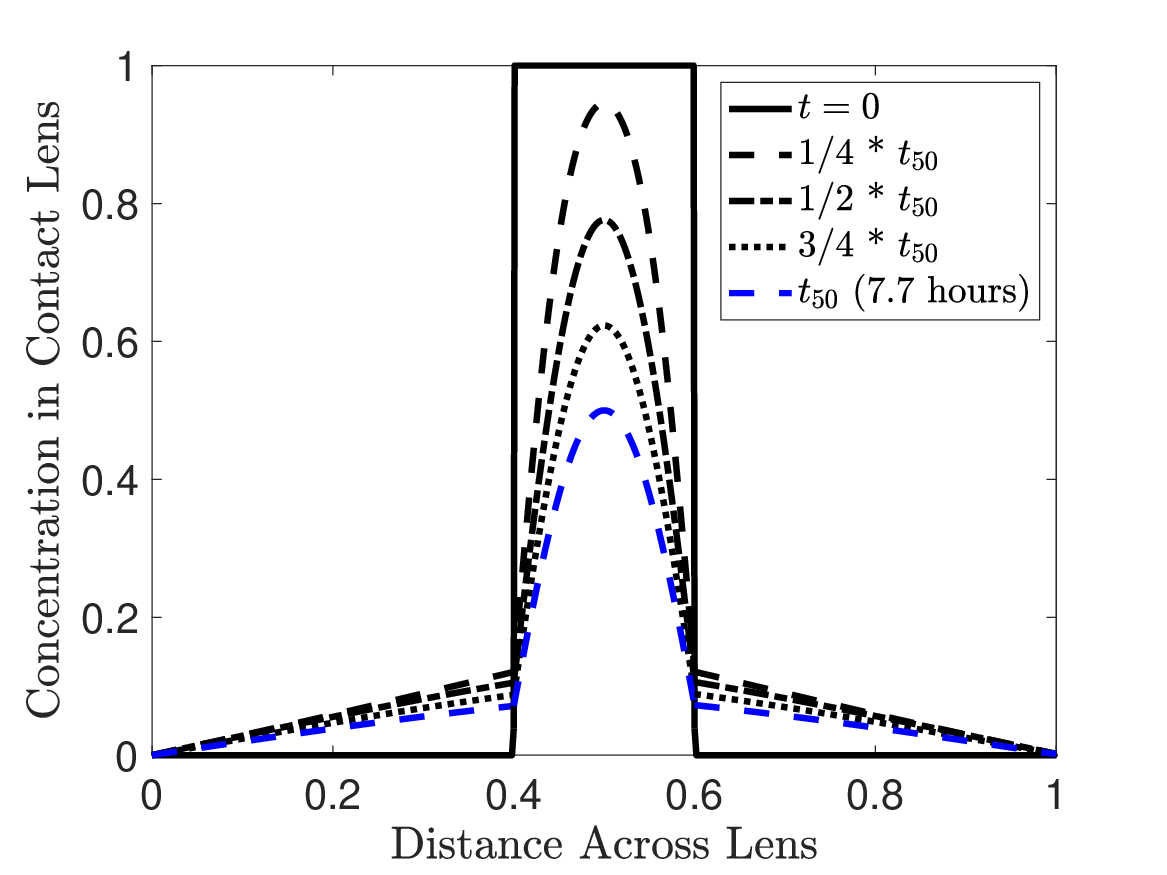}
\includegraphics[width=5.0cm]{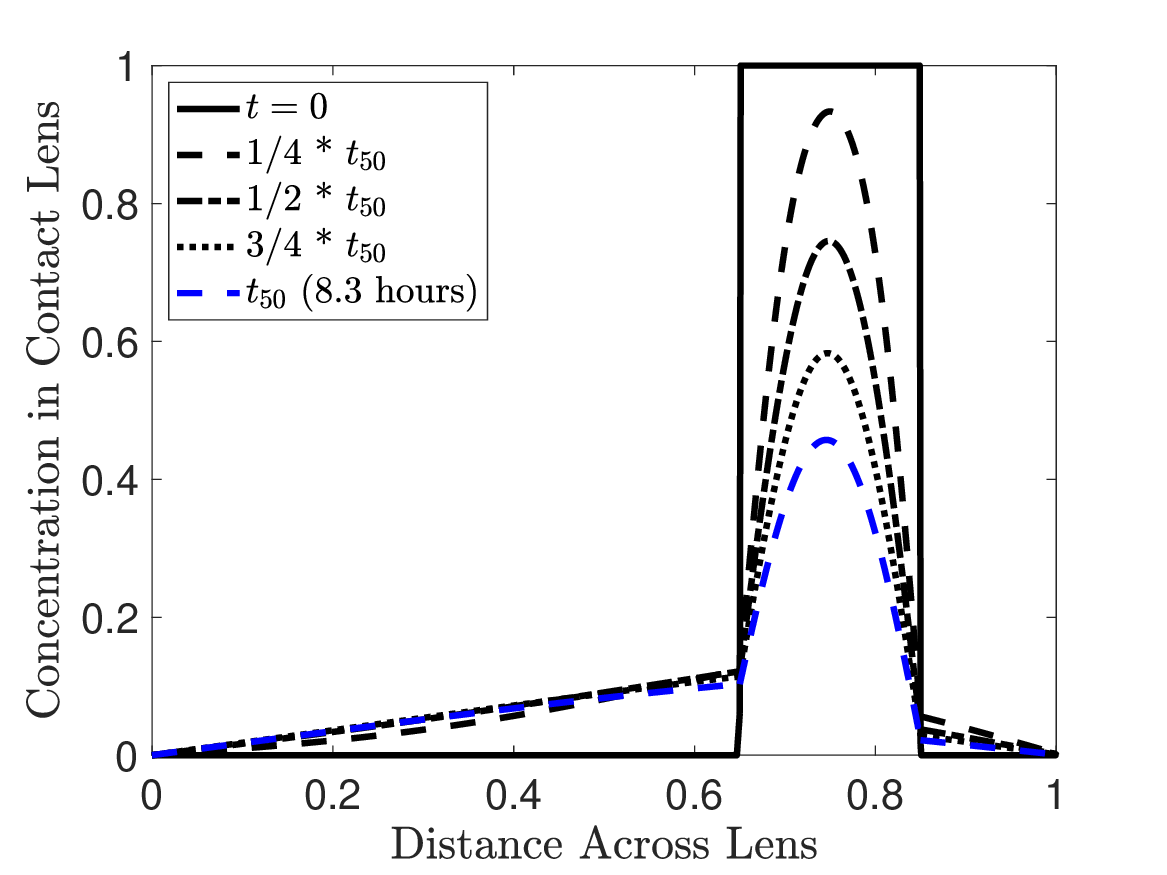} \\
\includegraphics[width=5.0cm]{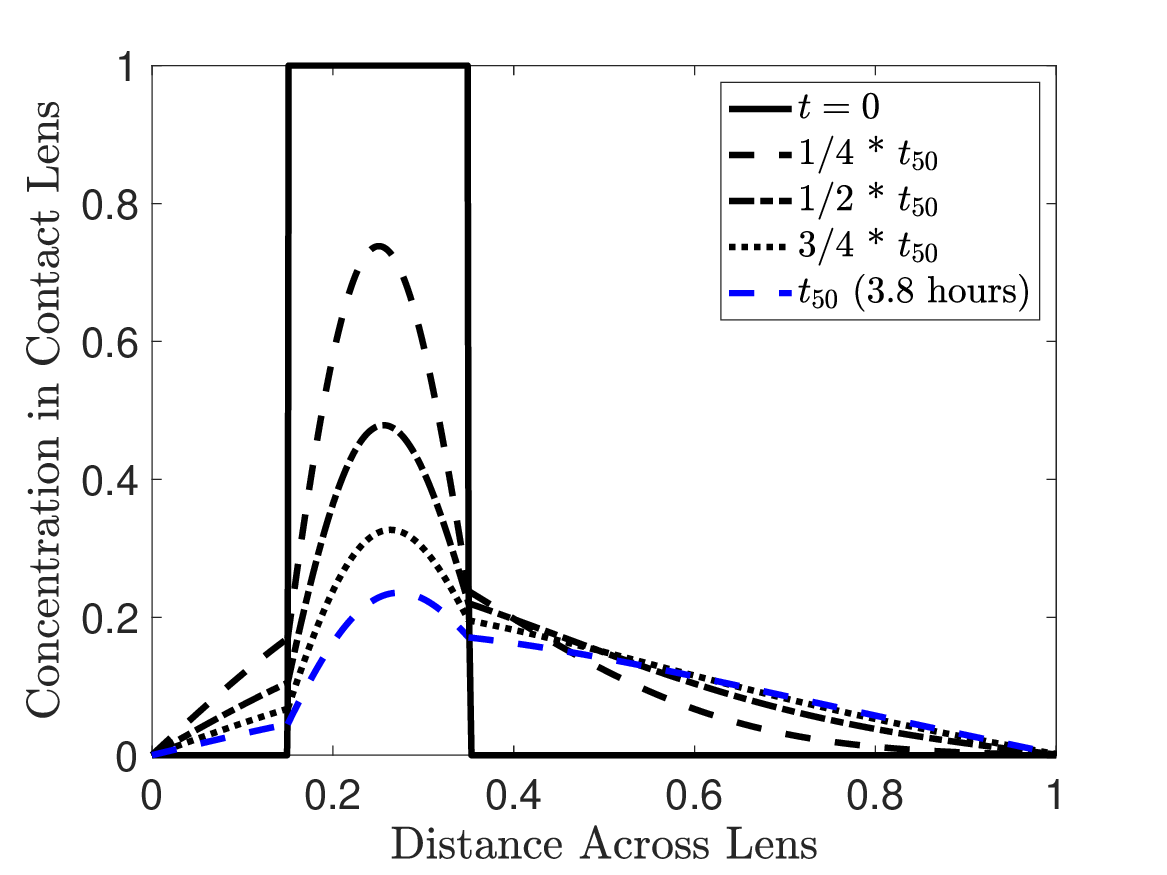}
\includegraphics[width=5.0cm]{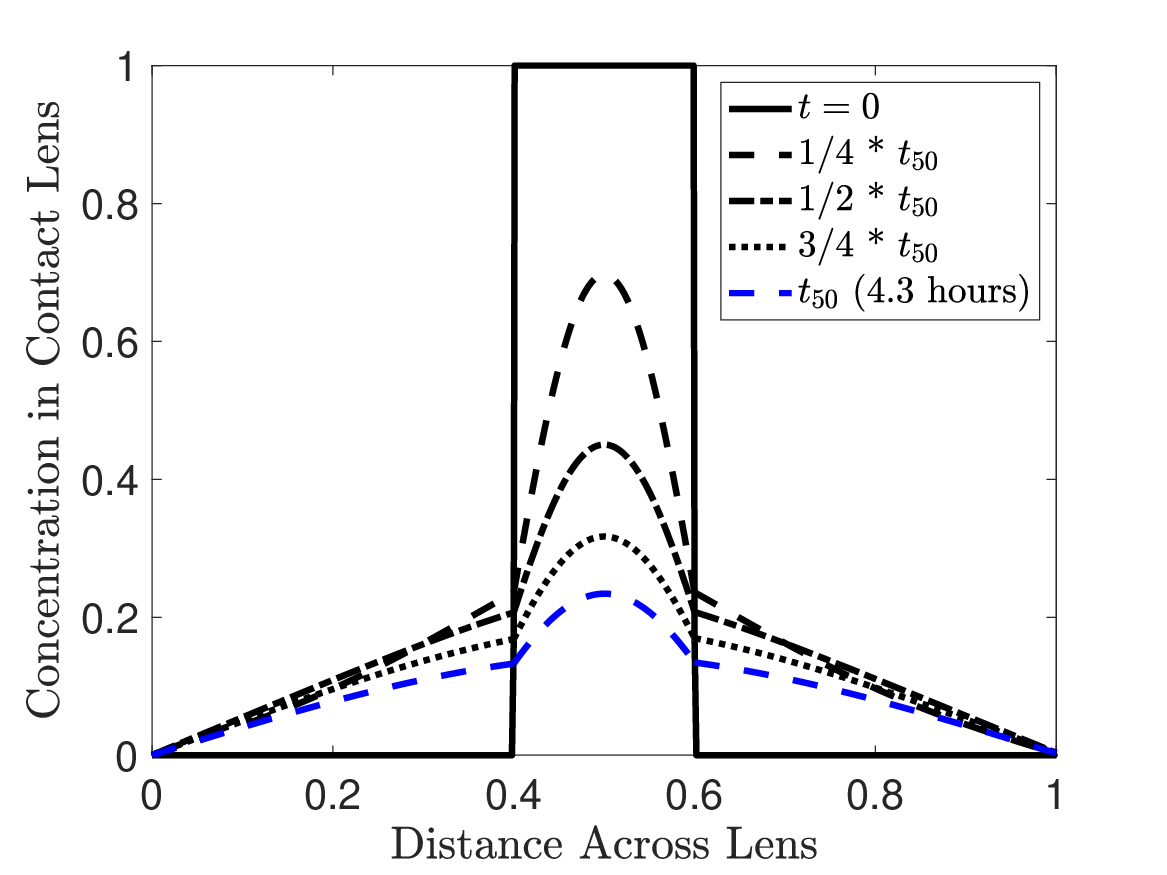}
\includegraphics[width=5.0cm]{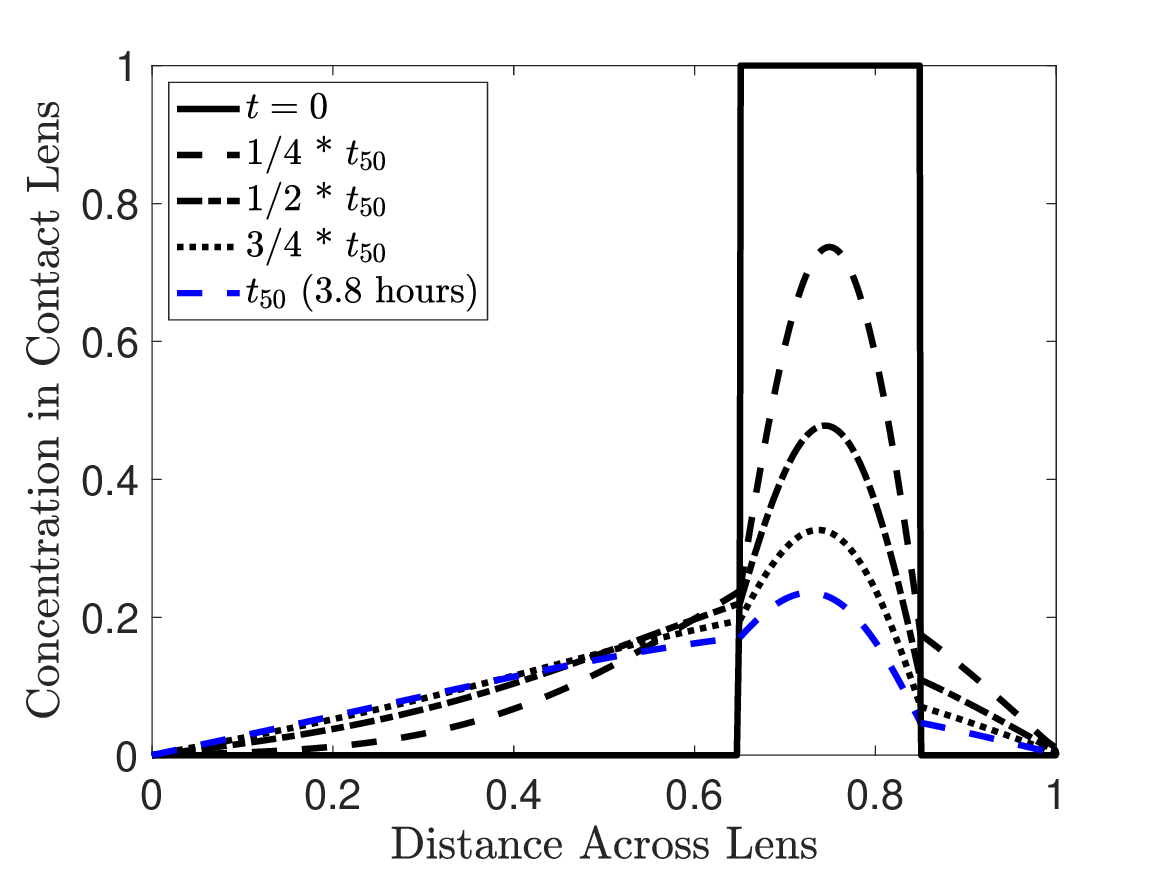}
\caption{
Eye wear with blinking setting: contact lens drug concentration profiles  from the post- to pre-lens side (posterior to anterior) for a range of equally-spaced time values, $t \in [0,t_{50}]$. 
All cases use $\Delta \bar{H} = 0.2$, $p=1$, and $\bar{k}_{\rm C}=0.4$.
The upper, middle, and lower rows use ${\cal D}=0.002$,
${\cal D}=0.02$, 
and  ${\cal D} = 0.1$.  
The left, middle, and right columns use $\bar{H}_{\rm mid}=0.25$, 
 $\bar{H}_{\rm mid} =0.5$, 
and  $\bar{H}_{\rm mid}=0.75$.  
The values for $t_{50}$ are reported in the legends and in Table~\ref{t50_table_kcZERO_kcNOTZERO}.
}
\label{fig_CL_Profiles_Hmid_variesP1_KCnotZERO}
\end{center}
\end{figure}

As a comparison to 
Figure~\ref{fig_CL_Profiles_Hmid_variesP1}, in Figure~\ref{fig_CL_Profiles_Hmid_variesP1_KCnotZERO} we show analogous results
with significantly enhanced drug transport out of the post-lens tear film associated with a nonzero corneal permeability, $\bar{k}_{\rm C} =0.4$ (but otherwise the same
parameter values as in Figure~\ref{fig_CL_Profiles_Hmid_variesP1}). The $t_{50}$ numbers
are also summarized for both $\bar{k}_{\rm C} =0$ and $\bar{k}_{\rm C} =0.4$
cases in Table~\ref{t50_table_kcZERO_kcNOTZERO}.
When $\bar{k}_{\rm C} =0.4$, $t_{50}$ is notably reduced compared to the
$\bar{k}_{\rm C}=0$ cases shown in Figure~\ref{fig_CL_Profiles_Hmid_variesP1}. 
In Figure~\ref{fig_CL_Profiles_Hmid_variesP1_KCnotZERO} we observe a more symmetric dependence of the concentration profiles on $\bar{H}_{\rm mid}$;  drug transport out of the post-lens tear film, both into the cornea and lost during a blink due to lens motion, is more balanced
with the drug transport out of the pre-lens tear film, with $p=1$ (when each blink zeros-out the pre-lens drug concentration).
The CL drug concentration profiles are
visually symmetric with $\bar{H}_{\rm mid}=0.5$ (center column plots)
and the results with $\bar{H}_{\rm mid}=0.25$ (left column plots) appear approximately as the mirror
images of those for $\bar{H}_{\rm mid}=0.75$ (right column plots).
The $t_{50}$ values also carry this approximate symmetry as well.  When $\bar{k}_{\rm C}=0.4$
and $p=1$ the $t_{50}$ numbers are approximately symmetric 
with respect to the midpoint positioning of the polymer insert.  
Like the $\bar{k}_{\rm C}=0$ case, $t_{50}$ depends in a nontrivial way on the diffusion coefficient ratio ${\cal D}$.  For the $\bar{k}_{\rm C}=0.4$ case,  the center position ($\bar{H}_{\rm mid}=0.5$) yields the larger $t_{50}$ value when ${\cal D}=0.1$; when ${\cal D}=0.02$ or $0.002$, the off-center values ($\bar{H}_{\rm mid} = 0.25$ and $0.75$) yield the larger value of $t_{50}$.

In the next section, we explore another important aspect of composite lens performance specifically relating to the practical requirement of storage of drug-loaded composite contact lenses during the interval between lens manufacture/drug-loading of the lens and actual
CL wear.

\section{Drug release from a composite contact lens: blister pack}
 \label{sec:blister}

  Contact lenses are often stored in ``blister packs''--small vials typically  in the 1-3 mL range \cite{hamilton2007patent}, but sometimes as small as 0.2 mL and as large as 5 mL. For a drug-eluting lens, we assume that the blister pack is loaded with drug. Depending on the relative concentrations over time, drug from the blister pack may diffuse into the lens, or drug may diffuse out of the lens into the blister pack. The diffusion direction will depend on the loading concentrations of the blister pack and drug-polymer film, lens geometric properties, and time (the flux direction may switch). Analyzing this scenario is important to answer questions related to storage considerations of drug-eluting CLs, such as, What blister pack settings preserve a significant amount of drug in the lens for a reasonable length of time? In this section, we will continue to use $t_{50}$ as a metric of therapeutic effectiveness. 

 \subsection{Dimensionless model formulation}

 This scenario can be viewed as a simplification of the eye model setting considered in Section \ref{sec:eye}, in which we ignore blinking, evaporation, osmosis, and eyelid absorption, and the 
 blister pack volume approaches that of the vial setting. 
 We use notation similar to Section \ref{sec:vial} to refer to the blister pack concentration $\bar{C}_{\rm B}$, which applies to the anterior (referred to with ``pre'') and posterior (``post'') regions of the blister pack in our model. Let $\bar{h}_B^{\rm pre}$ and $\bar{h}_B^{\rm post}$  represent the constant, dimensionless heights of the blister pack ``pre''- and ``post''-lens regions.  
 The dimensionless version of the model is very similar to that in Section \ref{sec:eye}; we note the deviations.
 For the blister pack compartment, we have
 \bea
 \left( \bar{h}_B^{\rm pre} + \bar{h}_B^{\rm post} \right)  \frac{d   \bar{C}_B }{d \bar{t}} & = & \bar{D}_2  \left( \left. \frac{\partial  \bar{C}_2^{\rm post}}{\partial \bar{z}} \right|_{\bar{z}=0} - \left. \frac{\partial  \bar{C}_2^{\rm pre}}{\partial \bar{z}} \right|_{\bar{z}=1} \right),
 \eea
 where the dimensionless hydrogel diffusion coefficient is the same as in Section \ref{sec:eye}: $
 \bar{D}_2 = D_2 \tau/H_{\rm cl}^2$.
 The blister pack dynamics are coupled to those in the CL via the model already presented in Section~\ref{sec-dimensionless_vial}.  
 At the anterior blister pack--contact lens interface, $\bar{z}=1$,
\bea
\bar{C}_2^{\rm pre}(1,\bar{t}) & = & k \bar{C}_B(\bar{t}).
\eea
The dynamics inside the CL remain the same as in Section~\ref{sec-dimensionless_vial}; see \eqref{eq:C2pre_diff}--\eqref{eq:C2post_diff}. 
At the posterior blister pack--contact lens interface, $\bar{z}=0$ we have
\bea
\bar{C}_2^{\rm post}(0,\bar{t}) & = & k \bar{C}_B(\bar{t}).
\eea
 Since the blister pack may have a loading concentration, the drug mass in the blister pack outside the lens, $\bar{\mathcal{M}}_B$, may start nonzero and increase or decrease over time. In dimensionless form, this is
 \begin{equation}
    \bar{\mathcal{M}}_B(\bar{t}) = \bar{C}_B^{\rm load} \bar{V}_B  +  (\bar{H}_2 - \bar{H}_1) - \left( \int_0^{\bar{H}_1} \bar{C}_2^{\rm post} (\bar{z},\bar{t}) d\bar{z} + \int_{\bar{H}_1}^{\bar{H}_2} \bar{C}_1(\bar{z},\bar{t}) d \bar{z} + \int_{\bar{H}_2}^{1} \bar{C}_2^{\rm pre}(\bar{z},\bar{t}) d \bar{z} \right),
    \label{eq:Mbar_blister}
 \end{equation}
 where  $\bar{C}_B^{\rm load} = C_B^{\rm load}/C_{\rm load}$ and $\bar{V}_B = V_B/A_{\rm poly} H_{\rm cl}$, where $C_B^{\rm load}$ is the blister pack loading concentration and $V_B$ is the blister pack volume. Here we have used the scaling $\bar{\cal M}_B = M_B/(C_{\rm load} A_{\rm poly} H_{\rm cl})$ for consistency with Section  \ref{sec:eye}. Note, \eqref{eq:Mbar_blister} only differs from \eqref{eq:calMbar} by the inclusion of the first term.

 \subsection{Numerical investigation}

 In this section, unless otherwise stated, the polymer insert thickness is $\Delta \bar{H}=0.2$, the partition coefficient is $k =2$, the blister pack loading ratio relative to the drug-polymer insert is $\bar{C}_{B}^{\rm load} = 0.01$, the blister pack volume is $V_B = 0.2$ mL, and $h_{\rm pre} = h_{\rm post} \approx 1.6$ mm based on $h_{\rm pre} = h_{\rm post} = \frac{1}{2} V_B/A_{\rm poly}$. Whereas the other parameter values are selected with regards to the literature \cite{ross2019topical,hamilton2007patent}, the relative blister pack  concentration, $\bar{C}_B^{\rm load}$, is chosen arbitrarily and will be investigated later. 
 
\begin{figure}[t!]
\begin{center}
\vskip 0.05in
\includegraphics[width=5.9cm]{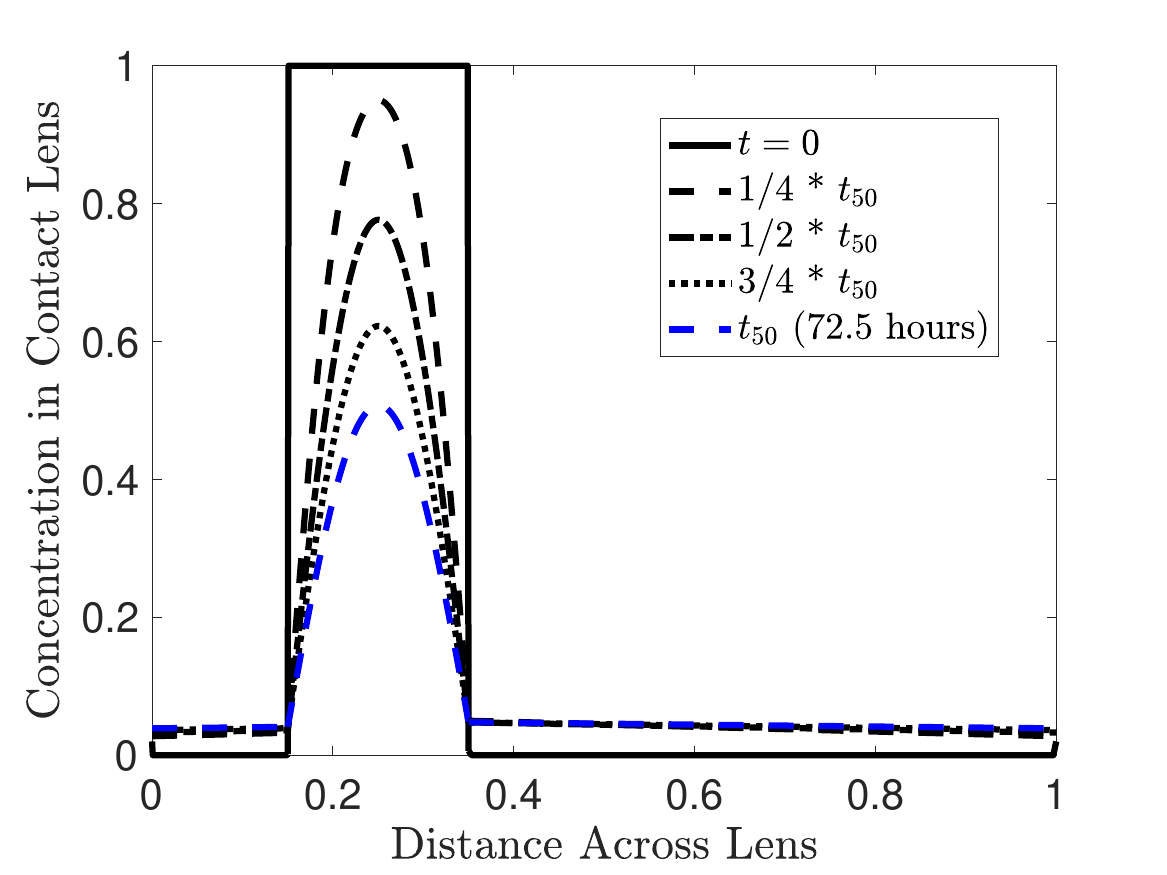}
\includegraphics[width=5.9cm]{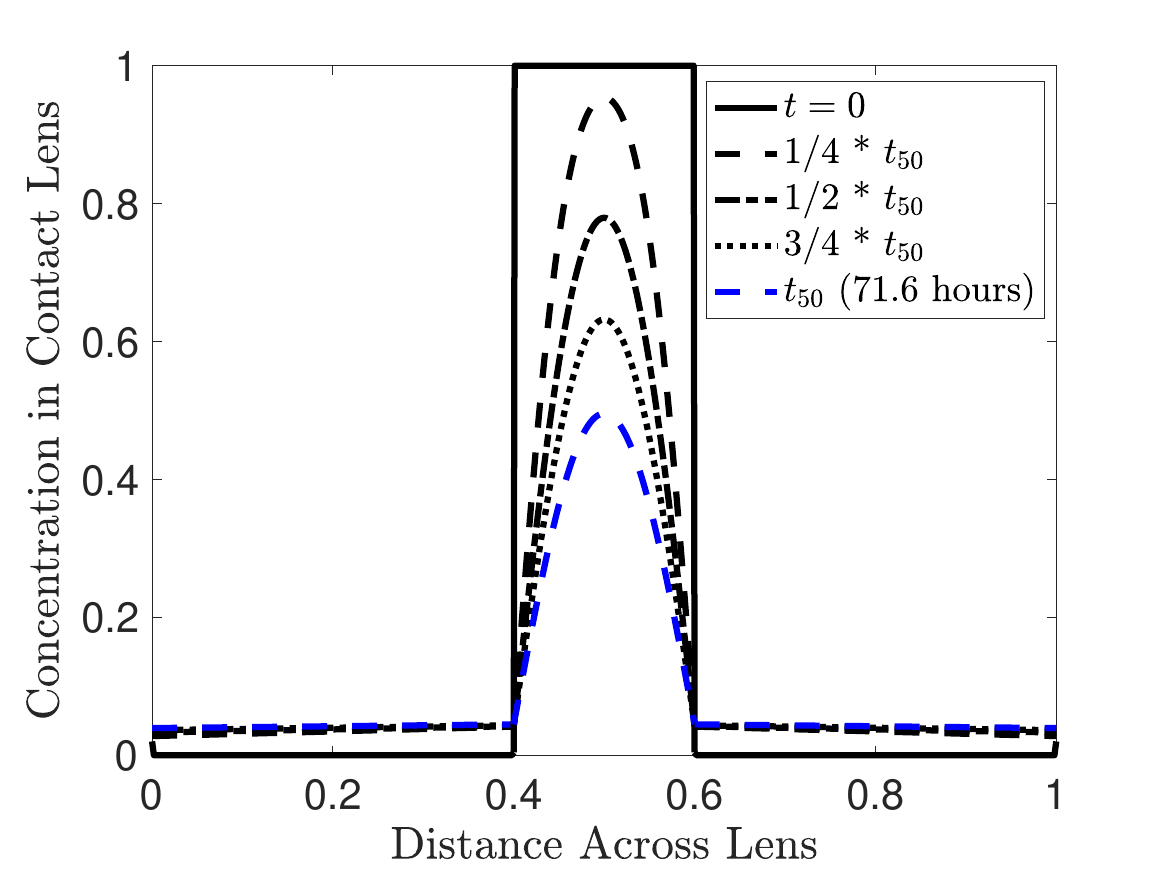} \\
\includegraphics[width=5.9cm]{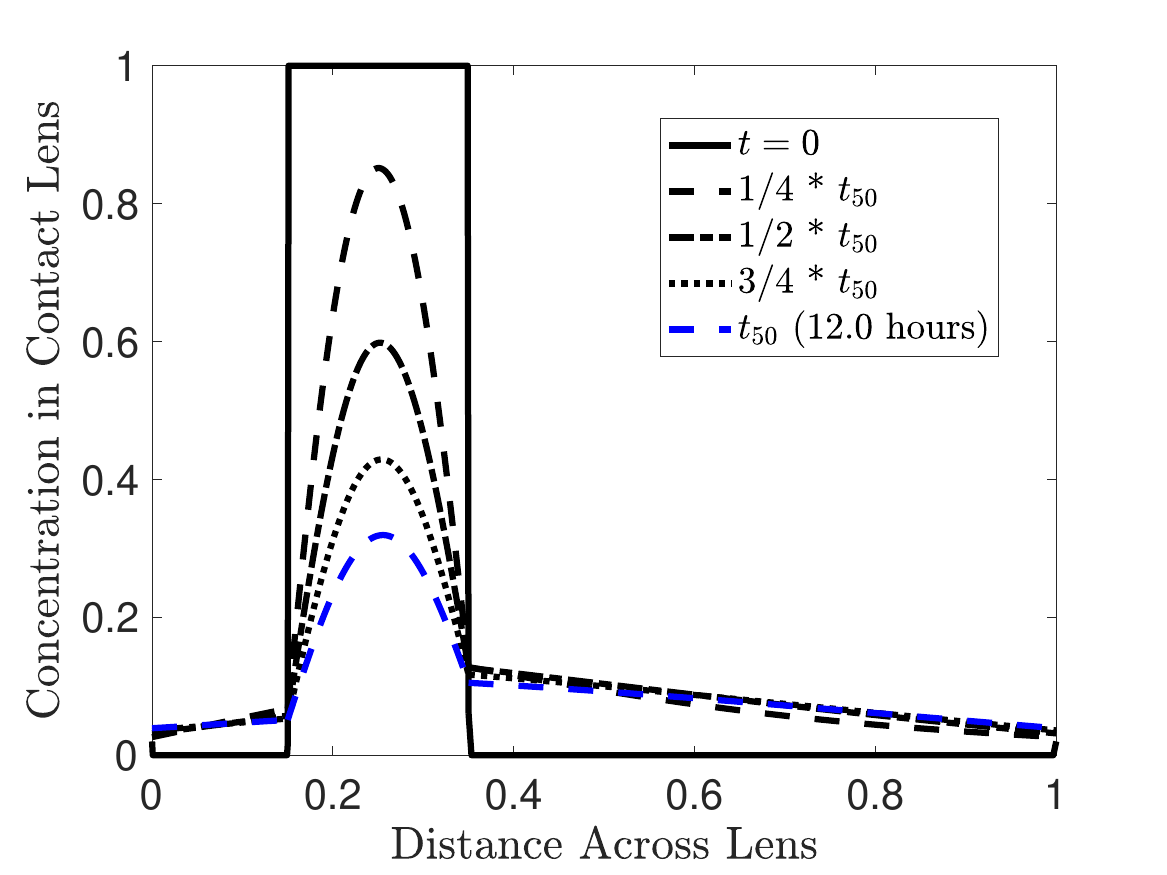}
\includegraphics[width=5.9cm]{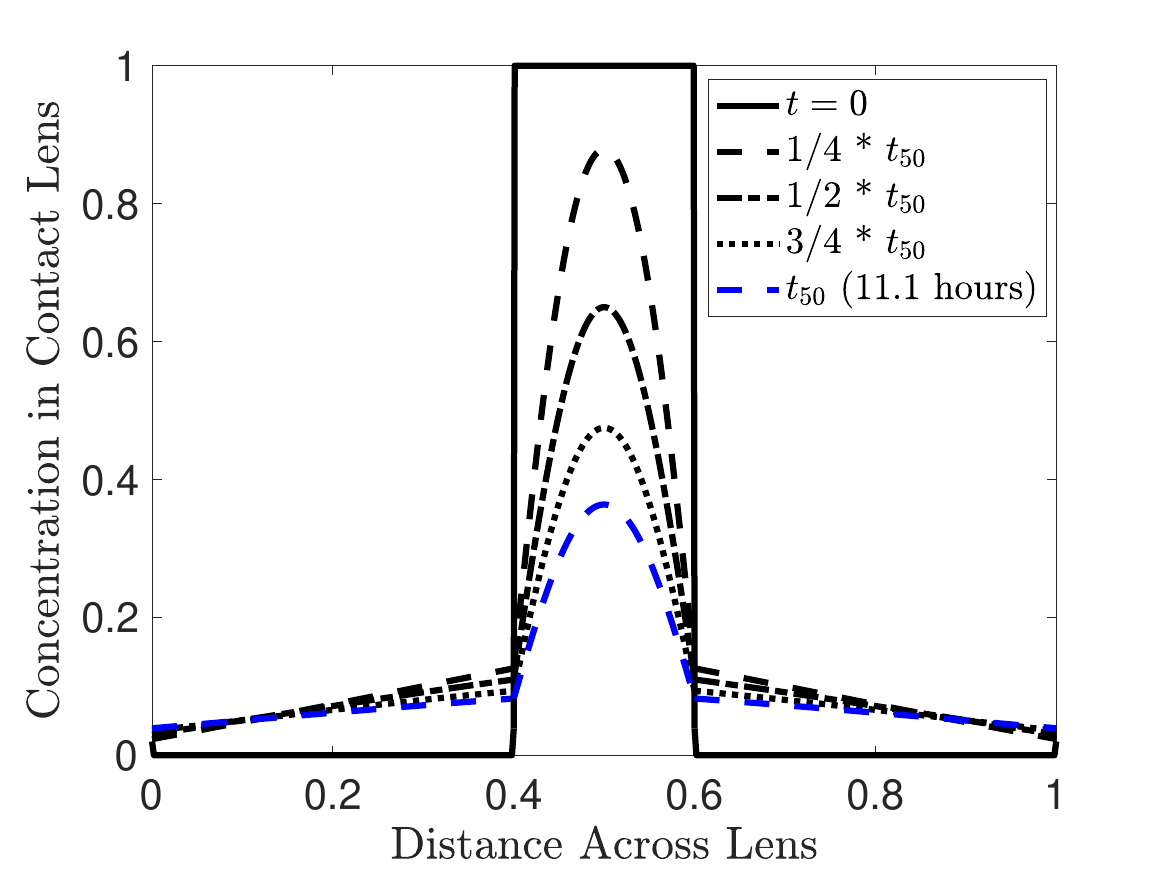} \\
\includegraphics[width=5.9cm]{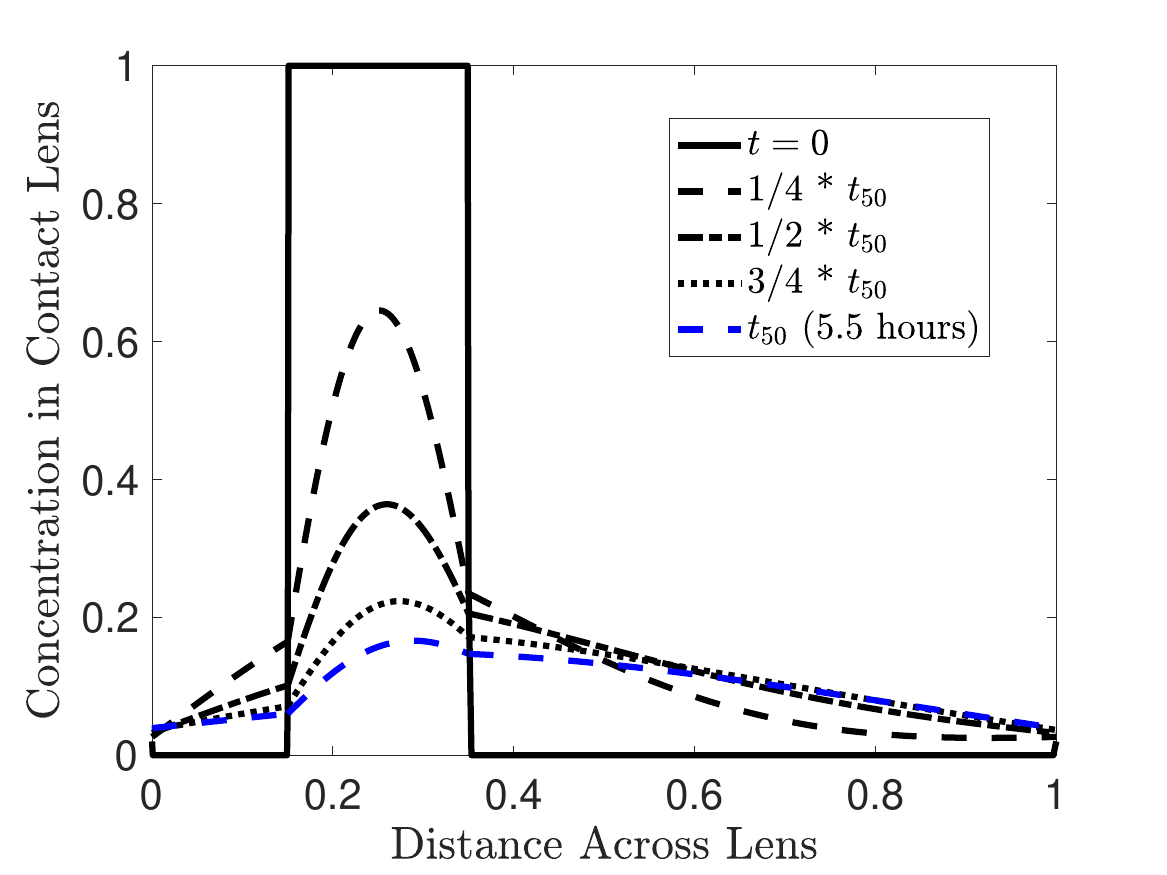}
\includegraphics[width=5.9cm]{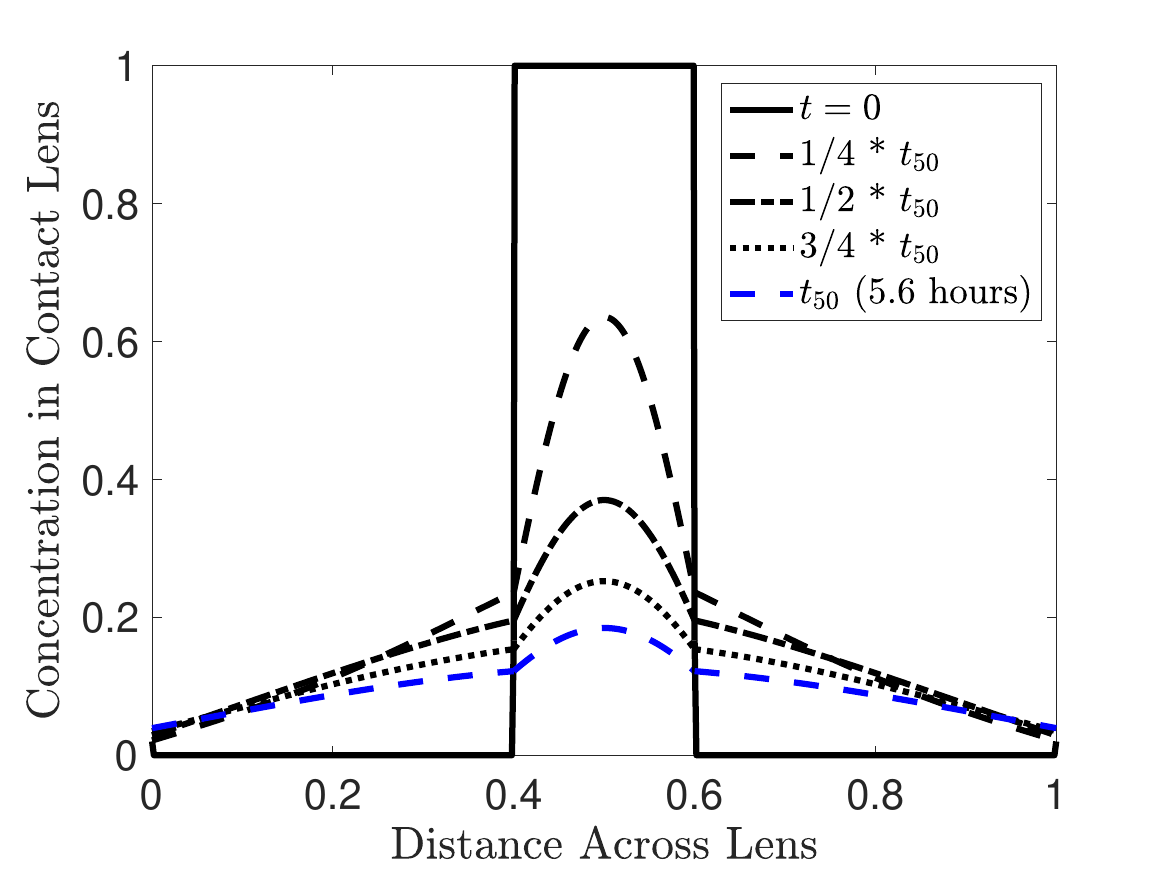}
\caption{
Blister pack setting: contact lens drug concentration profiles for a range of equally-spaced time values, $t \in [0,t_{50}]$. 
The upper, middle, and lower rows use ${\cal D}=0.002$,
 ${\cal D}=0.02$, 
and  ${\cal D} = 0.1$.  
The left and right columns use $\bar{H}_{\rm mid}=0.25$ and  
$\bar{H}_{\rm mid} =0.5$.  
}
\label{fig_CL_Profiles_Hmid_varies_blister}
\end{center}
\end{figure}

 Figure~\ref{fig_CL_Profiles_Hmid_varies_blister} shows CL drug concentration profiles for several parameter sets that vary the diffusion coefficient ratio $\cal{D}$ and polymer centerline positioning $\bar{H}_{\rm mid}$.
The top, middle, and bottom rows use
${\cal D}=0.002$, ${\cal D}=0.02$, and ${\cal D}=0.1$, respectively.
The left and right columns use
$\bar{H}_{\rm mid}=0.25$ and $\bar{H}_{\rm mid}=0.5$, respectively. The $\bar{H}_{\rm mid} = 0.75$ case is symmetric to that of $\bar{H}_{\rm mid}=0.25$ when $\bar{h}_{\rm pre} = \bar{h}_{\rm post}$ and thus is omitted.
 The $t_{50}$ times are indicated in the legends and in Table \ref{table:t50_blister}. Nonzero ``pre"- and ``post"-lens concentrations are maintained at all times due to the blister pack drug loading. Varying $\bar{H}_{\rm mid}$ has little effect on $t_{50}$ for all values of $\cal{D}$, as seen by comparing the values across each row of Table~\ref{table:t50_blister}, but increasing $\cal{D}$ has a strong negative effect on $t_{50}$, seen by comparing the values down each column of Table~\ref{table:t50_blister}. The strong, negative influence of $\cal{D}$ on $t_{50}$ is consistent with our observations in the vial and eye wear settings. Figure~\ref{fig_CL_Profiles_Hmid_varies_blister} exhibits qualitative similarities to the left and middle columns of the eye wear results in Figure~\ref{fig_CL_Profiles_Hmid_variesP1}. A discrepancy is that symmetry is observed in the drug concentration profile in the blister pack case for the centered drug-polymer film ($\bar{H}_{\rm mid} = 0.5$), which is not true for the eye wear case due to the different properties assigned to the pre- and post-lens~tear~films. 

\vspace{-3mm}

\begin{table}[h!]
\begin{center}
\begin{tabular}{ll|cc}
  &    &  \multicolumn{2}{c}{$\bar{H}_{\rm mid}$}  \\
    &   & 0.25 & 0.5 \\ \hline
& $0.002$ 
&  $72.5$ hrs
&  $71.6$ hrs \\                
${\cal D}$
& $0.02$
&  $12.0$ hrs
&  $11.1$ hrs \\                 
& $0.1$ 
&  $5.48$ hrs
&  $5.64$ hrs\\
   \hline 
 \end{tabular}\\
\end{center}
\caption{Blister pack setting: dimensional ${t}_{50}$ values (in hours) 
predicted by the model for 
the results shown in Figure~\ref{fig_CL_Profiles_Hmid_varies_blister}. 
}
\label{table:t50_blister}
\vspace{-2mm}
\end{table}

\begin{figure}[h]
     \centering
     \subfloat[][Concentration in lens]{\includegraphics[width=0.47\linewidth]{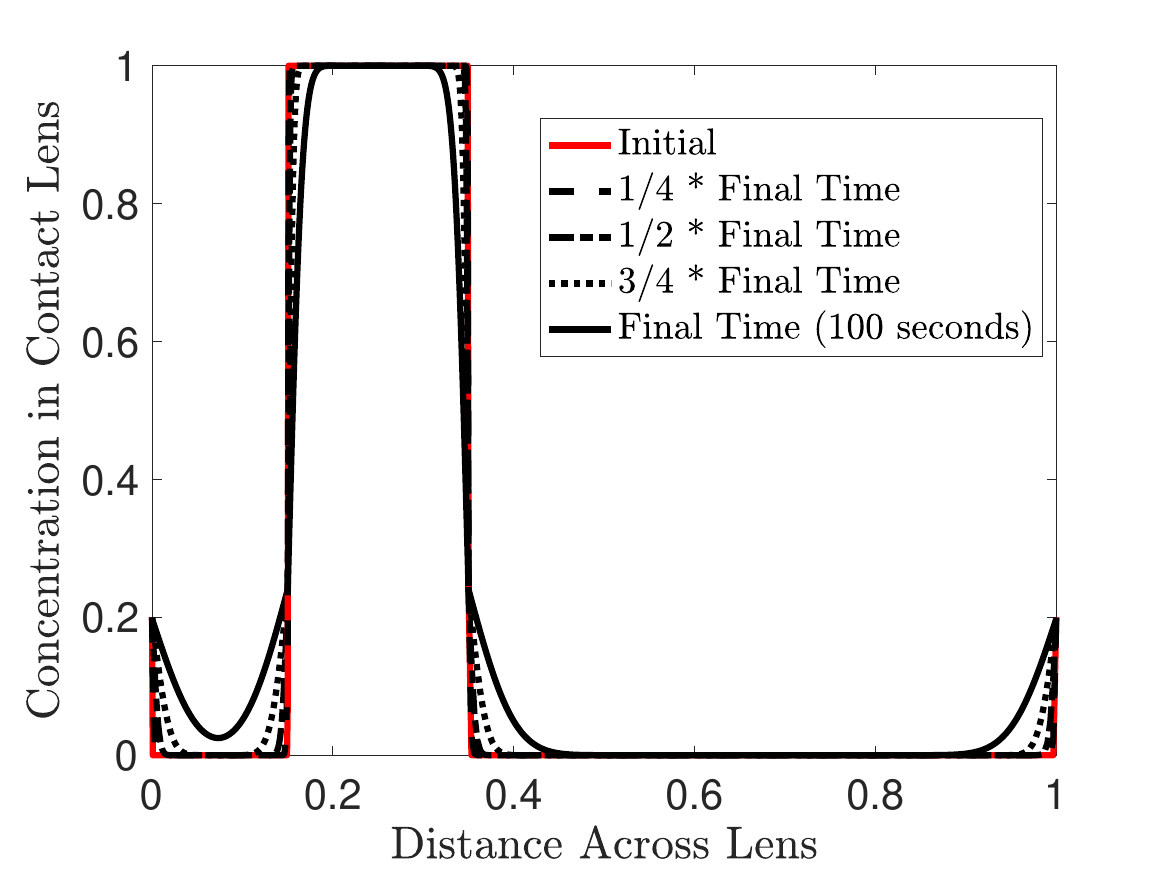}\label{fig:blister_short_cl}}
     \subfloat[][Blister pack drug mass]{\includegraphics[width=0.47\linewidth]{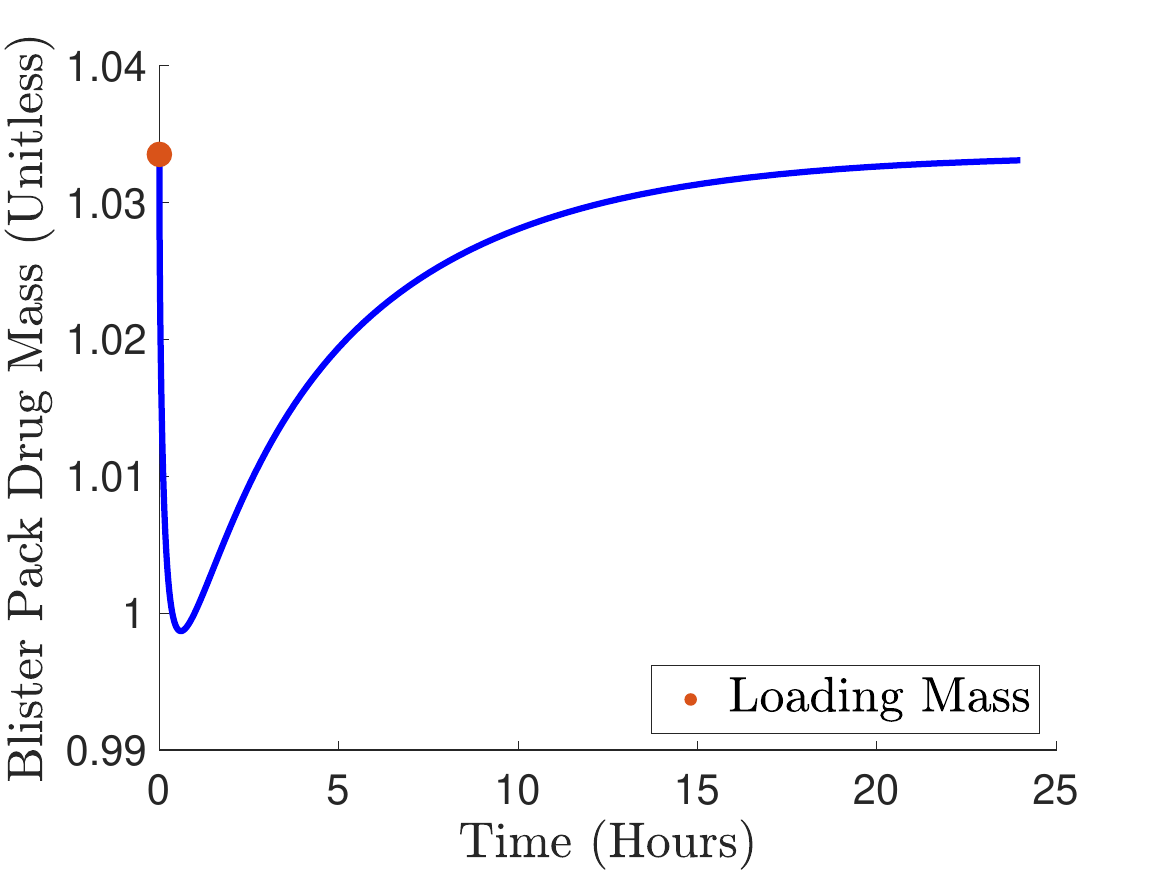}\label{fig:blister_short_drug}}
     \caption{Blister pack setting: (a) contact lens drug concentration profile  for a range of equally-spaced time values up to a short final time of 100 s and (b) blister pack drug mass $\bar{\mathcal{M}}_B$ over 24 hours. These results use $\bar{C}_B^{\rm load} = 0.1$,  $\mathcal{D} = 0.1$, and $\bar{H}_{\rm mid} = 0.25$. }
     \label{fig:blister_short_time}
 \end{figure}

Figure~\ref{fig:blister_short_cl} demonstrates an interesting and distinguishing feature of the blister pack environment: at short times (100 seconds, in this example), flux of drug is directed from the blister pack inwards to the initially drug-less hydrogel outer layers of the composite lens. Here, we have increased $\bar{C}_B^{\rm load}$ from 0.01 to 0.1 for a better visual demonstration, although the effect is also present for $\bar{C}_B^{\rm load} = 0.01$.  In the region between $\bar{z} = 0$ and $\bar{z} = 0.15$, there is a combination of drug diffusing into the hydrogel from both the blister pack and the drug-polymer film. In the region from $\bar{z} = 0.35$ to $\bar{z} = 1$, some drug has diffused out of the drug-polymer film and into the hydrogel by 100 seconds, and some drug has diffused into the hydrogel from the blister pack, but there is a region of the hydrogel that remains with effectively zero concentration.
Using the same parameters to simulate 24 hours in Figure~\ref{fig:blister_short_drug}, we visualize the blister pack drug mass over time. 
Within the first hour, the slope of the blister pack drug mass changes from negative to positive, indicating that diffusion of drug from the blister pack to the lens has stopped, and drug has started diffusing back from the lens into the blister pack. The blister pack drug mass approaches its initial value as time approaches~24~hours.

\begin{figure}[h!]
     \centering
     \subfloat[][One week, $\bar{C}_B^{\rm load} \approx 0.0328$]{\includegraphics[width=0.42\linewidth]{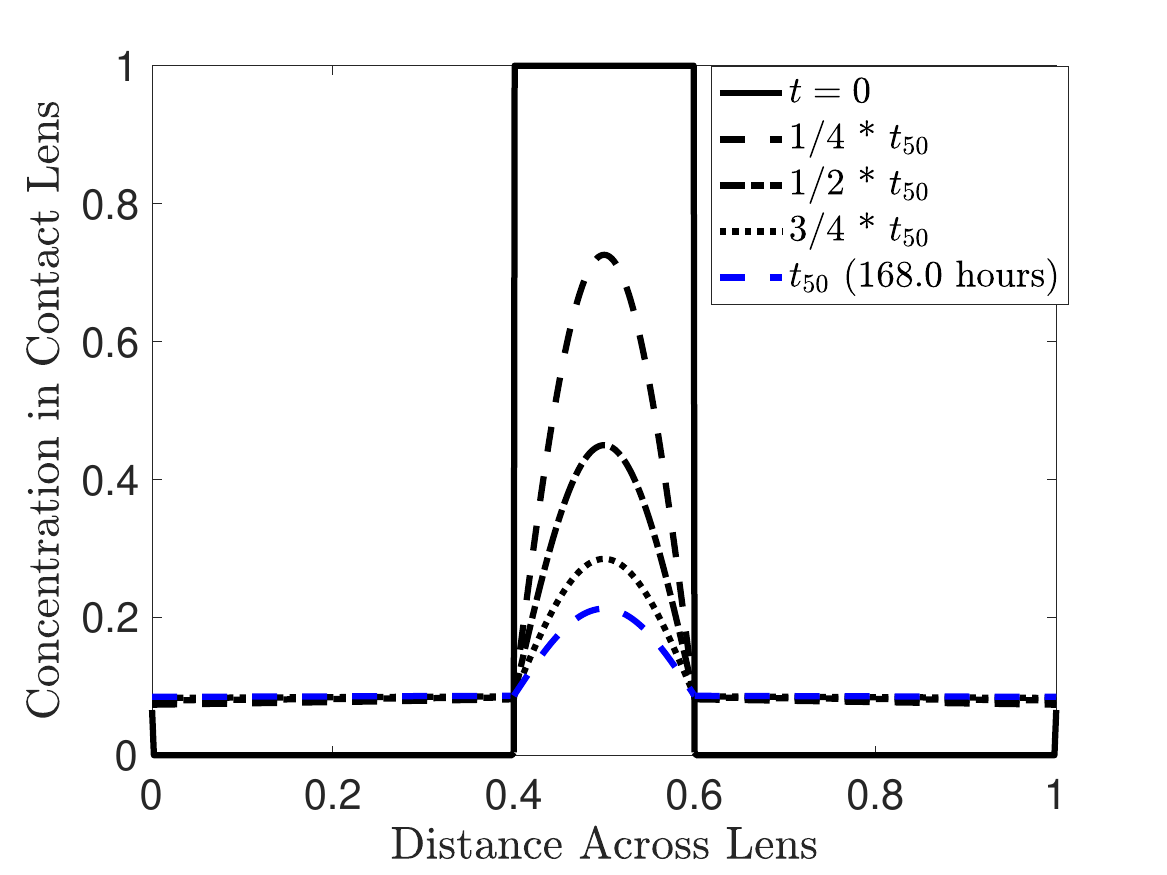} \label{fig:storage_week}}
     \subfloat[][One month, $\bar{C}_B^{\rm load} \approx 0.0403$]{\includegraphics[width=0.42\linewidth]{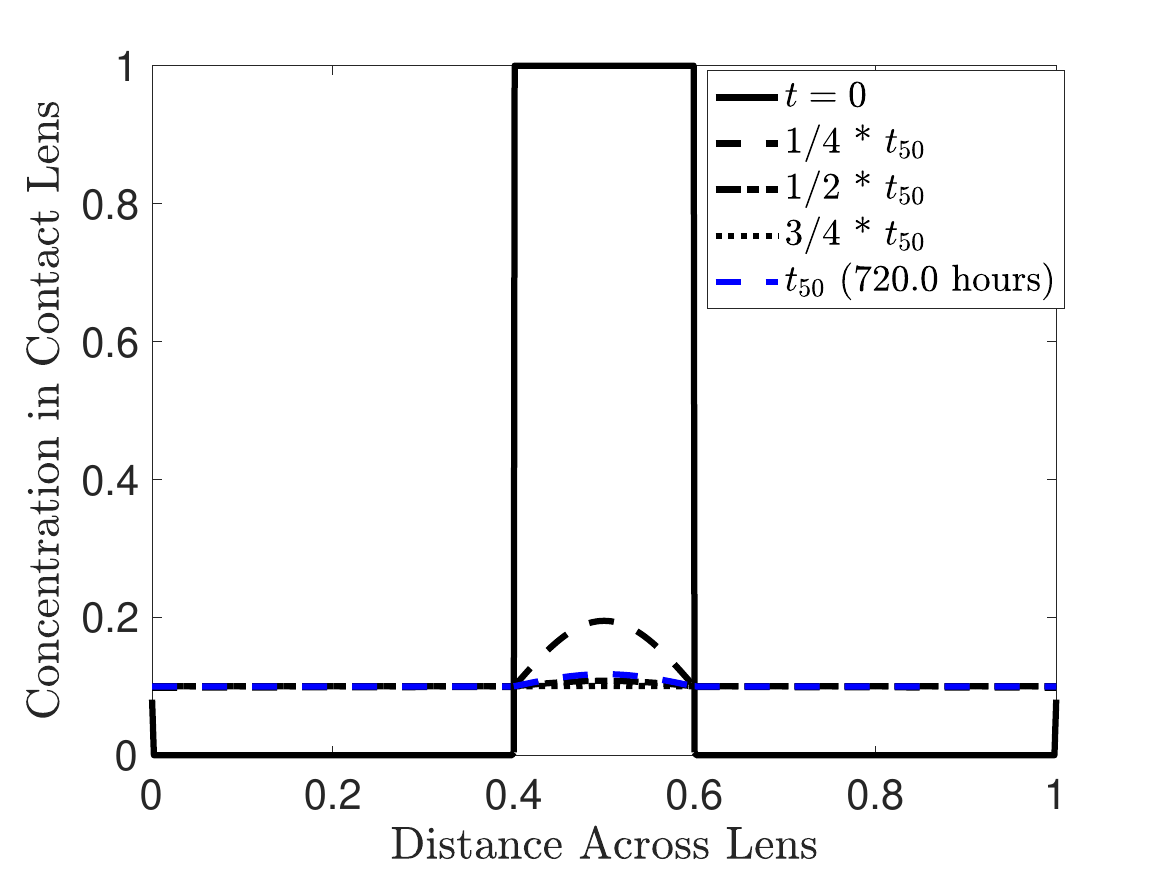}\label{fig:storage_month}}
     \caption{Blister pack setting: contact lens drug concentration profile for a range of equally-spaced time values, $t \in [0, t_{50}]$. These results use  $\mathcal{D} = 0.002$ and $\bar{H}_{\rm mid} = 0.5$.}
     \label{fig:storage_times}
 \end{figure}

 We now address an important storage question: What blister pack settings are required to preserve a sufficient amount of drug in the composite lens for a reasonable amount of time? We use one week (7 days) and one month (30 days) as time metrics. We   use $\mathcal{D} = 0.002$ to correspond to an experimental composite lens set-up \cite{ross2019topical}, fix $\bar{H}_{\rm mid} = 0.5$, and vary $\bar{C}_B^{\rm load}$, the  initial relative blister pack concentration, as a parameter that can be tuned relative to the drug-polymer film, which we assume to be fixed. By hand-tuning $\bar{C}_B^{\rm load}$, we find that $\bar{C}_B^{\rm load} \approx 0.0328$ (e.g.~about 3\% of the originally-loaded drug concentration in the polymer layer) maintains 50\% of drug in the composite lens after a week, and $\bar{C}_B^{\rm load} \approx 0.0403$ maintains 50\%  after a month. Figure~\ref{fig:storage_times} shows the drug concentration for both cases. The profile is nearly uniform in the hydrogel by one week (Figure~\ref{fig:storage_week}), with a small build-up of drug still preserved in the drug-polymer film; by a month, nearly the entire lens has uniform concentration (Figure~\ref{fig:storage_month}). 

\begin{figure}[h!]
     \centering
     \subfloat[][Concentration in lens]{\includegraphics[width=0.47\linewidth]{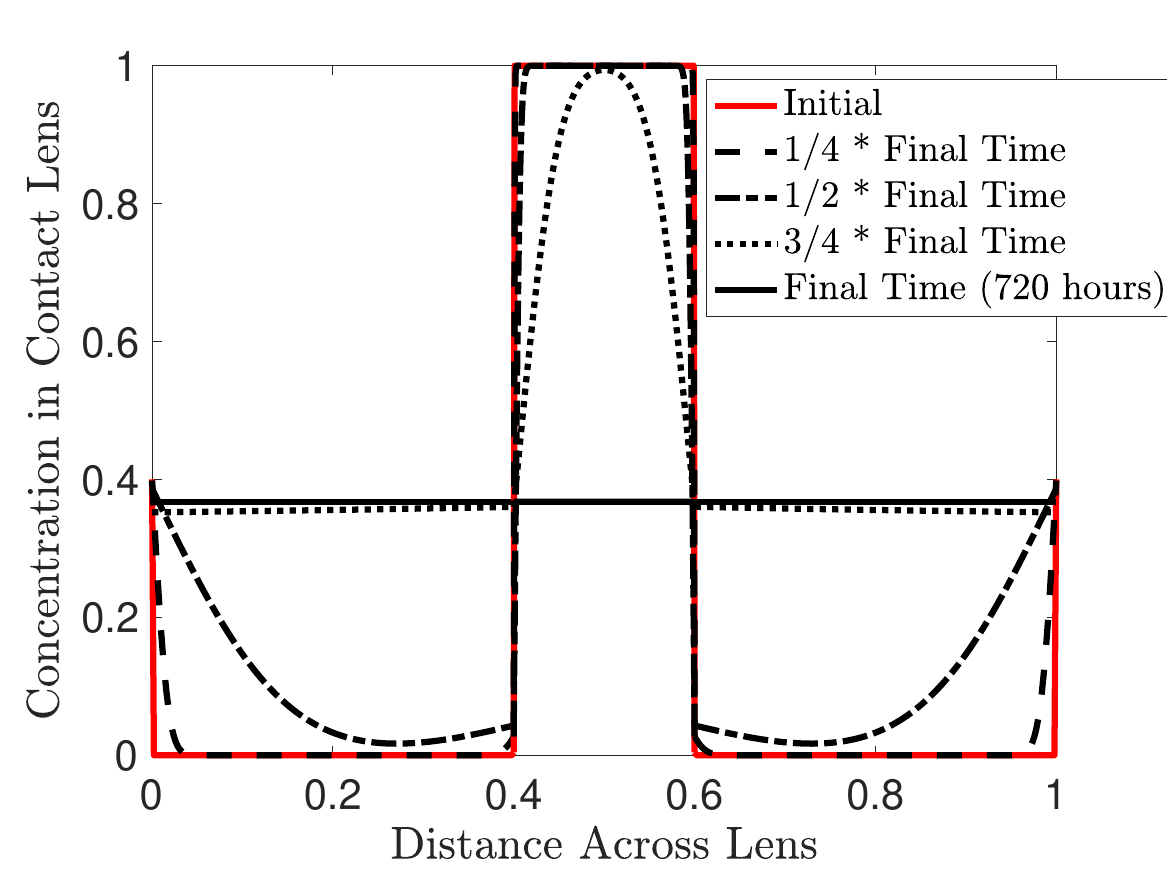}\label{fig:blister_large_CB_cl}} \hspace{1mm}
     \subfloat[][Blister pack drug mass]{\includegraphics[width=0.47\linewidth]{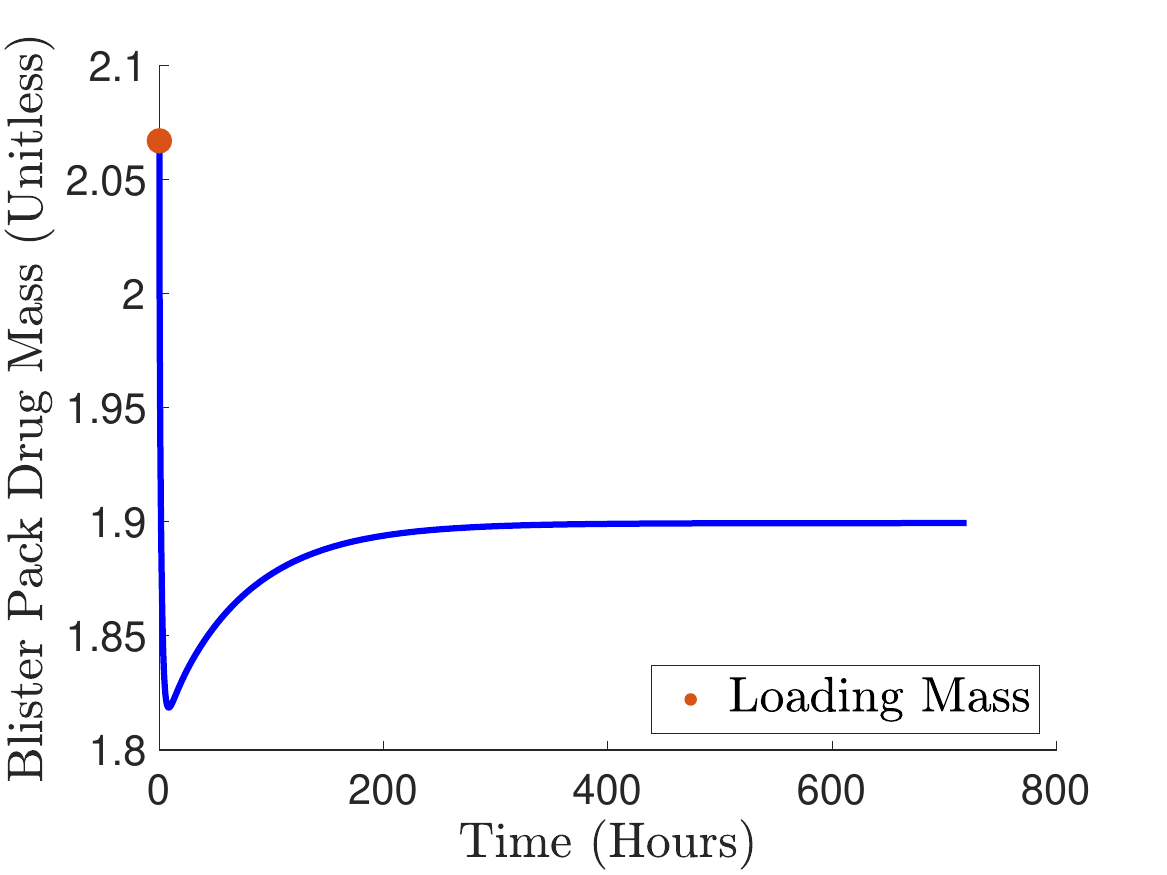}\label{fig:blister_large_CB_drug}}
     \caption{ Blister pack setting: (a) contact lens drug concentration profile  and (b) blister pack drug mass $\bar{\mathcal{M}}_B$ for a range of equally-spaced time values up to a month (30 days, 720 hours). These results use $\bar{C}_B^{\rm load} = 0.2$,  $\mathcal{D} = 0.002$, and $\bar{H}_{\rm mid} = 0.5$. }
     \label{fig:blister_large_CB}
 \end{figure}

 In Figure~\ref{fig:blister_large_CB} we investigate the effect of using a much larger $\bar{C}_B^{\rm load}$ than our nominal value of 0.01 and simulate dynamics over 30 days. Using $\bar{C}_B^{\rm load} = 0.2$,  the concentration in the lens approaches a uniform level just below a dimensionless value of 0.4 (Figure~\ref{fig:blister_large_CB_cl}), roughly double the starting concentration in the blister pack. In Figure~\ref{fig:blister_large_CB_drug}, $t_{50}$ is not  reached within 30 days, as there is a net transfer of drug mass into the lens from the blister pack since $\bar{\mathcal{M}}_B$ levels off at a value smaller than its initial condition. This suggests that a combination approach may be taken for storage and extended release purposes: if a composite lens is made such that the drug-polymer film has a much smaller diffusion coefficient than that of the surrounding hydrogel, and the lens is placed in a blister pack of sufficient concentration, a uniform long-term storage condition may be reached. This is still advantageous over simply soaking a conventional lens in the same concentration in a blister pack, because even for a uniform loading, due to the small $\mathcal{D}$, the drug-polymer film will release drug at a much slower rate into the hydrogel and thereby into the tear film. In fact, zero-order release kinetics may be attainable from this configuration.

 \section{Discussion}
 \label{sec:discussion}

\subsection{Comparison with drug loading in conventional contact lenses}

\begin{figure}
     \centering
     \subfloat[][Final time: 10 hours]{\includegraphics[width=0.48\linewidth]{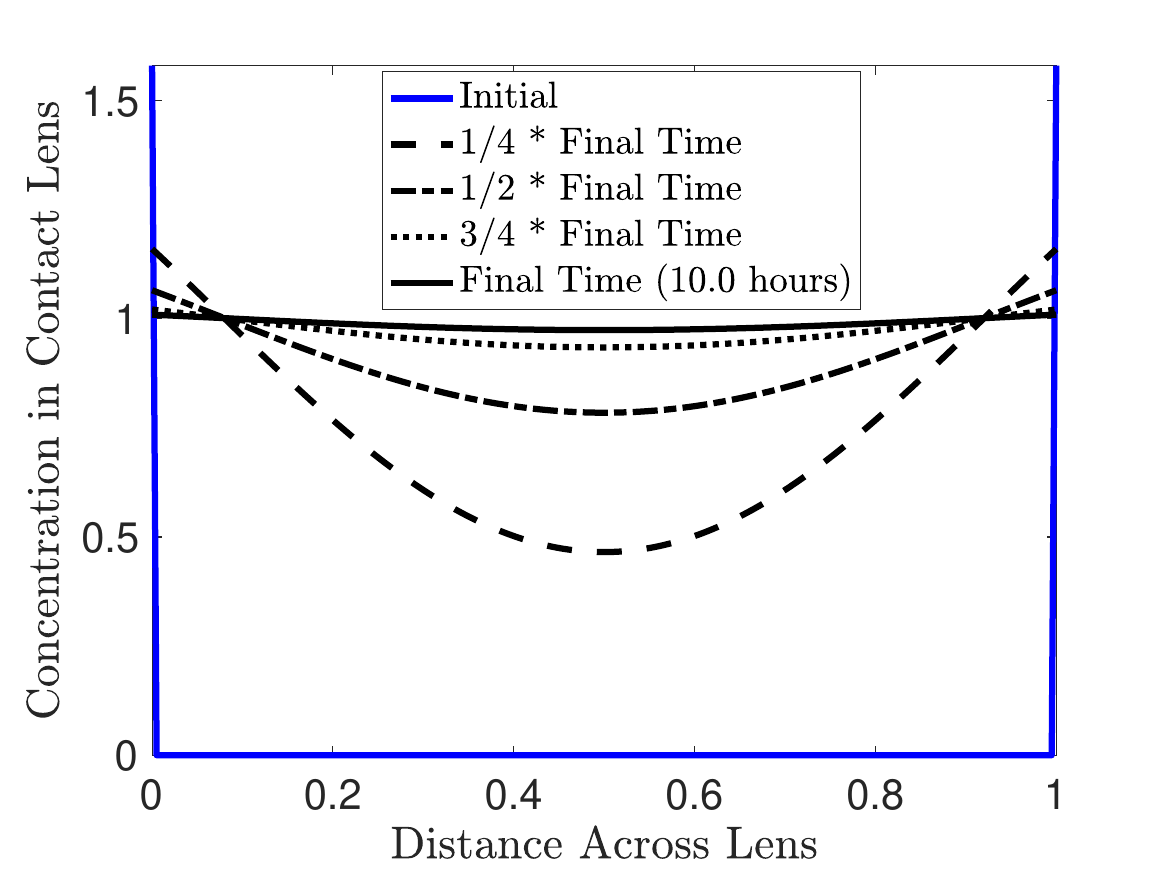}
     \label{fig:soak_10hr}}
      \subfloat[][Final time: 1 hour]{\includegraphics[width=0.48\linewidth]{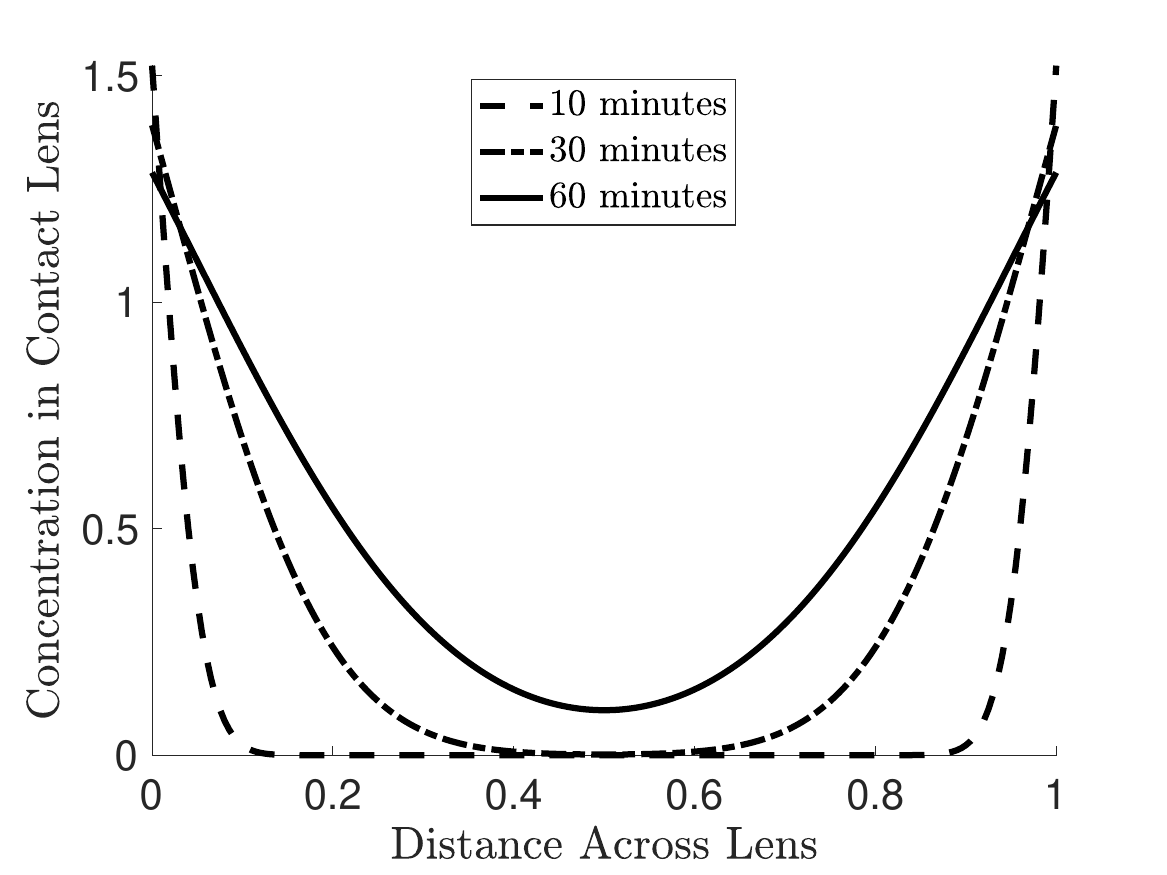}
      \label{fig:soak_1hr}}
     \caption{Vial soaking/uptake setting: contact lens drug concentration profile. (a): for a range of equally-spaced time values up to 10 hours, (b): for 10, 30, and 60 minutes for comparison with Figure 2(a) of \cite{Liu_etal_2013}. These results simulate a conventional lens.}
     \label{fig:soak}
 \end{figure}

 Using ideas from Section \ref{sec:vial_standard},  in Figure~\ref{fig:soak_10hr} we simulate the soak procedure whereby a conventional lens is loaded with drug in a blister pack setting. The final time is chosen so that the lens is nearly uniformly loaded with drug, but drug is still diffusing in from the blister pack. In Figure~\ref{fig:soak_1hr} we simulate the soak procedure for shorter times corresponding to those in Figure 2(a) of Liu \textit{et al}. \cite{Liu_etal_2013}. In their work \cite{Liu_etal_2013}, the authors compare transient intensity for the absorption of a molecular tracer in a gel slab with series expansion solutions to the diffusion equation (see their equations (1)--(4)). They report a diffusion coefficient of $D = 6.75 \times 10^{-12}$ m$^2$ s$^{-1}$ for the tracer used, and we estimate a gel slab thickness of 800 $\mu$m from their Figure 2(a), yielding a diffusive time scale of $H_{\rm gel}^2/D_{\rm tracer} \approx 26.3$ hrs, about 25\% shorter than ours ($H_{\rm cl}^2/D_2 \approx 35.7$ hours). Thus, we expect quicker uptake of the tracer than the drug in our simulation, which is consistent with the result shown in Figure~\ref{fig:soak_1hr}. At the 10 minute mark, a large portion ($\sim$80\%) of the lens remains at zero concentration, with uptake beginning only at the edges, whereas in Figure 2(a) of \cite{Liu_etal_2013}, only $\sim$25\% of the gel remains at zero intensity at 10 minutes. 

\subsection{Comparison with Gudnason \textit{et al.} (2017, 2018) and  Pimenta \textit{et al}. (2016)}
\label{sec-model_discussion}

Our vial setting composite-lens model shares similarities with the work of Pimenta \textit{et al}. \cite{pimenta2016diffusion} and Gudnason \textit{et al.} \cite{gudnason2017numerical,gudnason2018numerical}.  As noted in Section~\ref{sec:vial_model}, one model difference relates to the conditions applied at the internal boundaries.
In particular, our boundary conditions \eqref{eq:C_continuity_at_H1} at $z=H_1$ and \eqref{eq:C_continuity_at_H2} at $z=H_2$ can be viewed as limiting cases of similar ones in Pimenta \textit{et al}. \cite{pimenta2016diffusion} and Gudnason \textit{et al.} \cite{gudnason2017numerical,gudnason2018numerical}.
In our notation, the boundary condition (5) of Gudnason \textit{et al.} \cite{gudnason2017numerical} applied at our internal boundaries $z=H_1$ and $z=H_2$ reads
\begin{eqnarray}
\label{eq:general_bc1}
-D_2 \frac{\partial C_2^{\rm post}}{\partial z} (H_1,t) 
 = -D_1 \frac{\partial C_1}{\partial z} (H_1,t) & = & \alpha \Big(C_2^{\rm post}(H_1,t) - k_{\rm int} C_1(H_1,t)\Big),\\
 \label{eq:general_bc2}
 -D_1 \frac{\partial C_1}{\partial z} (H_2,t) = 
 -D_2 \frac{\partial C_2^{\rm pre}}{\partial z} (H_2,t) & = & \alpha \Big(C_1(H_2,t) - k_{\rm int} C_2^{\rm pre}(H_2,t)\Big), 
\end{eqnarray}
where $\alpha$ is a mass transfer rate coefficient (units of length per time) and $k_{\rm int}$ is a (dimensionless) partition coefficient associated with these internal boundaries; to simplify, 
we assume the same values of $\alpha$ and 
$k_{\rm int}$ apply at both interfaces.\footnote{
These boundary conditions, with $k_{\rm int}=1$, correspond to equations (3.6) of Pimenta \textit{et al}. \cite{pimenta2016diffusion}. We note a typographical error of a missing sign in their versions.}
Our model used simplifications corresponding to $k_{\rm int}=1$ (partition coefficients at internal layers set to unity, an assumption also made by Pimenta \textit{et al}. \cite{pimenta2016diffusion}), and the formal
limit $\alpha \gg D_1/H_{\rm cl}$, in which case
these general internal boundary conditions recover continuity of the concentrations at the internal boundaries, $z=H_1$ and $z=H_2$.   This corresponds to 
$\alpha \gg 5 \times 10^{-12}$m~s$^{-1}$ for parameters values used in our study.\footnote{Pimenta \textit{et al}. \cite{pimenta2016diffusion} report a significant role of the parameter $\alpha$ in their predictions, but give $\alpha$ in the range $[0.01, 0.5]$ without units and refer to $\alpha=1$ as the ``continuous interface'' case.   Based on the absence of units,  
we are unable to assess 
their value of the dimensionless quantity $\alpha H_{\rm cl}/D_1$ and therefore hesitate to make any quantitative comparisons with our results.   
}

As discussed in a slightly more general setting by Gudnason \textit{et al}. \cite{gudnason2017numerical}, one interpretation of the boundary conditions~(\ref{eq:general_bc1}) and~(\ref{eq:general_bc2}) is that they represent the presence of a thin intermediate layer, with thickness $\ell$ and diffusion coefficient $D_{\ell}$, separating (in our context) the hydrogel and polymer layers.  The concentration profile through a sufficiently thin intermediate layer can be approximated as linear, and the corresponding diffusive flux is $D_{\ell} \Delta C /\ell$, where $\Delta C$ is the jump in concentration across this thin layer, e.g.,~$\Delta C = C_2^{\rm post}(H_1,t) -  C_1(H_1,t)$ for the simplest case at $z=H_1$ if $k_{\rm int}=1$.  In this thin intermediate layer view, the mass transfer rate coefficient, $\alpha$, can be interpreted as the ratio of the thin layer diffusion coefficient to layer thickness: $\alpha = D_{\ell} /\ell$.

Gudnason \textit{et al}. \cite{gudnason2017numerical} make another important observation regarding diffusion in layered systems. 
If one considers the polymer layer with an adjacent thin layer 
with diffusion coefficient $D_{\ell}$ and thickness $\ell$,  
in the context of an equilibrium argument, it is possible to identify an effective diffusion coefficient, $D_1^{\rm eff}$, for this two layer system.
The form of this effective diffusion coefficient is a layer-width-weighted harmonic average of the polymer diffusivity $D_1$ and the thin layer diffusivity $D_{\ell}$, given by 
\begin{eqnarray}
\frac{H_{\rm total}}{D_1^{\rm eff}} & = & \frac{H_2 - H_1}{D_1} + \frac{\ell}{D_{\ell}} = \frac{H_2 - H_1}{D_1} + \frac{1}{\alpha}.
\end{eqnarray}
Here, $H_{\rm total} = H_2 - H_1 + \ell$ denotes the total thickness of the two layers, and in the last expression we have used $\alpha = D_{\rm \ell}/\ell$.  
Both the polymer layer diffusion coefficient $D_1$ and the mass transfer coefficient $\alpha$ contribute in very similar ways to this effective diffusivity.
Importantly, this expression is based on an equilibrium argument, and although similar relationships have been derived for time-dependent diffusion in layered systems (e.g., de la Cruz \& Gurevich \cite{delacruz1995layers}), in general this formula need not hold in the context of the present, transient, dynamics.  
That said, it gives a concrete example of the comparable roles played by $D_1$ and $\alpha$.
Gudnason \textit{et al.} \cite{gudnason2017numerical} note and illustrate the difficulty in distinguishing the role of 
$D_1$ and 
$\alpha$ (see their Section 4.1).

Given the overlapping roles played by polymer layer diffusivity $D_1$ and mass transfer rate coefficient $\alpha$, we chose to eliminate $\alpha$ from consideration and characterize the drug release from the polymer layer via the resulting continuous concentration boundary condition, and 
an ``effective'' diffusion coefficient of the polymer layer (in view of the above discussion).   This made sense in our comparison with the Ross \textit{et al.} \cite{ross2019topical} data, as we lacked access to independent measurements of $D_1$ and $\alpha$.
In contrast, the inner and outer layers in the experiments of Pimenta \textit{et al.} \cite{pimenta2016diffusion} were of the same material ($D_1=D_2$ in our notation), and so knowing their diffusion coefficient  independently, 
$\alpha$ could be estimated by fitting to their cumulative release curves.

With the above differences between our model and that of Pimenta \textit{et al}. \cite{pimenta2016diffusion} in mind, we  make observations of their experimental results and model predictions.  
In their experiments \cite{pimenta2016diffusion}, 
all three layers were composed of the same material and
the drug was loaded only in the inner layer.
Their Figure 6 
shows a slower release in the three-layer ``coated'' lens compared to the standard ``single'' lens (qualitatively similar to the trends observed in our Figure~\ref{fig:base_case}b with the Ross \textit{et al.} \cite{ross2019topical} data).  
The Pimenta \textit{et al.} model parameters used to compare to their experiments were $D_1=D_2$ and $\alpha^{-1} \neq 0$.
Our model comparisons that reveal a similar slower cumulative drug release trend use $D_1 < D_2$ and $\alpha^{-1} =0$ (e.g., ${\cal D}=0.002$ in the Ross \textit{et al.} comparison).

Pimenta \textit{et al}. \cite{pimenta2016diffusion} reported an additional sequence of experiments in a ``coated'' lens configuration for a fixed inner thickness of $0.4$ mm and three outer layer thicknesses: $0.2$, $0.4$, and $0.6$ mm.  
These results (see their Figure 7) 
show:
\begin{itemize}
    \item A -- fastest drug release: inner layer ($0.4$ mm) and thin outer layer ($0.2$ mm) corresponding to our
    $\bar{H}_{\rm mid}=0.5$, $\Delta \bar{H}=1/2$ and $H_{\rm cl}=0.8$ mm.
    \item B -- intermediate drug release: inner layer ($0.4$ mm) and medium outer layer ($0.4$ mm) corresponding to our
    $\bar{H}_{\rm mid}=0.5$, $\Delta \bar{H}=1/3$ and $H_{\rm cl}=1.2$ mm.
    \item C -- slowest drug release: inner layer ($0.4$ mm) and thick outer layer ($0.6$ mm) corresponding to our
    $\bar{H}_{\rm mid}=0.5$, $\Delta \bar{H}=1/4$ and $H_{\rm cl}=1.6$ mm.
\end{itemize}
Our prediction in \eqref{eq:t50_formula} for the 50\% release time metric, $t_{50} = \bar{t}_{50}(\bar{H}_{\rm mid}, \Delta \bar{H}, {\cal D}) (H_{\rm cl}^2/D_2)$,
shows that for fixed $D_2$, ${\cal D}$, and $\bar{H}_{\rm mid}$, changes in both $\Delta \bar{H}$ and $H_{\rm cl}$ influence $t_{50}$.
The dependence of $t_{50}$ on $H_{\rm cl}^2$ is explicit (and positively correlated), while that on the dimensionless inner layer thickness $\Delta \bar{H}$ is significantly more nuanced; without specific knowledge of the value of ${\cal D}$, we cannot predict if the influence of simultaneously changing $\Delta \bar{H}$ enhances or opposes the slowing trend in these experiments.

Overall trends observed by Pimenta \textit{et al}. \cite{pimenta2016diffusion} show a clear slowing of drug
release dynamics (e.g.~increase in $t_{50}$) with a decrease in the diffusion coefficient or with a decrease in the mass transfer coefficient.  These are trends consistent with Gudnason \textit{et al.} \cite{gudnason2017numerical}.
In our work, the analog role of the hydrogel diffusion coefficient, $D_2$, and the diffusion ratio ${\cal D}$ are consistent with these trends.
Other studies, such as Ferreira \textit{et al}. \cite{ferreira2011mathematical}, identified similar trends for their ``therapeutic effect'' metric, though they modeled a single, heterogeneous CL rather than a multi-layered system with homogeneous layers.

 \subsection{Limitations and extensions}

In our blister pack model simulations, we have used a relatively small blister pack volume (0.2 mL), but typical blister packs range from 1-3 mL. Differences in results are expected as the volume scales and the concentration is reduced accordingly.
  In terms of extensions, one could contrast $t_{50}$ for the composite lens with the conventional lens in the eye model setting.  
Further, one could significantly increase the partition coefficient $k$ on one or both sides of the lens or introduce mass transfer rate coefficients as proxies for the effect of a lens coating. Other performance metrics, such as targeted early time or sustained drug release kinetics, could be more thoroughly investigated alongside the 50\% therapeutic release time that was a primary focus in the present study.  In the future, the blister pack and eye wear with blinking models could be coupled to simulate single-day and multi-day wear to study long-term drug delivery conditions. This extension is a clear next step, as we report $t_{50}$ values in the eye wear setting for $\mathcal{D} = 0.002$ of up to 70 hours, which is long beyond actual continuous wear with blinking. In reality, continuous wear duration might be at most 16 hours, followed by storage (multi-day use), or obtaining a fresh lens (single-day use). Other lens properties could be explicitly characterized by our models, such as oxygen permeability and binding and disassociating of drug compounds. Our modeling efforts may be extendable to describe dissolvable wafer ocular inserts, which are another drug delivery method with the potential for zero-order release kinetics \cite{saettone1995ocular,mariz2022ocular}.

\section{Conclusions}
 \label{sec:conclusion}

 In this work, we model a composite contact lens (CL) in the vial, eye wear with blinking, and blister pack settings, and investigate the effect of adjustable  lens material and geometric properties on drug release kinetics via the 50\% therapeutic release time, $t_{50}$. 
 
 In the vial setting, we identified an expression 
 for $t_{50}$ as the product of the diffusion time scale $H_{\rm cl}^2/D_2$ and a dimensionless quantity, 
$\bar{t}_{50}(\bar{H}_{\rm mid}, \Delta \bar{H}, {\cal D})$. The latter is a function of drug-polymer film midline position $\bar{H}_{\rm mid}$, dimensionless polymer thickness $\Delta \bar{H}$, and polymer to hydrogel diffusion coefficient ratio ${\cal D}$.
 Our model predictions show that 
 $\cal{D}$ strongly negatively affects $\bar{t}_{50}$, while 
 $\bar{H}_{\rm mid}$ and
 $\Delta \bar{H}$ interact non-trivially. If $\cal{D}$ is small, $\Delta \bar{H}$ strongly positively influences $\bar{t}_{50}$, but 
 this trend is exactly reversed when ${\cal D}=1$.
 Also, when ${\cal D}$ is relatively large, moving $\bar{H}_{\rm mid}$ off the centerline $\bar{H}_{\rm mid}=0.5$ decreases the predicted $\bar{t}_{50}$, but for smaller ${\cal D}$ this trend can be reversed. 
 
 In the eye wear with blinking setting, in general, the CL drug concentration is asymmetrical even with a symmetrically-placed polymer layer  ($\bar{H}_{\rm mid}=0.5$); the details depend on 
 $\mathcal{D}$ and processes that influence pre- and post-lens drug clearance rates 
 (e.g.,~blinking and corneal absorption). We demonstrated that $t_{50}$ could increase or decrease as the polymer midline 
 $\bar{H}_{\rm mid}$ is moved posterior to anterior, but
 the observed trend depended on 
 ${\cal D}$ and the relative pre- and post-lens drug clearance rates. 
 For example, for small ${\cal D}$ and relatively slow post-lens drug loss 
 compared to the pre-lens tear film, $t_{50}$ could be increased by moving $\bar{H}_{\rm mid}$ closer to the pre-lens side and trapping drug on the post-lens side (e.g.~see Figure~\ref{fig_t50_blink}).
 We also gave an anterior--posterior invariance argument that recognized the opposite possibility: for small ${\cal D}$ and relatively fast post-lens drug loss compared to the pre-lens, $t_{50}$ could be increased by moving $\bar{H}_{\rm mid}$ closer to the post-lens side and trapping drug on the pre-lens side.

 In the blister pack storage setting, a composite lens pre-loaded with drug only in the interior polymer layer will lose a portion of the drug into the adjacent hydrogel layers.  Stored in a drug-loaded blister pack solution,
 however, the drug loss from the lens can be reduced or even eliminated.
 We showed that relatively small blister pack loading concentrations, e.g., less than 5\%
 of that of the drug-polymer film, are required to sustain 50\% of the drug in the lens for one month.

 There is a broad range of performance characteristics of the composite lens
for ophthalmic drug delivery during CL wear with blinking.  Rather than identifying any specific set of
optimal design parameters, the present study reveals that any such optimal design
depends not only on the clinically-desired drug delivery rates, but also on the lens wear conditions, perhaps measured in our model by parameters for blink-induced drug clearance rates, corneal and/or lid permeabilities, CL motion and fluid transport mechanisms in the post-lens tear film.  It is clear, however, that composite lens design over geometric parameters like $\bar{H}_{\rm mid}$ and $\Delta \bar{H}$, and material properties like ${\cal D}$, can be exploited to reach or at least move closer to specific drug delivery performance goals.

 Our work may aid in optimizing drug-eluting lens design to achieve desired release rates,
residence times, and zero-order release kinetics by identifying the dependence of $t_{50}$
on adjustable CL factors. Coupled with optimized blister pack storage conditions, this may be important for determining the feasibility of composite lenses as an alternate to commercially available drug-eluting conventional lenses  \cite{novack2023us}. This work may lead to a better
understanding of the mechanics of drug-eluting CLs.

\section*{Use of AI tools declaration}
The authors declare they have not used Artificial Intelligence (AI) tools in the creation of this article.

\section*{Conflict of interest}

The authors declare there is no conflict of interest.

\section*{Code availability}

Numerical simulations and analyses were conducted using MATLAB scripts will be made available at a sub-respository on \href{https://github.com/danielmanderson/}{https://github.com/danielmanderson/}.

\bibliographystyle{unsrt}
 
 \bibliography{CL_bib.bib}

\end{document}